\documentclass[a4paper,USenglish]{lipics-v2021}

\title{Oracular Byzantine Reliable Broadcast \\ \footnotesize{\normalfont{(Extended Version)}}}

\author{Martina Camaioni}{Ecole Polytechnique Fédérale de Lausanne (EPFL), Switzerland}{}{}{}
\author{Rachid Guerraoui}{Ecole Polytechnique Fédérale de Lausanne (EPFL), Switzerland}{}{}{}
\author{Matteo Monti}{Ecole Polytechnique Fédérale de Lausanne (EPFL), Switzerland}{}{}{}
\author{Manuel Vidigueira}{Ecole Polytechnique Fédérale de Lausanne (EPFL), Switzerland}{}{}{Supported in part by the Hasler Foundation (\#21084).}

\nolinenumbers
\hideLIPIcs
    
\authorrunning{M. Camaioni, R. Guerraoui, M. Monti, M. Vidigueira}

\Copyright{Martina Camaioni, Rachid Guerraoui, Matteo Monti, and Manuel Vidigueira}

\keywords{Byzantine reliable broadcast, Good-case complexity, Amortized complexity, Batching}

\ccsdesc[500]{Theory of computation~Distributed algorithms}

\usepackage{amsthm}
\usepackage{amsmath}
\usepackage{amsfonts}
\usepackage{amssymb}
\usepackage{mathrsfs}
\usepackage{cancel}
\usepackage[overload]{empheq}

\usepackage{xspace}
\usepackage{xparse}
\usepackage{xifthen}
\usepackage{tabto}
\usepackage{scalerel}
\usepackage{stackengine}
\usepackage{lscape}
\usepackage{xstring}
\usepackage{afterpage}

\usepackage{graphicx}
\usepackage{algorithmicx}
\usepackage{algorithm} 
\usepackage{algpseudocode}
\usepackage{listings}
\usepackage{styles/listings-pseudo}
\usepackage[scaled=0.85]{beramono}
\usepackage[T1]{fontenc}

\NewDocumentCommand\miniparagraph{m}{\medskip\textbf{#1.}}

\NewDocumentCommand{\mayberp}{o}{\IfValueTF{#1}{{\rp{#1}}}{}}
\NewDocumentCommand{\maybeqp}{o}{\IfValueTF{#1}{{\qp{#1}}}{}}
\NewDocumentCommand{\maybecp}{o}{\IfValueTF{#1}{{\cp{#1}}}{}}

\NewDocumentCommand{\maybefoot}{o}{\IfValueTF{#1}{{_{#1}}}{}}
\NewDocumentCommand{\maybesup}{o}{\IfValueTF{#1}{{^{#1}}}{}}

\NewDocumentCommand\rp{m}{{\left(#1\right)}}
\NewDocumentCommand\qp{m}{{\left[#1\right]}}
\NewDocumentCommand\cp{m}{{\left\{#1\right\}}}
\NewDocumentCommand\ap{m}{{\left\langle#1\right\rangle}}

\NewDocumentCommand\abs{m}{{\left|#1\right|}}

\NewDocumentCommand\floor{m}{{\left\lfloor #1 \right\rfloor}}
\NewDocumentCommand\ceil{m}{{\left\lceil #1 \right\rceil}}

\DeclareMathOperator*{\append}{\frown}

\NewDocumentCommand\sign{o}{
    \IfValueTF{#1}
    {{\text{sign}\rp{#1}}}
    {{\text{sign}}}
}

\NewDocumentCommand\identity{o}{
	\IfValueTF{#1}
	{{\text{Id}\rp{#1}}}
	{{\text{Id}}}
}

\NewDocumentCommand\indicator{o}{
	\IfValueTF{#1}
	{{\text{I}\qp{#1}}}
	{{\text{I}}}
}

\NewDocumentCommand\true{}{\mathtt{true}}
\NewDocumentCommand\false{}{\mathtt{false}}

\stackMath
\NewDocumentCommand\pagoda{m}{
    \savestack{\tmpbox}{\stretchto{
        \scaleto{
            \scalerel*[\widthof{\ensuremath{#1}}]{\kern+1.5pt\bigwedge\kern+1.5pt}
            {\rule[-\textheight/2]{1ex}{\textheight}}
        }{\textheight}
    }{0.5ex}}
    \stackon[1pt]{#1}{\tmpbox}
}

\NewDocumentCommand\suchthat{}{\mid}

\NewDocumentCommand\powerset{om}{
    \IfValueTF{#1}
    {{\mathbb{P}^{#1}\rp{#2}}}
    {{\mathbb{P}\rp{#2}}}
}

\NewDocumentCommand\choice{o}{
	\IfValueTF{#1}
	{{\mathfrak{C}\rp{#1}}}
	{\mathfrak{C}}
}

\NewDocumentCommand\emptyseq{}{\cancel{\square}}

\NewDocumentCommand\sorted{m}{{#1^{\nearrow}}}
\NewDocumentCommand\sort{m}{{\mathcal{S}\rp{#1}}}

\NewDocumentCommand\mapdomain{m}{{\mathcal{D}\rp{#1}}}
\NewDocumentCommand\mapcodomain{m}{{\mathcal{C}\rp{#1}}}

\theoremstyle{definition}
    \newtheorem{assumption}{Assumption}

\makeatletter
\newcommand{\assumptiontag}[1]{
    \let\oldtheassumption\theassumption
    \renewcommand{\theassumption}{#1}
    \g@addto@macro\endassumption{
        \addtocounter{assumption}{-1}
        \global\let\theassumption\oldtheassumption
    }
}
\makeatother

\NewDocumentCommand\any{}{\_}
\NewDocumentCommand\tail{}{..}

\NewDocumentCommand\caseiff{}{&\text{iff}\;}
\NewDocumentCommand\caseotherwise{}{&\text{otherwise}}

\input{headers/algorithms}

\NewDocumentCommand\csb{}{Client-Server Byzantine Reliable Broadcast\xspace}
\NewDocumentCommand\csbprefix{}{CSB}
\NewDocumentCommand\csbshort{}{\csbprefix\xspace}

\NewDocumentCommand\csbalname{}{Draft\xspace}
\NewDocumentCommand\csbal{}{\textsf{\csbalname}\xspace}

\NewDocumentCommand\cl{}{\csbprefix~Client\xspace}
\NewDocumentCommand\clab{}{\csbprefix Client}
\NewDocumentCommand\clin{}{cl}

\NewDocumentCommand\bkab{}{\csbprefix Broker}
\NewDocumentCommand\bkin{}{bk}

\NewDocumentCommand\sr{}{\csbprefix~Server\xspace}
\NewDocumentCommand\srab{}{\csbprefix Server}
\NewDocumentCommand\srin{}{sr}

\NewDocumentCommand\dir{}{Directory\xspace}
\NewDocumentCommand\dirshort{}{DIR\xspace}

\NewDocumentCommand\diral{}{\textsf{Dibs}\xspace}

\NewDocumentCommand\dirab{}{Directory}
\NewDocumentCommand\dirin{}{dir}

\NewDocumentCommand\dirsrab{}{DirectoryServer}
\NewDocumentCommand\dirsrin{}{dsr}

\NewDocumentCommand\csbalbatchingwindow{}{b}

\NewDocumentCommand\processes{}{\Pi}
\NewDocumentCommand\correctprocesses{}{{\processes_C}}
\NewDocumentCommand\faultyprocesses{}{{\processes_F}}

\NewDocumentCommand\process{}{\pi}
\NewDocumentCommand\processsecondary{}{\rho}

\NewDocumentCommand\clients{}{X}
\NewDocumentCommand\client{}{\chi}
\NewDocumentCommand\clientcount{}{c}

\NewDocumentCommand\brokers{}{B}
\NewDocumentCommand\broker{}{\beta}
\NewDocumentCommand\brokercount{}{k}

\NewDocumentCommand\servers{}{\Sigma}
\NewDocumentCommand\server{}{\sigma}
\NewDocumentCommand\servercount{}{n}
\NewDocumentCommand\faultcount{}{f}

\NewDocumentCommand\contexts{}{\mathbb{C}}
\NewDocumentCommand\messages{}{\mathbb{M}}

\NewDocumentCommand\signatures{}{{\mathbb{S}^1}}
\NewDocumentCommand\multisignatures{}{{\mathbb{S}^+}}

\NewDocumentCommand\auniverse{}{\mathcal{U}}
\NewDocumentCommand\aroots{}{\mathcal{R}}
\NewDocumentCommand\aproofs{}{\mathcal{P}}

\NewDocumentCommand\arootsym{}{\rho}
\NewDocumentCommand\aroot{m}{{\arootsym\rp{#1}}}

\NewDocumentCommand\aproofsym{}{\zeta}
\NewDocumentCommand\aproof{mm}{{\aproofsym\rp{#1, #2}}}

\NewDocumentCommand\averifysym{}{\nu}
\NewDocumentCommand\averify{mmmm}{{\averifysym\rp{#1, #2, #3, #4}}}

\NewDocumentCommand\domains{}{\mathbb{D}}
\NewDocumentCommand\ids{}{\mathbb{I}}

\NewDocumentCommand\directoryrecordsym{}{\mathtt{D}}
\NewDocumentCommand\directoryrecord{mo}{
    \IfValueTF{#2}
    {{\directoryrecordsym_{#1}\rp{#2}}}
    {{\directoryrecordsym_{#1}}}
}
\NewDocumentCommand\directorymap{m}{{\directoryrecordsym\rp{#1}}}

\NewDocumentCommand\directorycert{}{{\directoryrecordsym^{\infty}}}

\NewDocumentCommand\joinsym{}{\mathtt{j}}
\NewDocumentCommand\join{mm}{{\joinsym\rp{#1, #2}}}

\NewDocumentCommand\compresssym{}{{\mathtt{c}}}
\NewDocumentCommand\compress{m}{{\compresssym\rp{#1}}}

\NewDocumentCommand\expandsym{}{{\mathtt{e}}}
\NewDocumentCommand\expand{m}{{\expandsym\rp{#1}}}

\NewDocumentCommand{\bpayloadssym}{}{\mathtt{P}}
\NewDocumentCommand{\bpayloads}{oomo}{{\bpayloadssym\maybefoot[#1]\mayberp[#2]\qp{#3}\maybeqp[#4]}}

\NewDocumentCommand{\bcontextssym}{}{\mathtt{C}}
\NewDocumentCommand{\bcontexts}{oomo}{{\bcontextssym\maybefoot[#1]\mayberp[#2]\qp{#3}\maybeqp[#4]}}

\NewDocumentCommand{\bmessagessym}{}{\mathtt{M}}
\NewDocumentCommand{\bmessages}{oomo}{{\bmessagessym\maybefoot[#1]\mayberp[#2]\qp{#3}\maybeqp[#4]}}

\NewDocumentCommand{\bleavessym}{}{\mathtt{L}}
\NewDocumentCommand{\bleaves}{oomo}{{\bleavessym\maybefoot[#1]\mayberp[#2]\qp{#3}\maybeqp[#4]}}

\NewDocumentCommand{\bsignaturessym}{}{\mathtt{S}}
\NewDocumentCommand{\bsignatures}{oomo}{{\bsignaturessym\maybefoot[#1]\mayberp[#2]\qp{#3}\maybeqp[#4]}}

\NewDocumentCommand{\bmultisignaturessym}{}{\mathtt{Q}}
\NewDocumentCommand{\bmultisignatures}{oomo}{{\bmultisignaturessym\maybefoot[#1]\mayberp[#2]\qp{#3}\maybeqp[#4]}}

\NewDocumentCommand{\bcommittosym}{}{\mathtt{Y}}
\NewDocumentCommand{\bcommitto}{oomo}{{\bcommittosym\maybefoot[#1]\mayberp[#2]\qp{#3}\maybeqp[#4]}}

\NewDocumentCommand{\bcommittablesym}{}{\mathtt{Z}}
\NewDocumentCommand{\bcommittable}{oomo}{{\bcommittablesym\maybefoot[#1]\mayberp[#2]\qp{#3}\maybeqp[#4]}}

\NewDocumentCommand{\bwitnessessym}{}{\mathtt{W}}
\NewDocumentCommand{\bwitnesses}{oomo}{{\bwitnessessym\maybefoot[#1]\mayberp[#2]\qp{#3}\maybeqp[#4]}}

\NewDocumentCommand{\bcommitssym}{}{\mathtt{X}}
\NewDocumentCommand{\bcommits}{oomo}{{\bcommitssym\maybefoot[#1]\mayberp[#2]\qp{#3}\maybeqp[#4]}}

\NewDocumentCommand{\bexclusionssym}{}{\mathtt{E}}
\NewDocumentCommand{\bexclusions}{oomo}{{\bexclusionssym\maybefoot[#1]\mayberp[#2]\qp{#3}\maybeqp[#4]}}

\NewDocumentCommand{\bcompletionssym}{}{\mathtt{T}}
\NewDocumentCommand{\bcompletions}{oomo}{{\bcompletionssym\maybefoot[#1]\mayberp[#2]\qp{#3}\maybeqp[#4]}}

\NewDocumentCommand\bitstrings{}{\mathbb{S}}
\NewDocumentCommand\crop{mm}{{\rp{#1\vert_{#2}}}}

\NewDocumentCommand\intreprsym{m}{{{\tilde \iota}_{#1}}}
\NewDocumentCommand\intreadsym{m}{{{\tilde \iota}^{-1}_{#1}}}

\NewDocumentCommand\intrepr{mm}{{\intreprsym{#1}\rp{#2}}}
\NewDocumentCommand\intread{mm}{{\intreadsym{#1}\rp{#2}}}

\NewDocumentCommand\intencsym{m}{{\iota_{#1}}}
\NewDocumentCommand\intdecsym{m}{{\iota^{-1}_{#1}}}

\NewDocumentCommand\intenc{mmm}{{\intencsym{#1}\rp{#2, #3}}}
\NewDocumentCommand\intdec{mm}{{\intdecsym{#1}\rp{#2}}}

\NewDocumentCommand\varreprsym{}{{\tilde \nu}}
\NewDocumentCommand\varreadsym{}{{\tilde \nu^{-1}}}

\NewDocumentCommand\varrepr{m}{{\varreprsym\rp{#1}}}
\NewDocumentCommand\varread{m}{{\varreadsym\rp{#1}}}

\NewDocumentCommand\varencsym{}{\nu}
\NewDocumentCommand\vardecsym{}{{\nu^{-1}}}

\NewDocumentCommand\varparseable{}{\bitstrings^{\varencsym}}

\NewDocumentCommand\varenc{mm}{{\varencsym\rp{#1, #2}}}
\NewDocumentCommand\vardec{m}{{\vardecsym\rp{#1}}}

\NewDocumentCommand\varlensym{}{\lambda}
\NewDocumentCommand\varlen{m}{{\varlensym\rp{#1}}}

\NewDocumentCommand\enumeration{m}{{\mathcal{E}\rp{#1}}}

\NewDocumentCommand\partitionreprsym{}{\rho}
\NewDocumentCommand\partitionreadsym{}{\partitionreprsym^{-1}}
\NewDocumentCommand\partitionrepr{m}{{\partitionreprsym\rp{#1}}}
\NewDocumentCommand\partitionread{m}{{\partitionreadsym\rp{#1}}}

\begin{document}

\maketitle

\begin{abstract}
    Byzantine Reliable Broadcast (BRB) is a fundamental distributed computing primitive, with applications ranging from notifications to asynchronous payment systems. 
Motivated by practical consideration, we study \csb (\csbshort), a multi-shot variant of BRB whose interface is split between broadcasting \emph{clients} and delivering \emph{servers}. 
We present \csbal, an optimally resilient implementation of \csbshort. 
Like most implementations of BRB, \csbal guarantees both liveness and safety in an asynchronous environment. 
Under good conditions, however, \csbal achieves unparalleled efficiency. 
In a moment of synchrony, free from Byzantine misbehaviour, and at the limit of infinitely many broadcasting clients, a \csbal server delivers a $b$-bits payload at an asymptotic amortized cost of $0$ signature verifications, and $\rp{log_2\rp{\clientcount} + b}$ bits exchanged, where $\clientcount$ is the number of clients in the system. 
This is the information-theoretical minimum number of bits required to convey the payload ($b$ bits, assuming it is compressed), along with an identifier for its sender ($\log_2\rp{\clientcount}$ bits, necessary to enumerate any set of $\clientcount$ elements, and optimal if broadcasting frequencies are uniform or unknown). 
These two achievements have profound practical implications. 
Real-world BRB implementations are often bottlenecked either by expensive signature verifications, or by communication overhead. 
For \csbal, instead, the network is the limit: a server can deliver payloads as quickly as it would receive them from an infallible oracle.
\end{abstract}

\section{Introduction}
\label{section:introduction}

Byzantine reliable broadcast (BRB) is one of the most fundamental and versatile building blocks in distributed computing, powering a variety of Byzantine fault-tolerant (BFT) systems~\cite{cach02sintra,du18beat}. 
The BRB abstraction has recently been shown to be strong enough to process payments, enabling cryptocurrency deployments in an asynchronous environment~\cite{at2-19-podc}.
Originally introduced by Bracha~\cite{bra87asynchronous} to allow a set of processes to agree on a single message from a designated sender, BRB naturally generalizes to the multi-shot case, enabling higher-level abstractions such as Byzantine FIFO~\cite{reit94, cac11intro} and causal~\cite{Breliablecommunication1987,AUVOLAT202155} broadcast. 
We study a practical, multi-shot variant of BRB whose interface is split between broadcasting \emph{clients} and delivering \emph{servers}.
We call this abstraction \csb (\csbshort).

\miniparagraph{\csbshort in brief} 
Clients broadcast, and servers deliver, \emph{payloads} composed by a \emph{context} and a \emph{message}. 
This interface allows, for example, Alice to announce her wedding as well as will her fortune by respectively broadcasting
\begin{align*}
    \rp{\underbrace{\texttt{"My wife is"}}_{\text{context} \; c_w}, \underbrace{\texttt{"Carla"}}_{\text{message} \; m_w}} \;\;\;
    \rp{\underbrace{\texttt{"All my riches go to"}}_{\text{context} \; c_r}, \underbrace{\texttt{"Bob"}}_{\text{message} \; m_r}}
\end{align*}
\csbshort guarantees that: (\emph{Consistency}) no two correct servers deliver different messages for the same client and context; (\emph{Totality}) either all correct servers deliver a message for a given client and context, or no correct server does; (\emph{Integrity}) if a correct server delivers a payload from a correct client, then the client has broadcast that payload; and (\emph{Validity}) a payload broadcast by a correct client is delivered by at least one correct server. 
Following from the above example, Carla being Alice's wife does not conflict with Bob being her sole heir (indeed, $c_w \neq c_r$), but Alice would not be able to convince two correct servers that she married Carla and Diana, respectively.
Higher-level broadcast abstractions can be easily built on top of \csbshort. 
For example, using integer sequence numbers as contexts and adding a reordering layer yields Client-Server Byzantine FIFO Broadcast. 
For the sake of \csbshort, however, it is not important for contexts to be integers, or satisfy any property other than comparability. 
Throughout the remainder of this paper, the reader can picture contexts as opaque binary blobs.
Lastly, while the set of servers is known, \csbshort as presented does not assume any client to be known a priori.
The set of clients can be permissionless, with servers discovering new clients throughout the execution.

\miniparagraph{A utopian model} 
Real-world BRB implementations are often bottlenecked either by expensive signature verifications~\cite{Crain18_redbelly} or by communication overhead~\cite{br85acb,Malkhi97_WAN,Malkhi96ahigh-throughput}. 
With the goal of broadening those bottlenecks, simplified, more trustful models are useful to establish a (sometimes grossly unreachable) bound on the efficiency that an algorithm can attain in the Byzantine setting.
For example, in a utopian model where any agreed-upon process can be trusted to never fail (let us call it an \emph{oracle}), \csbshort can easily be implemented with great efficiency. 
Upon initialization, the oracle organizes all clients in a list, which it disseminates to all servers.
For simplicity, let us call \emph{id} a client's position in the list. 
To broadcast a payload $p$, a client with id $i$ simply sends $p$ to the oracle: the oracle checks $p$ for equivocation (thus ensuring consistency), then forwards $\rp{i, p}$ to all servers (thus ensuring validity and totality). 
Upon receiving $\rp{i, p}$, a server blindly trusts the oracle to uphold all \csbshort properties, and delivers $\rp{i, p}$. \textsf{Oracle-\csbshort} is clearly very efficient. 
On the one hand, because the oracle can be trusted not to attribute spurious payloads to correct clients, integrity can be guaranteed without any server-side signature verification.
On the other, in order to deliver $\rp{i, p}$, a server needs to receive just $\rp{\ceil{\log_2\rp{\clientcount}} + \abs{p}}$ bits, where $\clientcount$ denotes the total number of clients, and $\abs{p}$ measures $p$'s length in bits.
This is optimal assuming the rate at which clients broadcast is unknown\footnote{
Lacking an assumption on broadcasting rates, an adversarial scheduler could have all messages broadcast by the client with the longest id, which we cannot guarantee to be shorter than $\ceil{\log_2\rp{\clientcount}}$ bits.
} or uniform\footnote{
Should some clients be expected to broadcast more frequently than others, we could further optimize \textsf{Oracle-\csbshort} by assigning smaller ids to more active clients, possibly at the cost of having less active clients have ids whose length exceeds $\ceil{\log_2\rp{\clientcount}}$. Doing so, however, is beyond the scope of this paper.
}~\cite{basicinfotheory}.

\miniparagraph{Matching the oracle}
Due to its reliance on a single infallible process, \textsf{Oracle-\csbshort} is not a fault-tolerant distributed algorithm: shifting back to the Byzantine setting, a single failure would be sufficient to compromise all \csbshort properties.
Common sense suggests that Byzantine resilience will necessarily come at some cost: protocol messages must be exchanged to preserve consistency and totality, signatures must be produced and verified to uphold integrity and, lacking the totally-ordering power that only consensus can provide, ids cannot be assigned in an optimally dense way.
However, this paper proves the counter-intuitive result that an asynchronous, optimally-resilient, Byzantine implementation of \csbshort can asymptotically match the efficiency of \textsf{Oracle-\csbshort}. 
This is not just up to a constant, but identically.
In a synchronous execution, free from Byzantine misbehaviour, and as the number of concurrently broadcasting clients goes to infinity (we call these conditions the \emph{batching limit}\footnote{
The batching limit includes other easily achievable, more technical conditions that we omit in this section for the sake of brevity.
We formally define the batching limit in Section \ref{subsubsection:csbal.complexity.batchinglimit}
}
), our \csbshort implementation \csbal delivers a payload $p$ at an asymptotic\footnote{The asymptotic costs are reached quite fast, at rates comparable to $C^{-1}$ or $log(C)\cdot C^{-1}$.}, amortized cost of $0$ signature verifications\footnote{This does not mean that batches are processed in constant time: hashes and signature aggregations, for example, still scale linearly in the size of a batch. The real-world computational cost of such simple operations, however, is several orders of magnitude lower than that of signature verification.} and $\rp{\ceil{\log_2\rp{\clientcount}} + \abs{p}}$ bits exchanged per server, the same as in \textsf{Oracle-\csbshort} (we say that \csbal achieves \emph{oracular efficiency}).
At the batching limit a \csbal server is dispensed from nearly all signature verifications, as well as nearly all traffic that would be normally required to convey protocol messages, signatures, or client public keys. 
Network is the limit: payloads are delivered as quickly as they can be received.

\miniparagraph{\csbshort's common bottlenecks} 
To achieve oracular efficiency, we focus on three types of server overhead that commonly affect a real-world implementation of \csbshort:
\begin{itemize}
    \item \emph{Protocol overhead}. 
    Safekeeping consistency and totality typically requires some form of communication among servers. 
    This communication can be direct (as in Bracha's original, all-to-all BRB implementation) or happen through an intermediary (as in Bracha's signed, one-to-all-to-one BRB variant), usually employing signatures to establish authenticated, intra-server communication channels through a (potentially Byzantine) relay.
    \item \emph{Signature overhead}. 
    Upholding integrity usually requires clients to authenticate their messages using signatures. 
    For servers, this entails both a computation and a communication overhead. 
    On the one hand, even using well-optimized schemes, signature verification is often CPU-heavy enough to dominate a server's computational budget, dwarfing in particular the CPU footprint of much lighter, symmetric cryptographic primitives such as hashes and ciphers. 
    On the other hand, transmitting signatures results in a fixed communication overhead per payload delivered. 
    While the size of a signature usually ranges from a few tens to a few hundreds of bytes, this overhead is non-negligible in a context where many clients broadcast small messages.
    This is especially true in the case of payments, where a message reduces to the identifier of a target account and an integer to denote the amount of money to transfer.
    \item \emph{Identifier overhead}. 
    \csbshort's multi-shot nature calls for a sender identifier to be attached to each broadcast payload. 
    Classically, the client's public key is used as identifier. 
    This is convenient for two reasons.
    First, knowing a client's identifier is sufficient to authenticate its payloads. 
    Second, asymmetric keypairs have very low probability of collision. 
    As such, clients can create identities in the system without any need for coordination: locally generating a keypair is sufficient to begin broadcasting messages. 
    By cryptographic design, however, public keys are sparse, and their size does not change with the number of clients.
    This translates to tens to hundreds of bytes being invested to identify a client from a set that can realistically be enumerated by a few tens of bits. 
    Again, this communication overhead is heavier on systems where broadcasts are frequent and brief.
\end{itemize}
On the way to matching \textsf{Oracle-\csbshort}'s performance, we develop techniques to negate all three types of overhead: at the batching limit, a \csbal server delivers a payload wasting $0$ bits to protocol overhead, performing $0$ signature verifications, and exchanging $\ceil{\log_2\rp{\clientcount}}$ bits of identifier, the minimum required to enumerate the set of clients.
We outline our contributions below, organized in three (plus one) take-home messages (T-HMs).

\miniparagraph{T-HM1: The effectiveness of batching goes beyond total order}
In the totally ordered setting, batching is famously effective at amortizing protocol overhead~\cite{SANTOS2013170,Androulaki18}. 
Instead of disseminating its message to all servers, a client hands it over to (one or more)\footnote{In most real-world implementations, a client optimistically entrusts its payload to a single process, extending its request to larger portions of the system upon expiration of a suitable timeout.} batching processes. 
Upon collecting a large enough set of messages, a batching process organizes all messages in a batch, which it then disseminates to the servers.
Having done so, the batching process submits the batch's hash to the system's totally-ordering primitive. 
Because hashes are constant in length, the cost of totally ordering a batch does not depend on its size.
Once batches are totally ordered, so too are messages (messages within a batch can be ordered by any deterministic function), and equivocations can be handled at the application layer (for example, in the context of a cryptocurrency, the second request to transfer the same asset can be ignored by all correct servers, with no need for additional coordination).
At the limit of infinitely large batches, the relative overhead of the ordering protocol becomes vanishingly small, and a server can allocate virtually all of its bandwidth to receiving batches.
This strategy, however, does not naturally generalize to \csbshort, where batches lack total order.
As payloads from multiple clients are bundled in the same batch, a correct server might detect equivocation for only a subset of the payloads in the batch. 
Entirely accepting or entirely rejecting a partially equivocated batch is not an option. 
In the first case, consistency could be violated.
In the second case, a single Byzantine client could single-handedly ``poison'' the batches assembled by every correct batching process with equivocated payloads, thus violating validity.
In \csbal, a server can partially reject a batch, acknowledging all but some of its payloads. 
Along with its partial acknowledgement, a server provides a proof of equivocation to justify each exception. 
Having collected a quorum of appropriately justified partial acknowledgements, a batching process has servers deliver only those payloads that were not excepted by any server. 
Because proofs of equivocations cannot be forged for correct clients, a correct client handing over its payload to a correct batching process is guaranteed to have that payload delivered. 
In the common case where batches have little to no equivocations, servers exchange either empty or small lists of exceptions, whose size does not scale with that of the batch. 
This extends the protocol-amortizing power of batching to \csbshort and, we conjecture, other non-totally ordered abstractions.

\miniparagraph{T-HM2: Interactive multi-signing can slash signature overhead}
Traditionally, batching protocols are non-interactive on the side of clients. 
Having offloaded its message to a correct batching process, a correct client does not need to interact further for its message to be delivered: the batching process collects an arbitrary set of independently signed messages and turns to the servers to get each signature verified, and the batch delivered. 
This approach is versatile (messages are not tied to the batch they belong to) and reliable (a client crashing does not affect a batch's progress) but expensive (the cost of verifying each signature is high and independent of the batch's size). 
In \csbal, batching processes engage in an interactive protocol with clients to replace, in the good case, all individual signatures in a batch with a single, batch-wide \emph{multi-signature}. 
In brief, multi-signature schemes extend traditional signatures with a mechanism to \emph{aggregate} signatures and public keys: an arbitrarily large set of signatures for the same message\footnote{Some multi-signature schemes also allow the aggregation of signatures on heterogeneous messages. 
In that case, however, aggregation is usually as expensive as signature verification. 
Given our goal to reduce CPU complexity for servers, this paper entirely disregards heterogeneous aggregation schemes.} can be aggregated into a single, constant-sized signature; similarly, a set of public keys can be aggregated into a single, constant-sized public key. 
The aggregation of a set of signatures can be verified in constant time against the aggregation of all corresponding public keys. 
Unlike verification, aggregation is a cheap operation, reducing in some schemes to a single multiplication on a suitable field. 
Multi-signature schemes open a possibility to turn expensive signature verification into a once-per-batch operation.
Intuitively, if each client contributing to a batch could multi-sign the entire batch instead of its individual payload, all multi-signatures could be aggregated, allowing servers to authenticate all payloads at once.
However, as clients cannot predict how their payloads will be batched, this must be achieved by means of an interactive protocol. 
Having collected a set of individually-signed payloads in a batch, a \csbal batching process shows to each contributing client that its payload was included in the batch. 
In response, clients produce their multi-signatures for the batch's hash, which the batching process aggregates.
Clients that fail to engage in this interactive protocol (e.g., because they are faulty or slow) do not lose liveness, as their original signature can still be attached to the batch to authenticate their individual payload.
In the good case, all clients reply in a timely fashion, and each server has to verify a single multi-signature per batch. 
At the limit of infinitely large batches, this results in each payload being delivered at an amortized cost of $0$ signature verifications.
The usefulness of this interactive protocol naturally extends beyond \csbshort to all multi-shot broadcast abstractions whose properties include integrity.

\miniparagraph{T-HM3: Dense id assignment can be achieved without consensus}
In order to efficiently convey payload senders, \textsf{Oracle-\csbshort}'s oracle organizes all clients in a list, attaching to each client a successive integral identifier. 
Once the list is disseminated to all servers, the oracle can identify each client by its identifier, sparing servers the cost of receiving larger, more sparse, client-generated public keys. 
Id-assignment strategies similar to that of \textsf{Oracle-\csbshort} can be developed, in the distributed setting, building on top of classical algorithms that identify clients by their full public keys (we call such algorithms \emph{id-free}, as opposed to algorithms such as \csbal, which are \emph{id-optimized}).
In a setting where consensus can be achieved, the identifier density of \textsf{Oracle-\csbshort} is easily matched.
Upon initialization, each client submits its public key to an id-free implementation of Total-Order Broadcast (TOB).
Upon delivery of a public key, every correct process agrees on its position within the common, totally-ordered log. 
As in \textsf{Oracle-\csbshort}, each client can then use its position in the list as identifier within some faster, id-optimized broadcast implementation.
In a consensus-less setting, achieving a totally-ordered list of public keys is famously impossible~\cite{Defago04}.
This paper, however, proves the counter-intuitive result that, when batching is used, the density of ids assigned by a consensus-less abstraction can asymptotically match that of those produced by \textsf{Oracle-\csbshort} or consensus.
In \diral, our consensus-less id-assigning algorithm, a client requests an id from every server. 
Each server uses an id-free implementation of FIFO Broadcast to order the client's public key within its own log.
Having observed its public key appear in at least one log, the client publicly elects the server in charge of that log to be its \emph{assigner}.
Having done so, the client obtains an id composed of the assigner's public key and the client's position within the assigner's log. 
We call the two components of an id \emph{domain} and \emph{index}, respectively. 
Because the set of servers is known to (and can be enumerated by) all processes, an id's domain can be represented in $\ceil{\log_2\rp{\servercount}}$ bits, where $\servercount$ denotes the total number of servers. 
Because at most $\clientcount$ distinct clients can appear in the FIFO log of any server, indices are at most $\ceil{\log_2\rp{\clientcount}}$ bits long. 
In summary, \diral assigns ids to clients without consensus, at an additional cost of $\ceil{\log_2\rp{\servercount}}$ bits per id. 
Interestingly, even this additional complexity can be amortized by batching.
Having assembled a batch, a \csbal batching process represents senders not as a list of ids, but as a map, associating to each of the $\servercount$ domains the indices of all ids in the batch under that domain. 
At the limit of infinitely large batches ($C \gg N$), the bits required to represent the map's keys are entirely amortized by those required to represent its values.
This means that, while $\rp{\ceil{\log_2\rp{\servercount}} + \ceil{\log_2\rp{\clientcount}}}$ bits are required to identify a client in isolation, $\ceil{\log_2\rp{\clientcount}}$ bits are sufficient if the client is batched: even without consensus, \csbal asymptotically matches the id efficiency of \textsf{Oracle-\csbshort}.

\miniparagraph{Bonus T-HM: Untrusted processes can carry the system}
In THM1, we outlined how batching can be generalized to the consensus-less case, and discussed its role in removing protocol overhead. 
In THM2, we sketched how an interactive protocol between clients and batching processes can eliminate signature overhead. 
In employing these techniques, we shifted most of the communication and computation complexity of our algorithms from servers to batching processes. 
Batching processes verify all client signatures, create batches, verify and aggregate all client multi-signatures, then communicate with servers in an expensive one-to-all pattern, engaging server resources (at the batching limit) as little as an oracle would. 
Our last contribution is to observe that a batching process plays no role in upholding \csbshort's safety. 
As we discuss in detail throughout the remainder of this paper, a malicious batching process cannot compromise consistency (it would need to collect two conflicting quorums of acknowledgements), totality (any server delivering a batch has enough information to convince all others to do the same) or integrity (batches are still signed, and forged or improperly aggregated multi-signatures are guaranteed to be detected). 
Intuitively, the only damage a batching process can do to the system is to refuse to process client payloads\footnote{Or cause servers to waste resources, e.g., by transmitting improperly signed batches. Simple accountability measures, we conjecture, would be sufficient to mitigate these attacks in \csbal. 
A full discussion of Denial of Service, however, is beyond the scope of this paper.}.
This means that a batching process does not need to satisfy the same security properties as a server. 
\csbshort's properties cannot be upheld if a third of the servers are faulty. 
Conversely, \csbal has both liveness and safety as long as \emph{a single} batching process is correct.
This observation has profound practical implications. 
In the real world, scaling the resources of a permissioned, security-critical set of servers can be hard. 
On the one hand, reputable, dependable institutions partaking in the system might not have the resources to keep up with its demands.
On the other, more trusted hardware translates to a larger security cross-section. 
Trustless processes, however, are plentiful to the point that permissionless cryptocurrencies traditionally waste their resources, making them compete against each other in expensive proofs of Sybil-resistance~\cite{nakamoto2008bitcoin}. 
In this paper, we extend the classical client-server model with \emph{brokers}, a permissionless, scalable set of processes whose only purpose is to alleviate server complexity. 
Unlike servers, more than two-thirds of which we assume to be correct, all brokers but one can be faulty. 
In \csbal, brokers act as an intermediary between clients and servers, taking upon themselves the batching of payloads, verification and aggregation of signatures, the dissemination of batches, and the transmission of protocol messages. 

\miniparagraph{Roadmap}
We discuss related work in Section \ref{section:relatedwork}. We state our model and recall useful cryptographic background in Section \ref{section:model}. In Section \ref{section:csbaloverview}, we introduce our \csbshort implementation \csbal: we overview \csbal's protocol in Section \ref{subsection:csbaloverview.protocol}, and provide high-level arguments for \csbal's efficiency in Section \ref{subsection:csbaloverview.complexity}. We draw our conclusions and propose future work in Section \ref{section:conclusions}. We leave a fully formal analysis of our algorithms to Appendices \ref{appendix:csbal} and \ref{appendix:diral}: Appendix \ref{appendix:csbal} fully defines the \csb abstraction (\ref{subsection:csbal.interface}), proves \csbal's correctness (\ref{subsection:csbal.correctness}), and analyzes \csbal's complexity at the batching limit (\ref{subsection:csbal.complexity}); Appendix \ref{appendix:diral} introduces the \dir abstraction (\ref{subsection:diral.interface}) and proves the correctness of our \dir implementation \diral (\ref{subsection:diral.correctness}). For the sake of tidiness, we gather all pseudocode in Appendix \ref{appendix:pseudocode} (\ref{subsection:pseudocode.csbalclient}, \ref{subsection:pseudocode.csbalbroker} and \ref{subsection:pseudocode.csbalserver} for \csbal, \ref{subsection:pseudocode.diral} and \ref{subsection:pseudocode.diralserver} for \diral).
\section{Related Work}
\label{section:relatedwork}

Byzantine Reliable Broadcast (BRB) is a classical primitive of distributed computing, with widespread practical applications such as in State Machine Replication (SMR)~\cite{miller2016honey,cachin2002secure, cachin2010state}, Byzantine agreement~\cite{nayak2020improved,castro2002practical,kursawe2005optimistic,keidar2021all,stathakopoulou2022state}, blockchains~\cite{Androulaki18,crain2021red, danezis2022narwhal}, and online payments~\cite{at2-19-podc,Collins20, kuznetsov2021permissionless}.
In classical BRB, a system of $n$ processes agree on a single message from a single \emph{source} (one of the $n$ processes), while tolerating up to $f$ Byzantine failures ($f$ of the $n$ processes can behave arbitrarily).
A well known solution to asynchronous BRB with provably optimal resilience ($f < n/3$) was first proposed by Bracha~\cite{bracha1984asynchronous, bra87asynchronous} who introduced the problem.
Bracha's broadcast reaches $O(n^2)$ message complexity, and $O(n^2L)$ communication complexity (total number of transmitted bits between correct processes \cite{Yao79}), where $L$ is the length of the message.
Since $O(n^2)$ message complexity is provably optimal~\cite{Dolev85}, the main focus of BRB-related research has been on reducing its communication complexity. 
The best lower bound for communication complexity is $\Omega(nL + n^2)$, although it is unknown whether it is tight.
The $nL$ term comes from all processes having to receive the message (length $L$), while the $n^2$ term comes from each of the $n$ processes having to receive $\Omega(n)$ protocol messages to ensure agreement in the presence of $f = \Theta(n)$ failures~\cite{Dolev85}.
One line of research focuses on worst-case complexity, predominantly using error correcting codes~\cite{reed1960polynomial, ben1993asynchronous} or erasure codes~\cite{plank2006optimizing,hendricks2007verifying, CacTes05, alhaddad2021succinct}, and has produced various BRB protocols with improved complexity~\cite{alhaddad2021succinct,CacTes05, cachin2001secure, das2021asynchronous,nayak2020improved}, many of them quite recently.
The work of Das, Xiang and Ren~\cite{das2021asynchronous} achieves $O(nL + kn^2)$ communication complexity (specifically, $7nL + 2kn^2$), where $k$ is the security parameter (e.g., the length of a hash, typically $256$ bits).
As the authors note, the value of hidden constants (and $k$, which is sometimes considered as a constant in literature) is particularly important when considering practical implementations of these protocols.
Another line of research focuses on optimizing the good case performance of BRB, i.e., when the network behaves synchronously and no process misbehaves~\cite{cachin2001secure,castro2002practical,kursawe2005optimistic,ramasamy2005parsimonious, Abraham21}.
As the good case is usually the common case, in practice, the real-world communication complexity of these optimistic protocols matches that of the good case.
A simple and widely-used hash-based BRB protocol is given by Cachin \emph{et al.}~\cite{cachin2001secure}.
It replaces the echo and ready phase messages in Bracha's protocol with hashes, achieving $O(nL + kn^2)$ in the good case (specifically, $nL + 2kn^2$), and $O(n^2L)$ in the worst-case.
Considering practical throughput, some protocols also focus on the \emph{amortized} complexity per source message~\cite{castro2002practical, ramasamy2005parsimonious,malkhi1997high}.
Combining techniques such as \emph{batching}~\cite{castro2002practical} and threshold signatures~\cite{shoup2000practical}, at the limit (of batch size), BRB protocols reach $O(nL)$ amortized communication complexity in the good case~\cite{ramasamy2005parsimonious}.
At this point, the remaining problem lies in the hidden constants.
In the authenticated setting, batching-based protocols rely on digital signatures to validate (source) messages before agreeing to deliver them~\cite{ramasamy2005parsimonious}.
In reality, each source message in a batch includes its content, an identifier of the source (e.g., a $k$-sized public key), a sequence id (identifying the message), and a $k$-sized signature.
When considering systems where $L$ is small (e.g., online payments), these can take up a large fraction of the communication.
To be precise, the good-case amortized communication complexity would be $O(nL + kn)$.
In fact, message signatures (the $kn$ factor) are by far the main bottleneck in practical applications of BRB today~\cite{danezis2022narwhal,stathakopoulou2022state}, both in terms of communication and computation (signature verification), leading to various attempts at reducing or amortizing their cost~\cite{crain2021red, malkhi1997high}.
For example, Crain \emph{et al.}~\cite{crain2021red} propose \emph{verification sharding}, in which only $f + 1$ processes have to receive and verify all message signatures in the good case, which is a $3$-fold improvement over previous systems (on the $kn$ factor) where all $n$ processes verify all signatures.
However, by itself, this does not improve on the amortized cost of $O(nL + kn)$ per message.
When contrasting theoretical research with practical systems, it is interesting to note the gap that can surge between the theoretical model and reality.
The recent work of Abraham \emph{et al.}~\cite{Abraham21}, focused on the good-case latency of Byzantine broadcast, expands on some of these mismatches and argues about the practical limitations of focusing on the worst-case.
Another apparent mismatch lies in the classical model of Byzantine broadcast.
In many of the applications of BRB mentioned previously (e.g., SMR, permissioned blockchains, online payments), there is usually a set of \emph{servers} ($n$, up to $f$ of which are faulty), and a set of external clients ($\clients$) which are the true sources of messages.
The usual transformation from BRB's classical model into these practical settings maps the set of $n$ servers as the $n$ processes and simply excludes clients as system entities, e.g., assuming their messages are relayed through one of the servers.
Since the number of clients can be very large ($|\clients| \gg n$), clients are untrusted (which can limit their usefulness), and the focus is on the communication complexity of the \emph{servers}, this transformation seems reasonable and simplifies the problem.
However, it can also limit the search for more practical solutions.
In this paper, in contrast with the classical model of BRB, we explicitly include the set of clients $\clients$ in our system while focusing on the communication complexity surrounding the servers (i.e., the bottleneck).
Furthermore, we introduce \emph{brokers}, an untrusted set $\brokers$ of processes, only one of which is assumed to be correct, whose goal is to assist servers in their operation. By doing this, we can leverage brokers to achieve a good-case, amortized communication complexity (for servers, information received or sent) of $nL + o(nL)$.

\section{Model \& background}
\label{section:model}

\subsection{Model}
\label{subsection:model.model}

\miniparagraph{System and adversary} We assume an asynchronous message-passing system where the set $\processes$ of processes is the distinct union of three sets: \textbf{servers} ($\servers$), \textbf{brokers} ($\brokers$), and \textbf{clients} ($\clients$). We use $\servercount = \abs{\servers}$, $\brokercount = \abs{\brokers}$ and $\clientcount = \abs{\clients}$. 
Any two processes can communicate via reliable, FIFO, point-to-point links (messages are delivered in the order they are sent). 
Faulty processes are Byzantine, i.e., they may fail arbitrarily. Byzantine processes know each other, and may collude and coordinate their actions. 
At most $\faultcount$ servers are Byzantine, with $\servercount = 3 \faultcount + 1$. At least one broker is correct. 
All clients may be faulty. 
We use $\correctprocesses$ and $\faultyprocesses$ to respectively identify the set of correct and faulty processes.
The adversary cannot subvert cryptographic primitives (e.g., forge signatures). 
Servers and brokers\footnote{The assumption that brokers are permissioned is made for simplicity, and can be easily relaxed to the requirement that every correct process knows at least one correct broker.} are permissioned (every process knows $\servers$ and $\brokers$), clients are permissionless (no correct process knows $\clients$ a priori). 
We call \emph{certificate} a statement signed by either a plurality ($f+1$) or a quorum ($2f+1$) of servers.
Since every process knows $\servers$, any process can verify a certificate.

\miniparagraph{Good case} The algorithms presented in this paper are designed to uphold all their properties in the model above.
\csbal, however, achieves oracular efficiency only in the \emph{good case}.
In the good case, links are synchronous (messages are delivered at most one time unit after they are sent), all processes are correct, and the set of brokers contains only one element.
To take advantage of the good case, \csbal makes use of timers (which is uncommon for purely asynchronous algorithms).
A timer with timeout $\delta$ set at time $t$ rings: after time $\rp{t + \delta}$, if the system is synchronous; after time $t$, otherwise.
Intuitively, in the non-synchronous case, timers disregard their timeout entirely, and are guaranteed to ring only eventually.

\subsection{Background}
\label{subsection:model.background}

Besides commonly used hashes and signatures, the algorithms presented in this paper make use of two less often used cryptographic primitives, namely, multi-signatures and Merkle trees. We briefly outline their use below.
An in-depth discussion of their inner workings, however is beyond the scope of this paper.

\miniparagraph{Multi-signatures}
Like traditional signatures, multi-signatures~\cite{Barreto02} are used to publicly authenticate messages: a public / secret keypair $\rp{p, r}$ is generated locally; $r$ is used to produce a signature $s$ for a message $m$; $s$ is publicly verified against $p$ and $m$. Unlike traditional signatures, however, multi-signatures for the same message can be \emph{aggregated}. 
Let $\rp{p_1, r_1}, \ldots, \rp{p_n, r_n}$ be a set of keypairs, let $m$ be a message, and let $s_i$ be $r_i$'s signature for $m$.
$\rp{p_1, \ldots, p_n}$ and $\rp{s_1, \ldots, s_n}$ can be respectively aggregated into a constant-sized public key $\hat p$ and a constant-sized signature $\hat s$.
As with individually-generated multi-signatures, $\hat s$ can be verified in constant time against $\hat p$ and $m$. 
Aggregation is cheap and non-interactive: provided with $\rp{p_1, \ldots, p_n}$ (resp., $\rp{s_1, \ldots, s_n}$) any process can compute $\hat p$ (resp., $\hat s$).

\miniparagraph{Merkle trees}
Merkle trees~\cite{merkle1987digital} extend traditional hashes with compact \emph{proofs of inclusion}.
As with hashes, a sequence $\rp{x_1, \ldots, x_n}$ of values can be hashed into a preimage and collision-resistant digest (or \emph{root}) $r$.
Unlike hashes, however, a proof $p_i$ can be produced from $\rp{x_1, \ldots, x_n}$ to attest that the $i$-th element of the sequence whose root is $r$ is indeed $x_i$. 
In other words, provided with $r$, $p_i$ and $x_i$, any process can verify that the $i$-th element of $\rp{x_1, \ldots, x_n}$ is indeed $x_i$, without having to learn $\rp{x_1, \ldots, x_{i - 1}, x_{i + 1}, \ldots, x_n}$.
The size of a proof of inclusion for a sequence of $n$ elements is logarithmic in $n$. 
\section{\csbal: Overview}
\label{section:csbaloverview}

In this section, we provide an intuitive overview of our \csbshort implementation, \csbal, as well as high-level arguments for its efficiency. 
A full analysis of \csbal can be found in Appendix \ref{appendix:csbal}, where we introduce \csbal's full pseudocode, formally prove its correctness, and thoroughly study its good-case signature and communication complexity.

\subsection{Protocol}
\label{subsection:csbaloverview.protocol}

\begin{figure}
    \centering
    \includegraphics[width=\textwidth,height=\textheight,keepaspectratio]{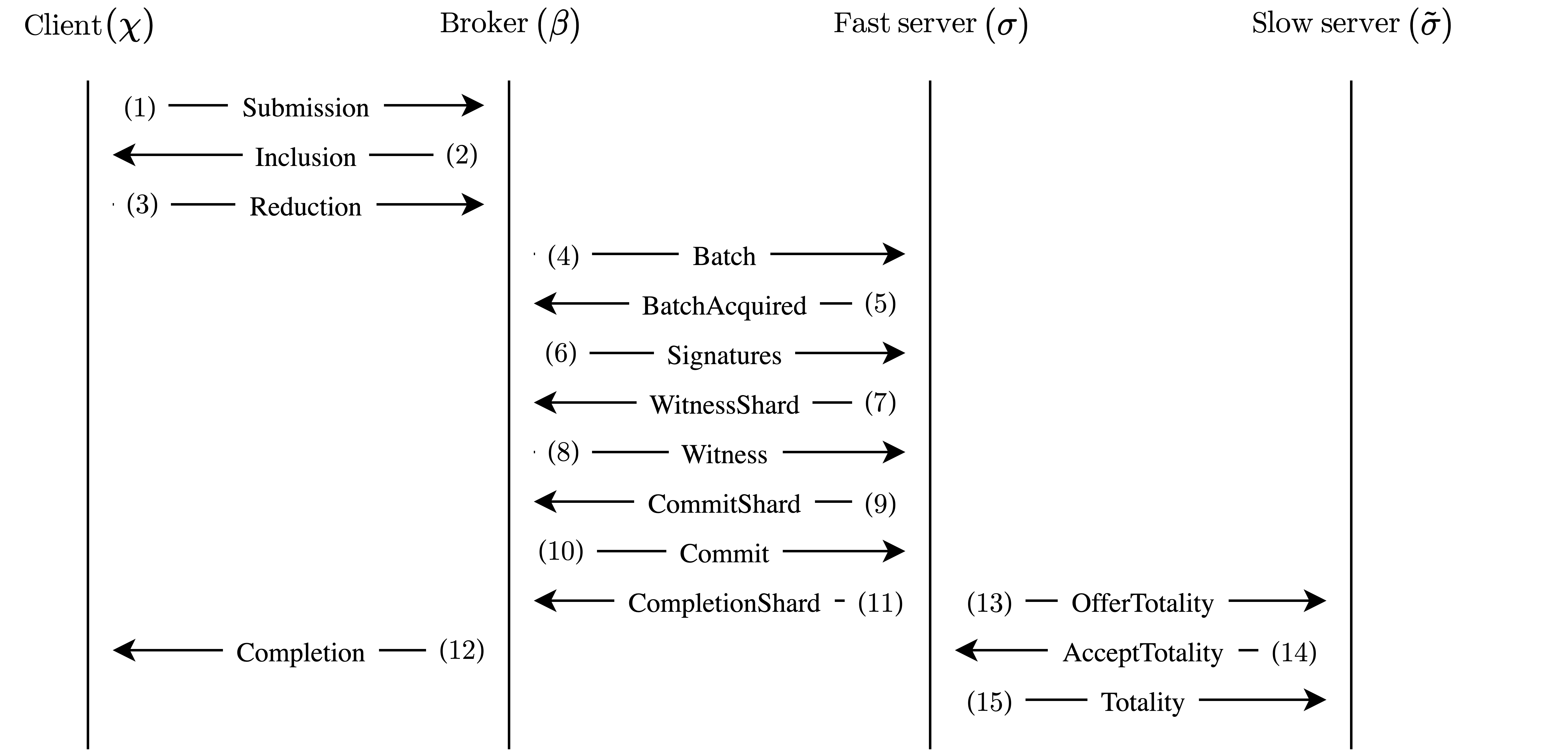}
    \caption{\csbal's protocol. 
    Having collected a batch of client payloads, a broker engages in an interactive protocol with clients to reduce the batch, replacing (most of) its individual payload signatures with a single, batch-wide multi-signature. 
    The broker then disseminates the batch to all servers, successively gathering a witness for its correctness and a certificate to commit (some of) its payloads. 
    Having had a plurality of servers deliver the batch, the broker notifies all clients with a suitable certificate. 
    In the bad case, servers can ensure totality without any help from the broker, propagating batches and commit certificates in an all-to-all fashion.}
    \label{figure:protocol}
\end{figure}

\miniparagraph{Dramatis personae} 
The goal of this section is to provide an intuitive understanding of \csbal's protocol.
In order to do this, we focus on four processes: a correct client $\client$, a correct broker $\broker$, a correct and fast server $\server$, and a correct but slow server $\tilde \server$. 
We follow the messages exchanged between $\client$, $\broker$, $\server$ and $\tilde \server$ as the protocol unfolds, as captured by Figure \ref{figure:protocol}.

\miniparagraph{The setting} 
$\client$'s goal is to broadcast a payload $p$. 
$\client$ has already used \csbal's underlying \dir abstraction (\dirshort) to obtain an id $i$. 
In brief, \dirshort guarantees that $i$ is assigned to $\client$ only, and provides $\client$ with an \emph{assignment certificate} $a$, which $\client$ can use to prove that its id is indeed $i$.
As we discussed in Section \ref{section:introduction}, \csbal uses \dirshort-assigned ids to identify payload senders. 
This is essential to \csbal's performance, as \dirshort guarantees \emph{density}: as we outline in Section \ref{subsection:csbaloverview.complexity}, $\ceil{\log_2\rp{\clientcount}}$ bits are asymptotically sufficient to represent each id in an infinitely large batch. 
We discuss the details of the \dirshort abstraction in Appendix \ref{appendix:diral}, along with our \dirshort implementation, \diral. 
Throughout the remainder of this paper, we say that a process $\process$ \emph{knows} an id $\hat i$ iff $\process$ knows the public keys to which $\hat i$ is assigned.

\miniparagraph{Building a batch}
In order to broadcast its payload $p$, $\client$ produces a signature $s$ for $p$, and then sends a \texttt{Submission} message to $\broker$ (fig. \ref{figure:protocol}, step 1). 
The \texttt{Submission} message contains $p$, $s$, and $\client$'s assignment certificate $a$. 
Upon receiving the \texttt{Submission} message, $\broker$ learns $\client$'s id $i$ from $a$, then verifies $s$ against $p$. 
Having done so, $\broker$ stores $\rp{i, p, s}$ in its \emph{submission pool}. 
For a configurable amount of time, $\broker$ fills its pool with submissions from other clients, before \emph{flushing} it into a \emph{batch}.
Let us use $\rp{i_1, p_1, s_1}, \ldots, \rp{i_\csbalbatchingwindow, p_\csbalbatchingwindow, s_\csbalbatchingwindow}$ to enumerate the elements $\broker$ flushes from the submission pool (for some $n$, we clearly have $\rp{i, p, s} = \rp{i_n, p_n, s_n}$). 
For convenience, we will also use $\client_j$ to identify the sender of $p_j$ (owner of $i_j$). 
Importantly, $\broker$ flushes the pool in such a way that $i_j \neq i_k$ for all $j \neq k$: for safety reasons that will soon be clear, \csbal's protocol prevents a client from having more than one payload in any specific batch. 
Because of this constraint, some payloads might linger in $\broker$'s pool. 
This is not an issue: $\broker$ will simply flush those payloads to a different batch at a later time. 
When building the batch, $\broker$ splits submissions and signatures, storing $\rp{i_1, p_1}, \ldots, \rp{i_\csbalbatchingwindow, p_\csbalbatchingwindow}$ separately from $s_1, \ldots, s_\csbalbatchingwindow$.

\begin{figure}[h!]
    \centering
    \includegraphics[width=0.9\textwidth,height=0.9\textheight,keepaspectratio]{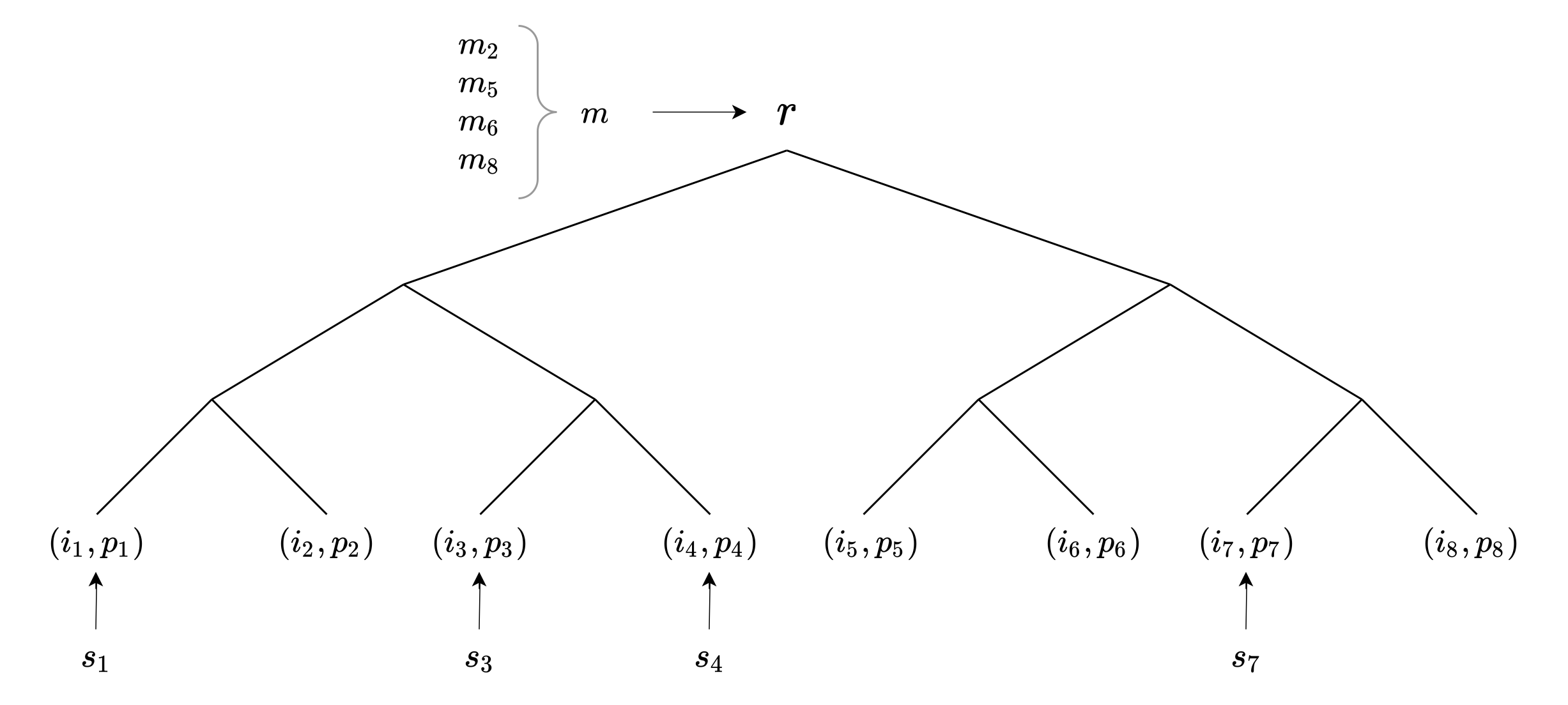}
    \caption{
    An example of partially reduced batch. 
    $B = 8$ submissions are organized on the leaves of a Merkle tree with root $r$. 
    Each submission $\rp{i_j, p_j}$ is originally authenticated by an individual signature $s_j$.
    Upon collecting a multi-signature $m_j$ for $r$, the broker drops $s_j$.
    Here the broker collected multi-signatures $m_2$, $m_5$, $m_6$ and $m_8$, leaving a \emph{straggler set} $S = \cp{\rp{i_1, s_1}, \rp{i_3, s_3}, \rp{i_4, s_4}, \rp{i_7, s_7}}$.
    Upon expiration of a suitable timeout, the broker aggregates $m_2$, $m_5$, $m_6$ and $m_8$ into a single multi-signature $m$. 
    As such, every payload in the batch is authenticated either by $m$ or by $S$.
    \label{figure:batch}
}
\end{figure}

\miniparagraph{Reducing the batch}
Having flushed submissions $\rp{i_1, p_1}, \ldots, \rp{i_\csbalbatchingwindow, p_\csbalbatchingwindow}$ and signatures $s_1, \ldots, s_\csbalbatchingwindow$, $\broker$ moves on to \emph{reduce} the batch, as exemplified in Figure \ref{figure:batch}. In an attempt to minimize signature overhead for servers, $\broker$ engages in an interactive protocol with clients $\client_1, \ldots, \client_\csbalbatchingwindow$ to replace as many signatures as possible with a single, batch-wide multi-signature.
In order to do so, $\broker$ organizes $\rp{i_1, p_1}, \ldots, \rp{i_\csbalbatchingwindow, p_\csbalbatchingwindow}$ in a Merkle tree with root $r$ (for brevity, we call $r$ the batch's root). $\broker$ then sends an \texttt{Inclusion} message to each $\client_j$ (fig. \ref{figure:protocol}, step 2). 
Each \texttt{Inclusion} message contains $r$, along with a proof of inclusion $q_j$ for $\rp{i_j, p_j}$. 
Upon receiving its \texttt{Inclusion} message, $\client$ checks $q_n$ against $r$. 
In doing so, $\client$ comes to two conclusions. First, $\client$'s submission $\rp{i, p} = \rp{i_n, p_n}$ is part of a batch whose root is $r$. 
Second, because no \csbal batch can contain multiple payloads from the same client, that batch does not attribute $\client$ any payload other than $p$. 
In other words, $\client$ can be certain that $\broker$ will not broadcast some spurious payload $p' \neq p$ in $\client$'s name: should $\broker$ attempt to do that, the batch would be verifiably malformed, and immediately discarded. 
This means $\client$ can safely produce a multi-signature $m$ for $r$: as far as $\client$ is concerned, the batch with root $r$ upholds integrity. 
Having signed $r$, $\client$ sends $m$ to $\broker$ by means of a \texttt{Reduction} message (fig. \ref{figure:protocol}, step 3).
Upon receiving $\client_j$'s \texttt{Reduction} message, $\broker$ checks $\client_j$'s multi-signature $m_j$ against $r$.
Having done so, $\broker$ discards $\client_j$'s original signature $s_j$.
Intuitively, with $m_j$, $\client_j$ attested its agreement with whatever payload the batch attributes to $\client_j$.
Because this is equivalent to individually authenticating $p_j$, $s_j$ is redundant and can be dropped. 
Upon expiration of a suitable timeout, $\broker$ stops collecting \texttt{Reduction} messages: clearly, if $\broker$ waited for every $\client_j$ to produce $m_j$, a single Byzantine client could prevent the protocol from moving forward by refusing to send its \texttt{Reduction} message.
$\broker$ aggregates all the multi-signatures it collected for $r$ into a single, batch-wide multi-signature $m$.
In the good case, every $\client_j$ is correct and timely. 
If so, $\broker$ drops all individual signatures, and the entire batch is authenticated by $m$ alone.

\miniparagraph{The perks of a reduced batch}
Having reduced the batch, $\broker$ is left with a sequence of submissions $\rp{i_1, p_1}, \ldots, \rp{i_\csbalbatchingwindow, p_\csbalbatchingwindow}$, a multisignature $m$ on the Merkle root $r$ of $\rp{i_1, p_1}, \ldots, \rp{i_\csbalbatchingwindow, p_\csbalbatchingwindow}$, and a \emph{straggler} set $S$ holding the individual signatures that $\broker$ failed to reduce.
More precisely, $S$ contains $\rp{i_j, s_j}$ iff $\broker$ did not receive a valid \texttt{Reduction} message from $\client_j$ before the reduction timeout expired. 
We recall that $m$'s size is constant, and $S$ is empty in the good case.
Once reduced, the batch is cheap to authenticate: it is sufficient to verify the batch's multi-signature against the batch's root, and each straggler signature against its individual payload.
More precisely, let $T$ denote the set of \emph{timely} clients ($\client_j$ is in $T$ iff $(i_j, \tail)$ is not in $S$). 
Let $t$ denote the aggregation of $T$'s public keys.
Provided with $\rp{i_1, p_1}, \ldots, \rp{i_\csbalbatchingwindow, p_\csbalbatchingwindow}$, $m$ and $S$, any process that knows $i_1, \ldots, i_\csbalbatchingwindow$ can verify that the batch upholds integrity by: (1) computing $r$ and $t$ from $\rp{i_1, p_1}, \ldots, \rp{i_\csbalbatchingwindow, p_\csbalbatchingwindow}$ and $S$; (2) using $t$ to verify $m$ against $r$; and (3) verifying each $s_j$ in $S$ against $p_j$.
In the good case, authenticating the batch reduces to verifying a single multi-signature.
This is regardless of the batch's size.

\miniparagraph{The pitfalls of a reduced batch}
As we discussed in the previous paragraph, reducing a batch makes it cheaper to verify its integrity.
Reduction, however, hides a subtle trade-off: once reduced, a batch gets easier to authenticate \emph{as whole}. 
Its \emph{individual payloads}, however, become harder to authenticate.
For the sake of simplicity, let us imagine that $\broker$ successfully dropped all the individual signatures it originally gathered from $\client_1, \ldots, \client_\csbalbatchingwindow$.
In order to prove that $\rp{\client = \client_n}$ broadcast $\rp{p = p_n}$, $\broker$ could naively produce the batch's root $r$, $\rp{i_n, p_n}$'s proof of inclusion $q_n$, and the batch's multi-signature $m$ for $r$. 
This, however, would not be sufficient to authenticate $p$: because the multi-signature $m_n$ that $\client$ produced for $r$ was aggregated with all others, $m$ can only be verified by the aggregation of \emph{all} $\client_1, \ldots, \client_\csbalbatchingwindow$'s public keys.
This makes authenticating $p$ as expensive as authenticating the entire batch: in order to verify $m$, \emph{all} $\rp{i_j, p_j}$ must be produced and checked against $r$, so that all corresponding public keys can be safely aggregated.

\miniparagraph{Witnessing the batch}
As we discuss next, proving the integrity of individual payloads is fundamental to ensure \csbal's validity. 
In brief, to prove that some $\client_k$ equivocated its payload $p_k = \rp{c_k, l_k}$, a server must prove to $\broker$ that $\client_k$ also issued some payload $p'_k = \rp{c_k, l'_k \neq l_k}$.
Lacking this proof, a single Byzantine server could, for example, claim without basis that $\client$ equivocated $p$.
This could trick $\broker$ into excluding $p$, thus compromising \csbal's validity.
As we discussed in the previous paragraph, however, proving the integrity of an individual payload in a reduced batch is difficult.
While we conjecture that purely cryptographic solutions to this impasse might be achievable in some schemes\footnote{For example, using BLS, $\broker$ could aggregate the public keys of $\client_1, \ldots, \client_{n - 1}, \client_{n + 1}, \ldots, \client_\csbalbatchingwindow$ into a public key $\tilde t_n$, then show that the aggregation of $\tilde t_n$ with $\client$'s public key correctly verifies $m$ against $r$.
Doing so, however, would additionally require $\broker$ to exhibit a proof that $\tilde t_n$ is not a rogue public key, i.e., that $\tilde t_n$ indeed results from the aggregation of client public keys.
This could be achieved by additionally having $\client_1, \ldots, \client_\csbalbatchingwindow$ multi-sign some hard-coded statement to prove that they are not rogues.
$\broker$ could aggregate such signatures on the fly, producing a rogue-resistance proof for $\tilde t_n$ that can be transmitted and verified in constant time. 
This, however, is expensive (and, frankly, at the limit of our cryptographic expertise).}, \csbal has $\broker$ engage in a simple protocol to further simplify the batch's authentication, replacing all client-issued (multi-)signatures with a single, server-issued certificate.
Having collected and reduced the batch, $\broker$ sends a \texttt{Batch} message to all servers (fig. \ref{figure:protocol}, step 4).
The \texttt{Batch} message only contains $\rp{i_1, p_1}, \ldots, \rp{i_\csbalbatchingwindow, p_\csbalbatchingwindow}$.
Upon receiving the \texttt{Batch} message, $\server$ collects in a set $U_\server$ all the ids it does not know ($i_j$ is in $U_\server$ iff $\server$ does not know $i_j$), and sends $U_\server$ back to $\broker$ by means of a \texttt{BatchAcquired} message (fig. \ref{figure:protocol}, step 5).
Upon receiving $\server$'s \texttt{BatchAcquired} message, $\broker$ builds a set $A_\server$ containing all id assignments that $\server$ is missing ($a_j$ is in $A_\server$ iff $i_j$ is in $U_\server$).
Having done so, $\broker$ sends a \texttt{Signatures} message to $\server$ (fig. \ref{figure:protocol}, step 6). 
The \texttt{Signatures} message contains the batch's multi-signature $m$, the straggler set $S$, and $A_\server$.
We underline the importance of sending id assignments upon request only.
Thinking to shave one round-trip off the protocol, $\broker$ could naively package in a single message all submissions, all (multi-)signatures, and all assignments relevant to the batch.
In doing so, however, $\broker$ would force each server to receive one assignment per submission, immediately forfeiting \csbal's oracular efficiency.
As we discuss in Section \ref{subsubsection:csbal.complexity.batchinglimit}, at the batching limit we assume that all servers already know all broadcasting clients.
In that case, both $U_\server$ and $A_\server$ are constant-sized, empty sets, adding only a vanishing amount of communication complexity to the protocol.
Upon receiving the \texttt{Signatures} message, $\server$ verifies and learns all assignments in $A_\server$.
Having done so, $\server$ knows $i_1, \ldots, i_\csbalbatchingwindow$.
As we outlined above, $\server$ can now efficiently authenticate the whole batch, verifying $m$ against the batch's root $r$, and each $i_j$ in $S$ against $p_j$.
Having established the integrity of the whole batch, $\server$ produces a \emph{witness shard} for the batch, i.e., a multi-signature $w_\server$ for $\qp{\texttt{Witness}, r}$, effectively affirming to have successfully authenticated the batch.
$\server$ sends $w_\server$ back to $\broker$ by means of a \texttt{WitnessShard} message (fig. \ref{figure:protocol}, step 7).
Having received a valid \texttt{WitnessShard} message from $f + 1$ servers, $\broker$ aggregates all witness shards into a \emph{witness} $w$.
Because $w$ is a plurality ($f+1$) certificate, at least one correct server necessarily produced a witness shard for the batch.
This means that at least one correct server has successfully authenticated the batch by means of client (multi-)signatures.
Because $w$ could not have been gathered if the batch was not properly authenticated, $w$ itself is sufficient to authenticate the batch, and $\broker$ can drop all (now redundant) client-generated (multi-)signatures for the batch.
Unlike $m$, $w$ is easy to verify, as it is signed by only $f + 1$, globally known servers.
Like $m$, $w$ authenticates $r$.
As such, any $p_j$ can now be authenticated just by producing $w$, and $\rp{i_j, p_j}$'s proof of inclusion $q_j$.
 
\miniparagraph{Gathering a commit certificate}
Having successfully gathered a witness $w$ for the batch, $\broker$ sends $w$ to all servers by means of a \texttt{Witness} message (fig. \ref{figure:protocol}, step 8). 
Upon receiving the \texttt{Witness} message, $\server$ moves on to check $\rp{i_1, p_1}, \ldots, \rp{i_\csbalbatchingwindow, p_\csbalbatchingwindow}$ for equivocations.
More precisely, $\server$ builds a set of \emph{exceptions} $E_\server$ containing the ids of all equivocating submissions in the batch ($i_j$ is in $E_\server$ iff $\server$ previously observed $\client_j$ submit a payload $p'_j$ that conflicts with $p_j$; we recall that $p_j$ and $p'_j$ conflict if their contexts are the same, but their messages are different).
$\server$ then produces a \emph{commit shard} for the batch, i.e., a multi-signature $c_\server$ for $\qp{\texttt{Commit}, r, E_\server}$, effectively affirming that $\server$ has found all submissions in the batch to be non-equivocated, \emph{except} for those in $E_\server$.
In the good case, every client is correct and $E_\server$ is empty.
Having produced $c_\server$, $\server$ moves on to build a set $Q_\server$ containing a \emph{proof of equivocation} for every element in $E_\server$.
Let us assume that $\server$ previously received from some $\client_k$ a payload $p'_k$ that conflicts with $p_k$.
$\server$ must have received $p'_k$ as part of some witnessed batch.
Let $r'_k$ identify the root of $p'_k$'s batch, let $w'_k$ identify $r'_k$'s witness, let $q'_k$ be $\rp{i_k, p'_k}$'s proof of inclusion in $r'_k$.
By exhibiting $\rp{r'_k, w'_k, p'_k}$, $\server$ can prove to $\broker$ that $\client_k$ equivocated: $p_k$ conflicts with $p'_k$, and $\rp{i_k, p'_k}$ is provably part of a batch whose integrity was witnessed by at least one correct server.
Furthermore, because correct clients never equivocate, $\rp{r'_k, w'_k, p'_k}$ is sufficient to convince $\broker$ that $\client_k$ is Byzantine.
For each $i_j$ in $E_\server$, $\server$ collects in $Q_\server$ a proof of equivocation $\rp{r'_j, w'_j, p'_j}$.
Finally, $\server$ sends a \texttt{CommitShard} message back to $\broker$ (fig. \ref{figure:protocol}, step 9).
The \texttt{CommitShard} message contains $c_\server$, $E_\server$ and $Q_\server$.
Upon receiving $\server$'s \texttt{CommitShard} message, $\server$ verifies $c_\server$ against $r$ and $E_\server$, then checks all proofs in $Q_\server$.
Having collected valid \texttt{CommitShard} messages from a quorum of servers $\server_1, \ldots, \server_{2f + 1}$, $\broker$ aggregates all commit shards into a \emph{commit certificate} $c$.
We underline that each $\server_j$ signed the same root $r$, but a potentially different set of exceptions $E_{\server_j}$.
Let $E$ denote the union of $E_{\server_1}, \ldots, E_{\server_{2f + 1}}$.
We call $E$ the batch's \emph{exclusion} set.
Because a proof of equivocation cannot be produced against a correct client, $\broker$ knows that all clients identified by $E$ are necessarily Byzantine.
In particular, because $\client$ is correct, $\rp{i = i_n}$ is guaranteed to not be in $E$.

\miniparagraph{Committing the batch}
Having collected a commit certificate $c$ for the batch, $\broker$ sends $c$ to all servers by means of a \texttt{Commit} message (fig. \ref{figure:protocol}, step 10).
Upon receiving the \texttt{Commit} message, $\server$ verifies $c$, computes the exclusion set $E$, then delivers every payload $p_j$ whose id $i_j$ is not in $E$.
Recalling that $c$ is assembled from a quorum of commit shards, at least $f + 1$ correct servers contributed to $c$.
This means that, if some $i_k$ is not in $E$, then at least $f + 1$ correct servers found $p_k$ not to be equivocated.
As in most BRB implementations~\cite{bra87asynchronous}, this guarantees that no two commit certificates can be gathered for equivocating payloads: \csbal's consistency is upheld.

\miniparagraph{The role of equivocation proofs}
As the reader might have noticed, $\broker$ does not attach any proof of equivocation to its \texttt{Commit} message.
Having received $\broker$'s commit certificate $c$, $\server$ trusts $\broker$'s exclusion set $E$, ignoring every payload whose id is in $E$.
This is not because $\server$ can trust $\broker$ to uphold validity.
On the contrary, $\server$ has \emph{no way} to determine that $\broker$ is not maliciously excluding the payload of a correct client.
Indeed, even if $\server$ were to verify a proof of exclusion for every element in $E$, a malicious $\beta$ could still censor a correct client simply by ignoring its \texttt{Submit} message in the first place.
Equivocation proofs are fundamental to \csbal's validity not because they \emph{force} malicious brokers to uphold validity, but because they \emph{enable} correct brokers to do the same.
Thanks to equivocation proofs, a malicious server cannot trick a correct broker into excluding the payload of a correct client.
This is enough to guarantee validity. 
As we discuss below, $\client$ successively submits $p$ to all brokers until it receives a certificate attesting that $p$ was delivered by at least one correct server.
Because we assume at least one broker to be correct, $\client$ is eventually guaranteed to succeed.

\miniparagraph{Notifying the clients}
Having delivered every payload whose id is not in the exclusion set $E$, $\server$ produces a \emph{completion shard} for the batch, i.e., a multi-signature $z_\server$ for $\qp{\texttt{Completion}, r, E}$, effectively affirming that $\server$ has delivered all submissions in the batch whose id is not in $E$.
$\server$ sends $z_\server$ to $\broker$ by means of a \texttt{CompletionShard} message (fig. \ref{figure:protocol}, step 11). 
Upon receiving $f + 1$ valid \texttt{CompletionShard} messages, $\broker$ assembles all completion shards into a \emph{completion certificate} $z$.
Finally, $\broker$ sends a \texttt{Completion} message to $\client_1, \ldots, \client_\csbalbatchingwindow$ (fig. \ref{figure:protocol}, step 12).
The \texttt{Completion} message contains $z$ and $E$.
Upon receiving the \texttt{Completion} message, $\client$ verifies $z$ against $E$, then checks that $i$ is not in $E$.
Because at least one correct server contributed a completion shard to $z$, at least one correct process delivered all payloads that $E$ did not exclude, including $p$.
Having succeeded in broadcasting $p$, $\client$ does not need to engage further, and can stop successively submitting $p$ to all brokers.

\miniparagraph{No one is left behind}
As we discussed above, upon receiving the commit certificate $c$, $\server$ delivers every payload in the batch whose id is not in the exclusion set $E$.
Having gotten at least one correct server to deliver the batch, $\broker$ is free to disengage, and moves on to assembling and brokering its next batch.
In a moment of asynchrony, however, all communications between $\broker$ and $\tilde \server$ might be arbitrarily delayed.
This means that $\tilde \server$ has no way of telling whether or not it will eventually receive batch and commit certificate: a malicious $\broker$ might have deliberately left $\tilde \server$ out of the protocol.
Server-to-server communication is thus required to guarantee totality.
Having delivered the batch, $\server$ waits for an interval of time long enough for all correct servers to deliver batch and commit certificate, \emph{should the network be synchronous and $\broker$ correct}.
$\server$ then sends to all servers an \texttt{OfferTotality} message (fig. \ref{figure:protocol}, step 13).
The \texttt{OfferTotality} message contains the batch's root $r$, and the exclusion set $E$.
In the good case, upon receiving $\server$'s \texttt{OfferTotality} message, every server has delivered the batch and ignores the offer.
This, however, is not the case for slow $\hat \server$, which replies to $\server$ with an \texttt{AcceptTotality} message (fig. \ref{figure:protocol}, step 14).
Upon receiving $\tilde \server$'s \texttt{AcceptTotality} message, $\server$ sends back to $\tilde \server$ a \texttt{Totality} message (fig. \ref{figure:protocol}, step 15).
The \texttt{Totality} message contains all submissions $\rp{i_1, p_1}, \ldots, \rp{i_\csbalbatchingwindow, p_\csbalbatchingwindow}$, id assignments for $i_1, \ldots, i_\csbalbatchingwindow$, and the commit certificate $c$.
Upon delivering $\server$'s \texttt{Totality} message, $\tilde \server$ computes $r$ from $\rp{i_1, p_1}, \ldots, \rp{i_\csbalbatchingwindow, p_\csbalbatchingwindow}$, checks $c$ against $r$, computes $E$ from $c$, and delivers every payload $p_j$ whose id $i_j$ is not in $E$.
This guarantees totality and concludes the protocol.

\subsection{Complexity}
\label{subsection:csbaloverview.complexity}

\miniparagraph{\dir density}
As we introduced in Section \ref{subsection:csbaloverview.protocol}, \csbal uses ids assigned by its underlying \dir (\dirshort) abstraction to identify payload senders. 
A \dirshort-assigned id is composed of two parts: a \emph{domain} and an \emph{index}.
Domains form a finite set $\domains$ whose size does not increase with the number of clients, indices are natural numbers.
Along with safety (e.g., no two processes have the same id) and liveness (e.g., every correct client that requests an id eventually obtains an id), \dirshort guarantees \emph{density}: the index part of any id is always smaller than the total number of clients $\clientcount$ (i.e., each id index is between $0$ and $\rp{\clientcount - 1}$).
Intuitively, this echoes the (stronger) density guarantee provided by \textsf{Oracle-\csbshort}, the oracle-based implementation of \csbshort we introduced in Section \ref{section:introduction} to bound \csbal's performance. 
In \textsf{Oracle-\csbshort}, the oracle organizes all clients in a list, effectively labeling each client with an integer between $0$ and $\rp{\clientcount - 1}$.
In a setting where consensus cannot be achieved, agreeing on a totally-ordered list of clients is famously impossible: a consensus-less \dirshort implementation cannot assign ids if $\abs{\domains} = 1$.
As we show in Appendix \ref{appendix:diral}, however, \dirshort can be implemented without consensus if servers are used as domains ($\domains = \servers$).
In our \dirshort implementation \diral, each server maintains an independent list of public keys.
In order to obtain an id, a client $\client$ has each server add its public key to its list, then selects a server $\server$ to be its \emph{assigner}.
In doing so, $\client$ obtains an id $\rp{\server, n}$, where $n \in 0..\rp{\clientcount - 1}$ is $\client$'s position in $\server$'s log.
In summary, a consensus-less implementation of \dirshort still guarantees that indices will be smaller than $\clientcount$, at the cost of a non-trivial domain component for each id.
This inflates the size of each individual id by $\ceil{\log_2\rp{\abs{\domains}}}$ bits.

\miniparagraph{Batching ids}
While \dirshort-assigned ids come with a non-trivial domain component, the size overhead due to domains vanishes when infinitely many ids are organized into a batch.
This is because domains are constant in the number of clients. 
Intuitively, as infinitely many ids are batched together, repeated domains become compressible.
When building a batch, a \csbal broker represents the set $I$ of sender ids not as a list, but as a map $\tilde i$.
To each domain, $\tilde i$ associates all ids in $I$ under that domain ($n$ is in $\tilde i\qp{d}$ iff $\rp{d, n}$ is in $I$).
Because $\tilde i$'s keys are fixed, as the size of $I$ goes to infinity, the bits required to represent $\tilde i$'s keys are completely amortized by those required to represent $\tilde i$'s values.
At the batching limit, the cost of representing each id in $\tilde i$ converges to that of representing its index only, $\ceil{\log_2\rp{\clientcount}}$.

\miniparagraph{Protocol cost}
As we discuss in Section \ref{subsubsection:csbal.complexity.batchinglimit}, at the batching limit we assume a good-case execution: links are synchronous, all processes are correct, and the set of brokers contains only one element. We additionally assume that infinitely many clients broadcast concurrently. Finally, we assume all servers to already know all broadcasting clients. Let $\broker$ denote the only broker. As all broadcasting clients submit their payloads to $\broker$ within a suitably narrow time window, $\broker$ organizes all submissions into a single batch with root $r$. Because links are synchronous and all clients are correct, every broadcasting client submits its multi-signature for $r$ in time. Having removed all individual signatures from the batch, $\broker$ is left with a single, aggregated multi-signature $m$ and an empty straggler set $S$. $\broker$ compresses the sender ids and disseminates the batch to all servers. As $m$ and $S$ are constant-sized, the amortized cost for a server to receive each payload $p$ is $\rp{\ceil{\log_2\rp{\clientcount}} + \abs{p}}$ bits. As $m$ authenticates the entire batch, a server authenticates each payload at an amortized cost of $0$ signature verifications. The remainder of the protocol unfolds as a sequence of constant-sized messages: because all broadcasting clients are known to all servers, no server requests any id assignment; witnesses are always constant-sized; and because all processes are correct, no client equivocates and all exception sets are empty. Finally, again by the synchrony of links, all offers of totality are ignored. In summary, at the batching limit a server delivers a payload at an amortized cost of $0$ signature verifications and $\rp{\ceil{\log_2\rp{\clientcount}} + \abs{p}}$ bits exchanged.

\miniparagraph{Latency}
As depicted in Figure \ref{figure:protocol}, the latency of \csbal is 10 message delays in the synchronous case (fast servers deliver upon receiving the broker’s Commit message), and at most 13 message delays in the asynchronous case (slow servers deliver upon receiving other servers’ Totality messages).
By comparison, the latency of the optimistic reliable broadcast algorithm by Cachin \emph{et al.}~\cite{cachin2001secure} is respectively 4 message delays (synchronous case) and 6 message delays (asynchronous case).
Effectively, \csbal trades oracular efficiency for a constant latency overhead.

\miniparagraph{Worst-case complexity}
In the worst case, a \csbal server delivers a $b$-bits payload by exchanging $O\rp{\rp{\log\rp{\clientcount} + b} \brokercount \servercount}$ bits, where $\clientcount$, $\brokercount$ and $\servercount$ respectively denote the number of clients, brokers and servers. In brief, the same id, payload and signature is included by each broker in a different batch (hence the $\brokercount$ term) and propagated in an all-to-all fashion (carried by \texttt{Totality} messages) across correct servers (hence the $\servercount$ term). By comparison, the worst-case communication complexity of Cachin \emph{et al.}'s optimistic reliable broadcast is $O\rp{l \servercount}$ per server, where $l$ is the length of the broadcast payload. A direct batched generalization of the same algorithm, however, would raise the worst-case communication to $O\rp{l \servercount^2}$ per server, similar to that of \csbal when $\servercount \sim \brokercount$. Both batched Bracha and \csbal can be optimized by polynomial encoding, reducing their per-server worst-case complexity to $O(l \servercount)$ and $O(\rp{\log\rp{\clientcount} + b} \brokercount)$ respectively. Doing so for \csbal, however, is beyond the scope of this paper.

\section{Conclusions}
\label{section:conclusions}

\miniparagraph{Our contributions}
In this paper we study \csb (\csbshort), a multi-shot variant of Byzantine Reliable Broadcast (BRB) whose interface is split between broadcasting clients and delivering servers. We introduce \textsf{Oracle-\csbshort}, a toy implementation of \csbshort that relies on a single, infallible oracle to uphold all \csbshort properties. Unless clients can be assumed to broadcast at a non-uniform rate, \textsf{Oracle-\csbshort}'s signature and communication complexities are optimal: in \textsf{Oracle-\csbshort}, a server delivers a payload $p$ by performing $0$ signature verifications, and exchanging $\rp{\ceil{\log_2\rp{\clientcount}} + \abs{p}}$ bits, where $\clientcount$ is the number of clients. We present \csbal, our implementation of \csbshort. \csbal upholds all \csbshort properties under classical BRB assumptions (notably asynchronous links and less than a third of faulty servers). When links are synchronous and all processes are correct, however, and at the limit of infinite concurrently broadcasting clients, \csbal's signature and communication complexities match those of \textsf{Oracle~\csbshort}.

\miniparagraph{Future work}
We hope to extend \csbal to allow multiple messages by the same client in the same batch.
We envision that this could be achieved by using other types of cryptographic accumulators or variants of Merkle trees, such as Merkle-Patricia trees~\cite{de2022overview}.
It would also be interesting to see if the worst-case performance of \csbal could be improved, e.g. by using error correction codes (ECC) or erasure codes, without significantly affecting its good-case performance.
Lastly, we hope to use \csbal's keys ideas to implement a total-order broadcast primitive, improving the scalability of existing SMR implementations.

\bibliographystyle{plainurl}
\bibliography{bibliography/reliable-broadcast}

\newpage

\tableofcontents

\newpage

\begin{appendix}
\section{\csbal: Full analysis}
\label{appendix:csbal}

\subsection{Interface}
\label{subsection:csbal.interface}

A \textbf{\csb (\csbshort) system} offers two interfaces, \cl (instance $\clin$) and \sr (instance $\srin$), exposing the following events:
\begin{itemize}
    \item \textbf{Request}: $\event{\clin}{Broadcast}[context, message]$: Broadcasts a message $message$ for context $context$ to all servers.
    \item \textbf{Indication} $\event{\srin}{Deliver}[client, context, message]$: Delivers the message $message$ broadcast by client $client$ for context $context$.
\end{itemize}
A \csb system satisfies the following properties:
\begin{enumerate}
    \item \textbf{No duplication}: No correct server delivers more than one message for the same client and context.
    \item \textbf{Integrity}: If a correct server delivers a message $m$ for context $c$ from a correct client $\client$, then $\client$ previously broadcast $m$ for $c$.
    \item\textbf{Consistency}: No two correct servers deliver different messages for the same client and context.
    \item\textbf{Validity}: If a correct client $\client$ broadcasts a message for context $c$, then eventually a correct server delivers a message for $c$ from $\client$.
    \item\textbf{Totality}: If a correct server delivers a message for context $c$ from a client $\client$, then eventually every correct server delivers a message for $c$ from $\client$.
\end{enumerate}
\subsection{Pseudocode}
\label{subsection:csbal.pseudocode}

The full code of our \csbshort implementation \csbal can be found can be found in Appendix \ref{appendix:pseudocode}. More in detail, Section \ref{subsection:pseudocode.csbalclient} implements \csbal's client, Section \ref{subsection:pseudocode.csbalbroker} implements \csbal's broker, and Section \ref{subsection:pseudocode.csbalserver} implements \csbal's server.
\subsection{Correctness}
\label{subsection:csbal.correctness}

In this section, we prove to the fullest extent of formal detail that \csbal implements a \csb system.

\subsubsection{No duplication}

In this section, we prove that \csbal satisfies no duplication. 

\begin{lemma}
\label{lemma:deliveredtracksdelivered}
Let $\server$ be a correct server, let $\client$ be a client, let $c$ be a context. If $\rp{\client, c} \notin delivered$ at $\server$, then $server$ did not deliver a message from $\client$ for $c$.
\begin{proof}
Upon initialization, $delivered$ is empty at $\server$ (line \ref{line:sdeliverinit}). Moreover, $\server$ adds $\rp{\client, c}$ to $delivered$ only by executing line \ref{line:sdeliveredupdate}. Immediately after doing so, $\server$ delivers a message from $\client$ for $c$ (line \ref{line:sdeliver}), and because $\server$ never removes elements from $delivered$, the lemma is proved.
\end{proof}
\end{lemma}

\begin{theorem}
\label{theorem:noduplication}
\csbal satisfies no duplication.
\begin{proof}
Let $\server$ be a correct server, let $\client$ be a client, let $c$ be a context. $\server$ delivers a message for $c$ from $\client$ only by executing line \ref{line:sdeliver}. $\server$ does so only if $\rp{\client, c} \notin delivered$ (line \ref{line:sdelivercheck}). By Lemma \ref{lemma:deliveredtracksdelivered}, we then have that $\server$ never delivers a message for $c$ from $\client$ more than once, and the theorem is proved.
\end{proof}
\end{theorem}

\subsubsection{Consistency}

\begin{notation}[Contexts, messages]
We use $\contexts$ and $\messages$ to respectively denote the set of contexts and messages broadcast in a \csb system.
\end{notation}

\begin{definition}[Ideal accumulator]
\label{definition:idealaccumulator}
An \textbf{ideal accumulator} is a tuple $(\auniverse, \aroots, \aproofs, \arootsym, \aproofsym, \averifysym)$ composed by
\begin{itemize}
    \item An \textbf{universe} $\auniverse$;
    \item A set of \textbf{roots} $\aroots$;
    \item A set of \textbf{proofs} $\aproofs$;
    \item A \textbf{root function} $\arootsym: \auniverse^{<\infty} \rightarrow \aroots$;
    \item A \textbf{proof function} $\aproofsym: \auniverse^{<\infty} \times \mathbb{N} \rightarrow \aproofs$;
    \item A \textbf{verification function} $\averifysym: \aroots \times \aproofs \times \mathbb{N} \times \auniverse \rightarrow \cp{\true, \false}$;
\end{itemize}
such that
\begin{align*}
    \forall z \in \auniverse^{<\infty}, \forall n \leq \abs{z}, &\;\;\averify{\aroot{z}}{\aproof{z}{n}}{n}{z_n} = \true \\
    \forall z \in \auniverse^{<\infty}, \forall n \leq \abs{z}, \forall q \neq z_n, \forall p \in \aproofs, &\;\;\averify{\aroot{z}}{p}{n}{q} = \false
\end{align*}
Let $r \in \aroots$, let $p \in \aproofs$, let $n \in \mathbb{N}$, let $x \in \auniverse$ such that $\averify{r}{p}{n}{x} = \true$. We say that $p$ is a proof for $x$ from $r$.
\end{definition}

\begin{lemma}
\label{lemma:rootisinjective}
Let $\rp{\auniverse, \aroots, \aproofs, \arootsym, \aproofsym, \averifysym}$ be an ideal accumulator. We have that $\arootsym$ is injective.
\begin{proof}
Let us assume by contradiction that $z, z' \in \auniverse^{<\infty}$ exist such that $z \neq z'$ and $\aroot{z} = \aroot{z'}$. Let $n \in \mathbb{N}$ such that $z_n \neq z'_n$. By Definition \ref{definition:idealaccumulator} we have
\begin{align*}
    \true &= \averify{\aroot{z}}{\aproof{z}{n}}{n}{z_n} = \\
    &= \averify{\aroot{z'}}{\aproof{z}{n}}{n}{z_n} = \false
\end{align*}
which immediately proves the lemma.
\end{proof}
\end{lemma}

Throughout the remainder of this document, we use Merkle hash-trees as a cryptographic approximation for an ideal accumulator
\begin{equation*}
    \rp{\rp{\ids \times \contexts \times \messages}, \aroots, \aproofs, \arootsym, \aproofsym, \averifysym}
\end{equation*}
on the universe of triplets containing an id, a context and a message. In brief: while a successfully verifiable, corrupted proof can theoretically be forged for a Merkle tree (Merkle tree roots are highly non-injective), doing so is unfeasible for a computationally bounded adversary. An in-depth discussion of Merkle trees is beyond the scope of this document.

\begin{lemma}
\label{lemma:sidssamelengthaspayloads}
Let $\server$ be a correct server. Let $\rp{r, b} \in batches$ at $\server$, let
\begin{align*}
    i &= b.ids \\
    p &= b.payloads
\end{align*}
We have $\abs{i} = \abs{p}$.
\begin{proof}
We start by noting that, upon initialization, $batches$ is empty at $\server$ (line \ref{line:sinitbatches}). Moreover, $\server$ adds $Batch\cp{ids: i, payloads: p, \tail}$ to $batches$ only by executing line \ref{line:sbatchesupdate}: $\server$ does so only if $\abs{i} = \abs{p}$ (lines \ref{line:sbatchcheck} and \ref{line:sbatchcheckfail}). 
\end{proof}
\end{lemma}

\begin{lemma}
\label{lemma:sidsnonrepeating}
Let $\server$ be a correct server. Let $\rp{r, b} \in batches$ at $\server$, let
\begin{align*}
    i &= b.ids
\end{align*}
The elements of $i$ are all distinct.
\begin{proof}
The proof of this lemma is identical to that of Lemma \ref{lemma:sidssamelengthaspayloads}, and we omit it for the sake of brevity.
\end{proof}
\end{lemma}

\begin{definition}[Join function]
\label{definition:joinfunction}
The \textbf{join function}
\begin{equation*}
    \joinsym: \cp{\rp{i, p} \suchthat i \in \ids^{<\infty}, p \in \rp{\contexts \times \messages}^{<\infty}, \abs{i} = \abs{p}} \rightarrow \rp{\ids \times \contexts \times \messages}^{<\infty}
\end{equation*}
is defined by
\begin{gather*}
    \abs{\join{i}{p}} = \rp{\abs{i} = \abs{p}} \\
    \join{i}{p}_n = \rp{i_n, c_n, m_n}, \;\text{where}\; \rp{c_n, m_n} = p_n
\end{gather*}
\end{definition}

\begin{lemma}
\label{lemma:servercomputesmerkletree}
Let $\server$ be a correct server. Let $\rp{r, b} \in batches$ at $\server$, let $l = \join{b.ids}{b.payloads}$. We have $r = \aroot{l}$.
\begin{proof}
We underline that, because by Lemma \ref{lemma:sidssamelengthaspayloads} we have $\abs{b.ids} = \abs{b.payloads}$, the definition of $l$ is well-formed. We start by noting that, upon initialization, $batches$ is empty at $\server$ (line \ref{line:sinitbatches}). Moreover, $\server$ sets 
\begin{equation*}
    batches\qp{r} = Batch\cp{ids: i, payloads: p, \tail}
\end{equation*}
only by executing line \ref{line:sbatchesupdate}. Immediately before doing so, $\server$ computes $r = \aroot{l}$ (lines \ref{line:scomputeleaves}, \ref{line:scomputetree} and \ref{line:scomputeroot}).
\end{proof}
\end{lemma}

\begin{lemma}
\label{lemma:matchingrootsmeansmatchingbatches}
Let $\server$, $\server'$ be correct servers. Let $\rp{r, b} \in batches$ at $\server$, let $\rp{r, b'} \in batches$ at $\server'$. We have $b = b'$.
\begin{proof}
Upon initialization, $batches$ is empty at both $\server$ and $\server'$. Moreover, $\server$ and $\server'$ respectively set
\begin{align*}
    batches\qp{r} &= Batch\cp{ids: i, payloads: p, tree: t} \\
    batches\qp{r} &= Batch\cp{ids: i', payloads: p', tree: t'} \\
\end{align*}
only by executing line \ref{line:sbatchesupdate}. By Lemma \ref{lemma:servercomputesmerkletree}, we have
\begin{equation*}
    \aroot{\join{i}{p}} = r = \aroot{\join{i'}{p'}}
\end{equation*}
which, by Lemma \ref{lemma:rootisinjective} and the injectiveness of $\joinsym$, proves $i = i'$ and $p = p'$. Because $\server$ (resp., $\server'$) computes $t$ (resp. $t'$) as a pure function of $i$ and $p$ (resp., $i' = i$ and $p' = p$) (line \ref{line:scomputetree}), we have $t = t'$. This proves $b = b'$ and concludes the lemma.
\end{proof}
\end{lemma}

\begin{lemma} 
\label{lemma:serverknowsallids}
Let $\server$ be a correct server, let $i$ be an id. Whenever $\server$ invokes $\dirin\qp{i}$, $\server$ knows $i$.
\begin{proof}
$\server$ invokes $\dirin\qp{i}$ only by executing lines \ref{line:ssignatureclientverification}, \ref{line:smultiverifyreduction} or \ref{line:stakekeycard}. If $\server$ invokes $\dirin\qp{i}$ by executing lines \ref{line:ssignatureclientverification} or \ref{line:smultiverifyreduction}, then $i \in I$ (lines \ref{line:ssignatureclientmissingcheck} and \ref{line:ssignatureclientmissingreturn}, line \ref{line:sdeterminemultisigners} respectively) for some $I \subseteq \ids$ such that, for all $i' \in I$, $\server$ knows $i'$ (lines \ref{line:sdircheck} and \ref{line:sdirreturn}). Similarly, if $\server$ invokes $\dirin\qp{i}$ by executing line \ref{line:stakekeycard}, then $i \in I$ (line \ref{line:sdeliverfrombrokerloop}) for some $I \subseteq \ids$ such, for all $i' \in I$, $\server$ knows $i'$ (lines \ref{line:sdircheckcommit} and \ref{line:sdirecheckcommitreturn}).
\end{proof}
\end{lemma}

\begin{lemma}
\label{lemma:serverknowssender}
Let $\server$ be a correct server. Let $\client$ be a client, let $c$ be a context, let $m$ be a message such that $\server$ delivers $\rp{\client, c, m}$. We have that $\server$ knows $\client$.
\begin{proof}
$\server$ delivers $\rp{\client, c, m}$ only by executing line \ref{line:sdeliver}. Immediately before doing so, $\server$ invokes $\dirin\qp{i}$ for some $i \in \ids$ to obtain $\client$ (line \ref{line:stakekeycard}). By Lemma \ref{lemma:serverknowsallids} we then have that $\server$ knows $i$ which, by Definition \ref{definition:directoryrecord}, proves that $\server$ knows $k$ as well, and concludes the lemma.
\end{proof}
\end{lemma}

\begin{lemma}
\label{lemma:deliverymeanssignature}
Let $\server$ be a correct server. Let $\client$ be a client, let $c$ be a context, let $m$ be a message such that $\server$ delivers $\rp{\client, c, m}$. Let $i = \directorymap{\client}$. Some $l \in \rp{\ids \times \contexts \times \messages}^{<\infty}$, $\server_1, \ldots, \server_{2f + 1} \in \servers$, $\epsilon_1, \ldots, \epsilon_{2f + 1} \subseteq \ids$ exist such that:
\begin{itemize}
    \item For some $k$, $l_k = \rp{i, c, m}$;
    \item For all $n$, $i \notin \epsilon_n$;
    \item For all $n$, $\server_n$ signed $\qp{\texttt{Commit}, \aroot{l}, \epsilon_n}$.
\end{itemize}
\begin{proof}
We underline the soundness of the lemma's statement. Because $\server$ delivers $\rp{\client, ..}$, by Lemma \ref{lemma:serverknowssender} $\server$ knows $\client$, and by Notation \ref{notation:directorymapping} $\directorymap{\client}$ is well-defined.

\medskip

We start by noting that some $b = batches\qp{r}$ exists (lines \ref{line:srootinbatches}, \ref{line:srootinbatchesreturn} and \ref{line:sbatchesname}) such that, for some $k$, we have $i = b.ids\qp{k}$ and $\rp{c, m} = b.payloads\qp{k}$ (line \ref{line:sdeliverfrombrokerloop}). Let $l = \join{b.ids}{b.payloads}$. By Definition \ref{definition:joinfunction} we immediately have $l_k = (i, c, m)$. Moreover, by Lemma \ref{lemma:servercomputesmerkletree}, we have $r = \aroot{l}$. 

\medskip

Moreover, some $S \subseteq \servers$ exists such that: 
\begin{itemize}
    \item $\abs{S} \geq 2f + 1$ (lines \ref{line:squorumcsignatureconflicts} and \ref{line:squorumcsignatureconflictsreturn});
    \item For all $\server' \in S$, some $\epsilon\rp{\server'}$ exists such that $\server'$ signed $\qp{\texttt{Commit}, r, \epsilon\rp{\server'}}$ (lines \ref{line:ssignerinit}, \ref{line:svaliditysignatureconflicts}, \ref{line:svaliditysignatureconflictsreturn} and \ref{line:ssignerextend}: $S$ is obtained by iteratively extending an initially empty set with certificate signers);
\end{itemize}
Let $\server_1, \ldots, \server_{2f + 1}$ be distinct elements of $S$, let $\epsilon_n = \epsilon\rp{\server_n}$.

\medskip

Finally, some $\epsilon \supseteq \epsilon_1 \cup \ldots \cup \epsilon_{2f + 1}$ exists (lines \ref{line:spatches} and \ref{line:sunionexclusion}) such that $i \notin \epsilon$ (line \ref{line:sdelivercheck}). For all $n$ we then obviously have $i \notin \epsilon_n$.

\medskip

In summary: $l_k = \rp{i, c, m}$; for all $n$, $\server_n$ signed $\qp{\texttt{Commit}, \aroot{l} = r, \epsilon_n}$; for all $n$, $i \notin \epsilon_n$. The lemma is therefore proved.
\end{proof}
\end{lemma}

\begin{lemma}
\label{lemma:messagesdoesntchange}
Let $\server$ be a correct server, let $i$ be an id, let $c$ be a context, let $m$ be a message. Let $t \in \mathbb{R}$ such that, at time $t$, $messages\qp{\rp{i, c}} = \rp{m, \any}$. At any time $t' > t$, we have $messages\qp{\rp{i, c}} = \rp{m, \any}$ as well.
\begin{proof}
After setting $messages\qp{\rp{i, c}} = \rp{m, \any}$, $\server$ sets $messages\qp{\rp{i, c}} = \rp{m, \any}'$ only by executing line \ref{line:updatevaraiablemessage}. It does so only if $m = m'$ (line \ref{line:sconflictcheck}).
\end{proof}
\end{lemma}

\begin{lemma}
\label{lemma:rememberwhatyousign}
Let $\server$ be a correct server. Let $i$ be an id, let $c$ be a context, let $m$ be a message. Let $l \in \rp{\ids \times \contexts \times \messages}^{<\infty}$, let $\epsilon \subseteq \ids$ such that:
\begin{itemize}
    \item For some $k$, $l_k = \rp{i, c, m}$;
    \item $i \notin \epsilon$;
    \item $\server$ signs $\qp{\texttt{Commit}, \aroot{l}, \epsilon}$.
\end{itemize}
We have that $\server$ sets $messages\qp{\rp{i, c}} = (m, \any)$.
\begin{proof}
Prior to signing $\qp{\texttt{Commit}, \aroot{l}, \epsilon}$ (line \ref{line:ssigncommit}), $\server$ retrieves $b' = batches\qp{\aroot{l}}$ (lines \ref{line:srootinbatchesw} and \ref{line:srootinbatchesreturnw} and \ref{line:sbatchesnamew}). Let $l' = \join{b'.ids}{b'.payloads}$. By Lemma \ref{lemma:servercomputesmerkletree} we have $\aroot{l'} = \aroot{l}$ which, by Lemma \ref{lemma:rootisinjective}, proves $l = l'$.

\medskip

Subsequently, $\server$ loops through all elements of $l = l'$ (line \ref{line:scheckidsummision}). For each $\rp{i_j, c_j, m_j}$ in $l$, $\server$ either adds $i_j$ to $\epsilon$ (line \ref{line:sconflictsupdate}), or sets $messages\qp{i_j, c_j} = \rp{m_j, \any}$ (line \ref{line:updatevaraiablemessage}). Because $\server$ does so in particular for $j = k$, and $i \notin \epsilon$, $\server$ sets $messages\qp{\rp{i, c}} = \rp{m, \any}$, and the lemma is proved.
\end{proof}
\end{lemma}

\begin{theorem}
\label{theorem:consistency}
\csbal satisfies consistency.
\begin{proof}
Let $\server, \server'$ be two correct servers. Let $\client$ be a client, let $c$ be a context, let $m, m'$ be messages such that $\server$ and $\server'$ deliver $\rp{\client, c, m}$ and $\rp{\client, c, m'}$ respectively. By Lemma \ref{lemma:serverknowssender}, $\server$ knows $\client$. Let $i = \directorymap{\client}$. By Lemma \ref{lemma:deliverymeanssignature}, and noting that at most $f$ servers are Byzantine, some $l \in \rp{\ids \times \contexts \times \messages}^{<\infty}$, $\server_1, \ldots, \server_{f + 1} \in \servers$, and $\epsilon_1, \ldots, \epsilon_{f + 1}$ exist such that: for some $k$, $l_k = \rp{i, c, m}$; for all $n$, $\server_n$ is correct; for all $n$, $i \notin \epsilon_n$; for all $n$, $\server_n$ signed $\qp{\texttt{Commit}, \aroot{l}, \epsilon_n}$. By Lemmas \ref{lemma:rememberwhatyousign} and \ref{lemma:messagesdoesntchange}, $messages\qp{\rp{i, c}}$ is permanently set to $\rp{m, \any}$ at $\server_1, \ldots, \server_{f + 1}$.

\medskip

By an identical reasoning, some $\server'_1, \ldots, \server'_{f + 1}$ exist such that for all $n$, $\server'_n$ is correct and permanently sets $messages\qp{\rp{i, c}} = \rp{m', \any}$. Because $\rp{\server_1, \ldots, \server_{f + 1}}$ and $\rp{\server'_1, \ldots, \server'_{f + 1}}$ intersect in at least one server, we have $m = m'$, and the theorem is proved.
\end{proof}
\end{theorem}

\subsubsection{Totality}

\begin{notation}[Ordering]
Let $X$ be a set endowed with a total order relationship. We use $\sorted{X} \subset X^{<\infty}$ to denote the set of finite, non-decreasing sequences on $X$. Let $Y \subseteq X$. We use $\sort{Y}$ to denote \textbf{sorted} $Y$, i.e., the sequence containing all elements of $Y$ in ascending order.
\end{notation}

\begin{definition}[Id compression and expansion]
\label{definition:idcompressionandexpansion}
The \textbf{id compression function}
\begin{equation*}
    \compresssym: \sorted{\ids} \rightarrow \rp{\domains \rightarrow \powerset{\mathbb{N}}}
\end{equation*}
is defined by
\begin{equation*}
    n \in \compress{i}\rp{d} \Longleftrightarrow \exists j \suchthat i_j = \rp{d, n}
\end{equation*}
The \textbf{id expansion function}
\begin{equation*}
    \expandsym: \rp{\domains \rightarrow \powerset{\mathbb{N}}} \rightarrow \sorted{\ids}
\end{equation*}
is defined by
\begin{equation*}
    \expand{f} = \sort{\cp{\rp{d, n} \suchthat n \in f\rp{d}}}
\end{equation*}
\end{definition}

\begin{lemma}
\label{lemma:expandandcompresscancelout}
Let $i \in \sorted{\ids}$. We have $\expand{\compress{i}} = i$.
\begin{proof}
It follows immediately from Definition \ref{definition:idcompressionandexpansion}.
\end{proof}
\end{lemma}

\begin{lemma}
\label{lemma:importbatch}
Let $\server, \server'$ be correct processes, let $\rp{r, b} \in batches$ at $\server$, let 
\begin{align*}
    i &= b.ids \\
    p &= b.payloads
\end{align*}
Upon evaluating $handle\_batch\rp{\compress{i}, p}$, $\server'$ sets $batches\qp{r} = b$.
\begin{proof}
We start by noting that, upon initialization, $batches$ is empty at $\server$ (line \ref{line:sinitbatches}). Moreover $\server$ adds $\rp{r, b}$ to $batches$ only by executing line \ref{line:sbatchesupdate}. $\server$ does so only if $i$ has no duplicates, and $\abs{i} = \abs{p}$ (line \ref{line:sbatchcheck}).

\medskip

Upon invoking $handle\_batch\rp{\compress{i}, p}$, $\server'$ verifies that $i = \expand{\compress{i}}$ (line \ref{line:sexpandprocedure}, see Lemma \ref{lemma:expandandcompresscancelout}) has no duplicates and satisfies $\abs{i} = \abs{p}$ (line \ref{line:sbatchcheck}). Having done so, $\server'$ sets $batches\qp{r} = b'$ for some $b'$. However, by Lemma \ref{lemma:matchingrootsmeansmatchingbatches} we have $b = b'$, and the lemma is proved.



\end{proof}
\end{lemma}

\begin{lemma}
\label{lemma:importcommit}
Let $\server, \server'$ be correct processes, let $\rp{\rp{r, \epsilon}, \zeta} \in commits$ at $\server$, let $\rp{r, b} \in batches$ at $\server'$ such that for all $i \in b.ids$, $\server'$ knows $i$. Upon evaluating $handle\_commit\rp{r, \zeta}$, $\server'$ sets $commits\qp{\rp{r, \epsilon}} = \zeta$.
\begin{proof}
We start by noting that, upon initialization, $commits$ is empty at $\server$ (lines \ref{line:sinitcommits} and \ref{line:svaliditysignatureconflictsreturn}). Moreover $\server$ adds $\rp{\rp{r, \epsilon}, \zeta}$ to $commits$ only by executing line \ref{line:scommitsupadte}. $\server$ does so only if, for all $\rp{c, z} \in \zeta$ (line \ref{line:spatches}), $z$ verifies correctly against $\qp{\texttt{Commit}, r, c}$ (line \ref{line:svaliditysignatureconflicts}) and
\begin{equation*}
    \abs{\bigcup_{\rp{\any, z} \in \zeta} z.signers()} \geq 2f + 1
\end{equation*}
(lines \ref{line:squorumcsignatureconflicts} and \ref{line:squorumcsignatureconflictsreturn}). We additionally have
\begin{equation*}
    \epsilon = \bigcup_{\rp{c, \any} \in \zeta} c
\end{equation*}
(line \ref{line:sunionexclusion}).

\medskip

Upon invoking $handle\_commits\rp{r, \zeta}$, by hypothesis $\server'$ verifies that $r \in batches$ (line \ref{line:srootinbatches}) and, for all $i \in b.ids$ (line \ref{line:sbatchesname}), $i \in \dirin$ (line \ref{line:sdircheckcommit}). Moreover, because $verify$ is a pure procedure, $\server'$ passes the checks at lines \ref{line:svaliditysignatureconflicts} and \ref{line:squorumcsignatureconflicts}. Next, $\server'$ computes 
\begin{equation*}
    \epsilon' = \bigcup_{\rp{c, \any} \in \zeta} c = \epsilon
\end{equation*}
(line \ref{line:sunionexclusion}) and sets $commits\qp{\rp{r, \epsilon'}} = \zeta$ (line \ref{line:scommitsupadte}), concluding the lemma.
\end{proof}
\end{lemma}

\begin{lemma}
\label{lemma:committedbatchishad}
Let $\server$ be a correct process, let $\rp{r, \any} \in commits$ at $\server$. We have $r \in batches$ at $\server$.
\begin{proof}
Upon initialization, $commits$ is empty at $\server$ (line \ref{line:sinitcommits}). Moreover, $\server$ adds $\rp{r, \any}$ to $commits$ only by executing line \ref{line:scommitsupadte}. $\server$ does so only if $r \in batches$ (lines \ref{line:srootinbatches} and \ref{line:srootinbatchesreturn}).
\end{proof}
\end{lemma}

\begin{lemma}
\label{lemma:committedbatchisknown}
Let $\server$ be a correct process, let $\rp{r, \any} \in commits$ at $\server$. Let $b = batches\qp{r}$ at server. For all $i \in b.ids$, $\server$ knows $i$.
\begin{proof}
We start by noting that, by Lemma \ref{lemma:committedbatchishad}, the definition of $b$ is well-formed. Upon initialization, $commits$ is empty at $\server$ (line \ref{line:sinitcommits}). Moreover, $\server$ adds $\rp{r, \any}$ to $commits$ only by executing line \ref{line:scommitsupadte}. $\server$ does so only if, for all $i$ in $b.ids$, $\server$ knows $i$ (lines \ref{line:sbatchesname} and \ref{line:sdircheckcommit} and \ref{line:sdirecheckcommitreturn}).
\end{proof}
\end{lemma} 

\begin{lemma}
\label{lemma:allmessagesincommitsaredelivered}
Let $\server$ be a correct process, let $\rp{r, \epsilon} \in commits$ at $\server$. Let $b = batches\qp{r}$ at $\server$, let
\begin{align*}
    i &= b.ids \\
    p &= b.payloads    
\end{align*}
For all $j$, let $\rp{c_j, \any} = p_j$. For all $j$ such that $i_j \notin \epsilon$, $\server$ delivered a message from $\directorymap{i_j}$ for $c_j$.
\begin{proof}
We underline that, by Lemma \ref{lemma:committedbatchisknown}, $\directorymap{i_j}$ is well-defined for all $j$. Upon initialization, $commits$ is empty at $\server$ (line \ref{line:sinitcommits}). Moreover, $\server$ adds $\rp{r, \epsilon}$ to $commits$ only by executing line \ref{line:scommitsupadte}. Upon doing so, $\server$ retrieves $b$ from $batches$ (line \ref{line:sbatchesname}) and delivers a message from $\directorymap{i_j}$ for $c_j$ (lines \ref{line:stakekeycard} and \ref{line:sdeliver}) for all $j$ such that: $i_j \notin \epsilon$; and $\server$ did not previously deliver a message from $\directorymap{i_j}$ for $c_j$ (line \ref{line:sdelivercheck}, see Lemma \ref{lemma:deliveredtracksdelivered}).
\end{proof}
\end{lemma}

\begin{theorem}
\csbal satisfies totality.
\begin{proof}
Let $\server, \server'$ be correct servers. Let $\client$ be a client, let $c$ be a context such that $\server$ delivers a message from $\client$ for $c$. $\server$ delivers a message from $\client$ for $c$ only by executing line \ref{line:sdeliver}. Immediately before doing so, $\server$ sets $commits\qp{r, \epsilon} = \zeta$ for some $r$, $\epsilon$ and $\zeta$ such that $\directorymap{\client} \notin \epsilon$ (lines \ref{line:scommitsupadte}, \ref{line:stakekeycard} and \ref{line:sdelivercheck}). Let $b = batches\qp{r}$ at $\server$ (by Lemma \ref{lemma:committedbatchishad}, $r \in batches$ at $\server$), let
\begin{align*}
    i &= b.ids \\
    p &= b.payloads
\end{align*}
For all $j$, let $\rp{c_j, \any} = p_j$. For some $k$ we have $i_k = \directorymap{\client}$ and $c_k = c$ (lines \ref{line:sbatchesname}, \ref{line:sdeliverfrombrokerloop} and \ref{line:stakekeycard}).

\medskip

Immediately after delivering a message from $\client$ for $c$, $\server$ sets a timer for $\qp{\texttt{OfferTotality}, r, \epsilon}$ (line \ref{line:soffertotality}). When the timer eventually rings (line \ref{line:soffertotalityring}), $\server$ sends an $\qp{\texttt{OfferTotality}, r, \epsilon}$ message to all servers, including $\server'$ (lines \ref{line:sforallserver} and \ref{line:ssendoffertotality}). Upon eventually delivering $\qp{\texttt{OfferTotality}, r, \epsilon}$ (line \ref{line:sdeliveroffertotality}), $\server'$ checks if $\rp{r, \epsilon} \in commits$ (line \ref{line:scommitsdelivercheck}). Let us assume that $\rp{r, \epsilon} \in commits$ at $\server'$. Let $b' = batches\qp{r}$ at $\server'$ (by Lemma \ref{lemma:committedbatchishad}, $r \in batches$ at $\server'$). By Lemma \ref{lemma:matchingrootsmeansmatchingbatches}, we have $b' = b$. As a result, by Lemma \ref{lemma:allmessagesincommitsaredelivered}, $\server'$ already delivered a message from $\directorymap{i_k} = \client$ for $c_k = c$. Throughout the remainder of this proof, we assume $\rp{r, \epsilon} \notin commits$ at $\server'$.

\medskip

Upon verifying that $\rp{r, \epsilon} \notin commits$ (line \ref{line:scommitsdelivercheck}), $\server'$ sends back to $\server$ an $\qp{\texttt{AcceptTotality}, r, \epsilon}$ message (line \ref{line:ssendaccepttotality}). Upon delivering $\qp{\texttt{AcceptTotality}, r, \epsilon}$ (line \ref{line:sdeliveraccepttotality}), $\server$ verifies that $r \in batches$ and $\rp{r, \epsilon} \in commits$ (line \ref{line:sbatchinbatches}), exports an array $a$ of assignments for all elements of $i$ (line \ref{line:sexportassigments}) (by Lemma \ref{lemma:committedbatchisknown} $\server$ knows all elements of $i$), and sends $\qp{\texttt{Totality}, r, a, \rp{\compress{i}, p}, \zeta}$ to $\server'$ (line \ref{line:sserverbacth})

\medskip

Upon delivering $\qp{\texttt{Totality}, r, a, \rp{\compress{i}, p}, \zeta}$ (line \ref{line:sdelivertotality}), $\server'$ imports all elements of $a$ (line \ref{line:simportassigmentstotality}). Having done so, $\server'$ knows all elements of $i$. Then, $\server'$ invokes $handle\_batch\rp{\compress{i}, p}$ (line \ref{line:shandlebatchdeliver}). By Lemma \ref{lemma:importbatch}, having done so $\server'$ sets $batches\qp{r} = b$. Finally, $\server'$ invokes $handle\_commit\rp{r, \zeta}$ (line \ref{line:shandlecommitsdeliver}). By Lemma \ref{lemma:importcommit}, having done so $\server'$ sets $commits\qp{\rp{r, \epsilon}} = \zeta$. By Lemma \ref{lemma:allmessagesincommitsaredelivered}, $\server'$ delivers a message from $\directorymap{\client} = i_k$ for $c = c_k$.

\medskip

In summary, if $\server$ delivers a message from $\client$ for $c$, then $\server'$ eventually delivers a message from $\client$ for $c$ as well, and the theorem is proved.
\end{proof}
\end{theorem}

\subsubsection{Integrity}

In this section, we prove that \csbal satisfies integrity.

\begin{lemma}
\label{lemma:delivermessagecheckingthewitness}
Let $\server$ be a correct server. Let $\client$ be a client, let $c$ be a context, let $m$ be a message such that $\server$ delivers $m$ from $\client$ for $c$. Some correct server $\server'$ and some $\rp{r, b} \in batches$ at $\server'$ exists such that $\server'$ signed $\qp{\texttt{Witness, r}}$ and, with
\begin{align*}
    i &= b.ids \\
    p &= b.payloads
\end{align*}
some $k$ exists such that $i_k = \directorymap{\client}$ and $p_k = \rp{c, m}$.
\begin{proof}
We start by noting that $\server$ delivers $m$ from $\client$ for $c$ only by executing line \ref{line:sdeliver}. When $\server$ does so, some $\rp{r, b} \in batches$ exists at $\server$ (line \ref{line:sbatchesname}) such that, with
\begin{align*}
    i &= b.ids \\
    p &= b.payloads
\end{align*}
some $k$ exists such that $i_k = \directorymap{\client}$ and $p_k = \rp{c, m}$ (lines \ref{line:sdeliverfrombrokerloop} and \ref{line:stakekeycard}). Moreover, a set $S \subseteq \servers$ exists such that $\abs{S} \geq 2f + 1$ (lines \ref{line:squorumcsignatureconflicts} and \ref{line:squorumcsignatureconflictsreturn}) and, for all $\tilde{\server} \in S$, $\tilde{\server}$ signed a $\qp{\texttt{Commit}, r, \any}$ message (lines  \ref{line:ssignerinit}, \ref{line:svaliditysignatureconflicts}, \ref{line:svaliditysignatureconflictsreturn} and \ref{line:ssignerextend}). Noting that at most $f$ processes are Byzantine, at least one element $\server^*$ of $S$ is correct.

\medskip

$\server^*$ signs $\qp{\texttt{Commit}, r, \any}$ only upon executing line \ref{line:ssigncommit}. $\server^*$ does so only if at least $f + 1$ servers signed $\qp{\texttt{Witness}, r}$ (lines \ref{line:switnesssignaturecheck} and \ref{line:switnesssignaturecheckreturn}). Noting that at most $f$ processes are Byzantine, at least one correct server $\server'$ signed $\qp{\texttt{Witness}, r}$.

\medskip

$\server'$ signs $\qp{\texttt{Witness}, r}$ only by executing line \ref{line:ssignwitness}. It does so only if some $\rp{r, b'} \in batches$ exists at $\server'$ (lines \ref{line:scheckbatcheshandlesignature}, \ref{line:scheckbatcheshandlesignaturereturn}). By Lemma \ref{lemma:matchingrootsmeansmatchingbatches}, we have $b' = b$.

\medskip

In summary, some correct server $\server'$ exists such that $\server'$ signed $\qp{\texttt{Witness}, r}$ and $\rp{r, b} \in batches$ at $\server'$, with $i_k = \directorymap{\client}$ and $p_k = \rp{c, m}$: the lemma is proved. 
\end{proof}
\end{lemma}

\begin{lemma}
\label{lemma:clientsubmissionsarebroadcast}
Let $\client$ be a correct client. Let $\rp{c, s} \in submissions$ at $\client$, let
\begin{equation*}
    m = s.message
\end{equation*}
We have that $\client$ broadcast $m$ for $c$.
\begin{proof}
Upon initialization, $submissions$ is empty at $\client$ (line \ref{line:csubmissionsinit}); $\client$ adds $\rp{c, s}$ to $submissions$ (line \ref{line:csubmissionsupdate}) only upon broadcasting $m$ for $c$ (line \ref{line:cbrodcastmessage}).
\end{proof}
\end{lemma}

\begin{lemma}
\label{lemma:witnessedbatchissigned}
Let $\server$ be a correct server, let $l \in \sorted{\rp{\ids \times \contexts \times \messages}}$ such that $\server$ signs $\qp{\texttt{Witness}, \aroot{l}}$. For all $j$, let $\rp{i_j, c_j, m_j} = l_j$, let $\client_j = \directorymap{i_j}$. For all $k$ such that $\client_k$ is correct, $\client_k$ broadcast $m_k$ for $c_k$.
\begin{proof}
We start by noting that $\server$ signs $\qp{\texttt{Witness}, \aroot{l}}$ only by executing line \ref{line:ssignwitness}. Immediately before doing so, $\server$ loads a batch $b = batches\qp{\aroot{l}}$ (line \ref{line:sbatchload}). By Lemmas \ref{lemma:servercomputesmerkletree} and \ref{lemma:rootisinjective}, we have
\begin{equation*}
    l = \join{b.ids}{b.payloads}
\end{equation*}
Moreover, by Lemma \ref{lemma:sidsnonrepeating}, all elements of $i$ are distinct. 

\medskip

Next, $\server$ uses a set $H \subseteq \ids$ to partition the elements of $i$. For each $h \in H$ (line \ref{line:ssignsturesloop}), $\server$ verifies that: for some $j$, we have $h = i_j$ (lines \ref{line:ssignatureclientmissingcheck} and \ref{line:ssignatureclientverificationreturn}); and $\client_j$ signed $\qp{\texttt{Message}, c_j, m_j}$ (lines \ref{line:stakepayloads}, \ref{line:ssignatureclientverification} and \ref{line:ssignatureclientverificationreturn}). Finally, for all $j$ such that $i_j \notin H$ (line \ref{line:sdeterminemultisigners}), $\server$ verifies that $\client_j$ signed $\qp{\texttt{Reduction}, \aroot{l}}$ (lines \ref{line:smultiverifyreduction} and \ref{line:smultiverifyreductionreturn}).

\medskip

In summary, for all $j$, we have that either $\client_j$ signed $\qp{\texttt{Message}, c_j, m_j}$, or $\client_j$ signed $\qp{\texttt{Reduction}, \aroot{l}}$.

\medskip

Let $k$ such that $\client_k$ is correct. Let us assume that $\client_k$ signed $\qp{\texttt{Message}, c_k, m_k}$. $\client_k$ signs $\qp{\texttt{Message}, c_k, m_k}$ only by executing line \ref{line:csignaturemessage}. It does so only upon broadcasting $m_k$ for $c_k$ (line \ref{line:cbrodcastmessage}). Throughout the remainder of this proof, we assume that $\client_k$ signed $\qp{\texttt{Reduction}, \aroot{l}}$.

\medskip

$\client_k$ signs $\qp{\texttt{Reduction}, \aroot{l}}$ only by executing line \ref{line:csendreduction}. It does so only if, for some $p \in \aproofs$, $n \in \mathbb{N}$, $c' \in \contexts$, $m' \in \messages$, we have
\begin{equation}
\label{equation:witnessedbatchissigned.verify}
    \averify{\aroot{l}}{p}{n}{\rp{i_k, c', m'}} = \true
\end{equation}
(line \ref{line:cverifymerkleproof}), and $\client_k$ previously broadcast $m'$ for $c'$ (line \ref{line:cexistsubmissionforinclusion}, see Lemma \ref{lemma:clientsubmissionsarebroadcast}). By Equation \ref{equation:witnessedbatchissigned.verify} and Definition \ref{definition:idealaccumulator}, we must have $l_n = \rp{i_k, \ldots}$. As a result, because $i$ is non-repeating, we immediately have $n = k$, $c' = c_k$ and $m' = m_k$. We then have that $\client_k$ previously broadcast $m_k$ for $c_k$, and the lemma is proved.
\end{proof}
\end{lemma}

\begin{theorem}
\csbal satisfies integrity.
\begin{proof}
Let $\server$ be a correct server. Let $\client$ be a correct client, let $c$ be a context, let $m$ be a message such that $\server$ delivers $m$ from $\client$ for $c$. By Lemma \ref{lemma:delivermessagecheckingthewitness}, some correct server $\server'$ and some $\rp{r, b} \in batches$ at $\server'$ exist such that $\server'$ signed $\qp{\texttt{Witness}, r}$ and, with
\begin{align*}
    i &= b.ids \\
    p &= b.payloads
\end{align*}
some $k$ exists such that $i_k = \directorymap{\client}$ and $p_k = \rp{c, m}$. Let $l = \join{i}{p}$. We trivially have $l_k = \rp{\directorymap{\client}, c, m}$. By Lemma \ref{lemma:servercomputesmerkletree} we have $r = \aroot{l}$. For all $j$, let $\rp{c_j, m_j} = p_j$. By Lemma \ref{lemma:witnessedbatchissigned}, we have that $\rp{\client = \directorymap{\directorymap{\client}} = \directorymap{i_k}}$ broadcast $\rp{m = m_k}$ for $\rp{c = c_k}$, and the theorem is proved.

\end{proof}
\end{theorem}

\subsubsection{Validity}

In this section, we prove that \csbal satisfies validity.

\begin{lemma}
\label{lemma:clientknowsitself}
Let $\client$ be a correct client. Upon initialization, $\client$ knows $\client$.
\begin{proof}
It follows immediately from lines \ref{line:csignup} and \ref{line:csignupcomplete}, and the signup validity of \dir.
\end{proof}
\end{lemma}

\begin{lemma}
\label{lemma:clientdoesnotstopuntildelivered}
Let $\client$ be a correct client, let $c$ be a context such that $\client$ broadcast a message for $c$. If $c \notin submissions$ at $\client$, then at least one correct server delivered a message from $\client$ for $c$.
\begin{proof}
Let $i = \directorymap{\client}$ (by Lemma \ref{lemma:clientknowsitself}, $i$ is well-defined). Let $m$ be the message $\client$ broadcast for $c$. We start by noting that $\client$ adds $c$ to $submissions$ (line \ref{line:csubmissionsupdate}) upon broadcasting a message for $c$ (line \ref{line:cbrodcastmessage}). As a result, $\client$ satisfies $c \notin submissions$ only upon removing $\rp{c, s}$ from $submissions$ (line \ref{line:csubmissionsremove} only). $\client$ does so only if
\begin{equation*}
    s.included\_in \cap completed \neq \emptyset
\end{equation*}
Let $r \in s.included\_in \cap completed$.

\medskip

Noting that $s.included\_in$ is initially empty at $\client$ (line \ref{line:csubmissionsupdate}), $\client$ must have added $r$ to $s.included\_in$ (line \ref{line:cupdateincludedin} only). Noting that $\client$ permanently sets $s.message$ to $m$ upon initialization (line \ref{line:csubmissionsupdate} only, see line \ref{line:cbrodcastmessage}), $\client$ adds $r$ to $s.included\_in$ (line \ref{line:cupdateincludedin}) only if, for some $p \in \aproofs$ and $n \in \mathbb{N}$, we have
\begin{equation}
\label{equation:clientdoesnotstopuntildelivered.inclusionproof}
    \averify{r}{p}{n}{\rp{i, c, m}} = \true
\end{equation}
(line \ref{line:cverifymerkleproof}).

\medskip

Noting that $completed$ is initially empty at $\client$ (line \ref{line:ccompleteddefine}), $\client$ must have added $r$ to $completed$ (line \ref{line:ccomplrtrdupdate} only). $\client$ does so only if some $\epsilon$ exists such that $i \notin \epsilon$ and at least $f + 1$ servers signed $\qp{\texttt{Completion}, r, \epsilon}$ (line \ref{line:cverifyplurality}). 

\medskip

Noting that at most $f$ servers are Byzantine, some correct server $\server$ exists such that $\server$ signed $\qp{\texttt{Completion}, r, \epsilon}$ (line \ref{line:scompletionsign} only). $\server$ does so only if $r \in batches$ (lines \ref{line:srootinbatches} and \ref{line:srootinbatchesreturn}). Let $b = batches\qp{r}$ at $\server$, let $l = \join{b.ids}{b.payloads}$, let $\rp{i_j, c_j, m_j} = l_j$, let $\client_j = \directorymap{i_j}$. By Lemma \ref{lemma:servercomputesmerkletree}, we have $r = \aroot{l}$, and by Equation \ref{equation:clientdoesnotstopuntildelivered.inclusionproof} and Definition \ref{definition:idealaccumulator} we have $l_n = \rp{i, c, m}$. 

\medskip

Immediately before signing $\qp{\texttt{Completion}, r, \epsilon}$, $\server$ loops through $l$ (line \ref{line:sdeliverfrombrokerloop}) and delivers $m_j$ from $\client_j$ for $c_j$ (lines \ref{line:stakekeycard} and \ref{line:sdeliver}) if and only if $i_j \notin \epsilon$ and $\server$ did not previously deliver a message from $\client_j$ for $c_j$ (line \ref{line:sdelivercheck}, see Lemma \ref{lemma:deliveredtracksdelivered}). Recalling that $\rp{i = i_n} \notin \epsilon$, $\server$ delivers a message from $\rp{\client = \client_j}$ for $\rp{c = c_j}$ before or upon signing $\qp{\texttt{Completion}, r, \epsilon}$, and the lemma is proved.
\end{proof}
\end{lemma}

\begin{lemma} 
\label{lemma:brodcastissubmittedtocorrect}
Let $\client$ be a correct client, let $c$ be a context, let $m$ be a message such that $\client$ broadcast $m$ for $c$. Eventually, either a correct server delivers a message from $\client$ for $c$, or a correct broker delivers a $\qp{\texttt{Submission}, a, \rp{c, m, s}}$ message, where $a$ is an assignment for $\client$ and $s$ is $\client$'s signature for $\qp{\texttt{Message}, c, m}$.
\begin{proof}
Upon broadcasting $m$ for $c$ (line \ref{line:cbrodcastmessage}), $\client$ sets
\begin{equation*}
    submissions\qp{c} = Submission \cp{message: m, signature: s, submitted\_to: \emptyset, ..}
\end{equation*}
where $s$ is a signature for $\qp{\texttt{Message}, c, m}$ (lines \ref{line:csignaturemessage} and \ref{line:csubmissionsupdate}). $\client$ never updates $submissions\qp{c}.message$ or $submissions\qp{c}.signature$. Subsequently, $\client$ invokes $submit\rp{c}$ (line \ref{line:csubmitinvoke}).

\medskip

Upon executing $submit\rp{c}$ (line \ref{line:cproceduresubmit}), $\client$ immediately returns if and only if $c \notin submissions$, or no $\broker \in \rp{\brokers \setminus submissions\qp{c}.submitted\_to}$ exists (line \ref{line:csubmissioncheck}). Otherwise, $\client$ sends a $\qp{\texttt{Submission}, a, \rp{c, m, s}}$ message to $\broker$ (where $a$ is an assignment for $\client$) (lines \ref{line:cdirexport} and \ref{line:csendsubmit}), then adds $\broker$ to $submissions\qp{c}.submitted\_to$ (line \ref{line:csubmissionupdate}), and finally schedules $submit\rp{c}$ for eventual re-execution (lines \ref{line:ctimerset}, \ref{line:ctimerring} and \ref{line:cinvokesubmitring}).

\medskip

Noting that $\client$ updates $submissions\qp{c}.submitted\_to$ only by executing line \ref{line:csubmissionupdate}, $\client$ keeps re-executing $submit\rp{c}$ until either
\begin{itemize}
    \item $c \notin submissions$: by Lemma \ref{lemma:clientdoesnotstopuntildelivered}, at least one correct server delivered a message from $\client$ for $c$; or
    \item $\client$ sent to all brokers a $\qp{\texttt{Submission}, a, \rp{c, m, s}}$ message, where $a$ is an assignment for $\client$
\end{itemize}
The lemma follows immediately from the observation that $\brokers$ is finite and the assumption that at least one broker is correct.
\end{proof}
\end{lemma}

\begin{lemma}
\label{lemma:poolalwaysflushes}
Let $\broker$ be a correct broker. If $pool \neq \cp{}$ at $\broker$, then eventually $pool = \cp{}$ at $\broker$.
\begin{proof}
We start by noting that the $\qp{\texttt{Flush}}$ timer is pending at $\broker$ if and only if $collecting = \true$ at $\broker$. Indeed, upon initialization, we have $collecting = \false$ at $\broker$. Moreover, $\broker$ sets $collecting = \true$ (line \ref{line:bcollectingtrue}) only upon setting the $\qp{\texttt{Flush}}$ timer (line \ref{line:btimerset} only). Finally, $\broker$ resets $collecting = \false$ (line \ref{line:bcollectingfalse} only) only upon ringing $\qp{\texttt{Flush}}$ (line \ref{line:btimerring}).

\medskip

Upon ringing $\qp{\texttt{Flush}}$ (line \ref{line:btimerring}), $\broker$ resets $pool = \cp{}$ (line \ref{line:bemptypool}). As a result, whenever $pool \neq \cp{}$ at $\broker$, we either have:
\begin{itemize}
    \item $collecting = \true$ at $\broker$: $\qp{\texttt{Flush}}$ is pending at $\broker$ and $\broker$ will eventually reset $pool = \cp{}$; or
    \item $collecting = \false$ at $\broker$: $\broker$ will eventually detect that $pool \neq \cp{}$ and $collecting = \false$ (line \ref{line:bpoolcollectingcheck}) and set $collecting = true$ - the above will then apply.
\end{itemize}
\end{proof}
\end{lemma}

\begin{lemma}
\label{lemma:firstpendingendsupinpool}
Let $\broker$ be a correct broker, let $i$ be an id such that
\begin{equation*}
    pending\qp{i} = \qp{s_1, \ldots, s_S}
\end{equation*}
at $\broker$, with $S \geq 1$. We eventually have
\begin{align*}
    pool\qp{i} &= s_1 \\
    pending\qp{i} &= \qp{s_2, \ldots, s_S, \ldots}
\end{align*}
at $\broker$.
\begin{proof}
We start by noting that, by Lemma \ref{lemma:poolalwaysflushes}, if $i \in pool$ then eventually $pool = \cp{}$ at $\broker$. Noting that $\broker$ adds elements to $pool$ only by executing line \ref{line:bpoolupdate}, we then have that eventually $\broker$ detects $pending\qp{i} \neq \qp{}$ and $i \notin pool$ (line \ref{line:bexistsinpanding}). Upon doing so, $\broker$ sets $pending\qp{i} = \qp{s_2, \ldots, s_S, \ldots}$ (line \ref{line:bpandingupdate}) and $pool\qp{i} = s_1$ (line \ref{line:bpoolupdate}).
\end{proof}
\end{lemma}

\begin{lemma}
\label{lemma:allpendingendupinpool}
Let $\broker$ be a correct broker, let $i$ be an id, let $s$ be a submission such that $\broker$ pushes $s$ to $pending\qp{i}$. We eventually have $pool\qp{i} = s$ at $\broker$.
\begin{proof}
It follows immediately by induction on Lemma \ref{lemma:firstpendingendsupinpool}, and the observation that $\broker$ only pushes elements to the back of $pending$.
\end{proof}
\end{lemma}

\begin{lemma}
\label{lemma:submissionsaddedtopool}
Let $\broker$ be a correct broker. Let $\client$ be a client, let $a$ be an assignment for $\client$, let $c$ be a context, let $m$ be a message, let $s$ be $\client$'s signature for $\qp{\texttt{Message}, c, m}$. If $\broker$ delivers a $\qp{\texttt{Submission}, a, \rp{c, m, s}}$, then $\broker$ eventually sets
\begin{equation*}
    pool\qp{i} = Submission\cp{context: c, message: m, \tail}
\end{equation*}
where $i = \directorymap{\client}$.
\begin{proof}
Upon delivering $\qp{\texttt{Submission}, a, \rp{c, m, s}}$ (line \ref{line:bdeliversubmit}), $\broker$ imports $a$ (line \ref{line:bimportsubmit}). Because $s$ successfully verifies against $\qp{\texttt{Message}, c, m}$ and $\broker$ knows $\client$ (line \ref{line:bverifyclientsignature}), $\broker$ pushes $Submission\cp{context: c, message: m, \tail}$ to $pending\qp{i}$ (line \ref{line:bpushback}). The lemma immediately follows from Lemma \ref{lemma:allpendingendupinpool}.
\end{proof}
\end{lemma}

\begin{definition}[Map]
Let $X$, $Y$ be sets. A \textbf{map} $m: X \rightarrowtail Y$ is a subset of $\rp{X \times Y}$ such that
\begin{equation*}
    \forall \rp{x, y}, \rp{x', y'} \neq \rp{x, y} \in m, x \neq x'
\end{equation*}
We use
\begin{align*}
    \mapdomain{m} = \cp{x \in X \suchthat \rp{x, \any} \in m} \\
    \mapcodomain{m} = \cp{y \in Y \suchthat \rp{\any, y} \in m}
\end{align*}
to respectively denote the \textbf{domain} and \textbf{codomain} of $m$.
\end{definition}

\begin{notation}[Sorting maps]
Let $X$, $Y$ be sets such that $X$ is endowed with a total order relationship, let $m: X \rightarrowtail Y$. With a slight abuse of notation, we use $\sort{\mapcodomain{m}}$ to denote the codomain of $m$, sorted by $m$'s domain. For example, let $m = \cp{\rp{1, \square}, \rp{2, \circ}, \rp{3, \triangle}}$, we have $\sort{\mapcodomain{m}} = \rp{\square, \circ, \triangle}$.
\end{notation}

\begin{definition}[Broker variables]
\label{definition:brokervariables}
Let $\broker$ be a correct broker, let $t \in \mathbb{R}$, let $r$ be a root. 
\begin{itemize}
    \item We use $\bpayloads[\broker][t]{r}: \ids \rightarrowtail \rp{\messages \times \contexts}$ to denote, if it exists, the value of $batches\qp{r}.payloads$ at $\broker$ at time $t$; we use $\bpayloads[\broker][t]{r} = \bot$ otherwise.
    
    \item We define $\bcontexts[\broker][t]{r}: \ids \rightarrowtail \contexts$ and $\bmessages[\broker][t]{r}: \ids \rightarrowtail \messages$ by
    \begin{equation*}
        \rp{\bcontexts[\broker][t]{r}[i], \bmessages[\broker][t]{r}[i]} = \bpayloads[\broker][t]{r}[i]
    \end{equation*}
    if $\bpayloads[\broker][t]{r}[i] \neq \bot$; we use $\bcontexts[\broker][t]{r}[i] = \bmessages[\broker][t]{r}[i] = \bot$ otherwise.
    
    \item We define $\bleaves[\broker][t]{r} \in \sorted{\rp{\ids \times \contexts \times \messages}}$ by
    \begin{equation*}
        \bleaves[\broker][t]{r} = \sort{\cp{\rp{i, c, m} \suchthat \rp{i, \rp{c, m}} \in \bpayloads[\broker][t]{r}}}
    \end{equation*}
    if $\bpayloads[\broker][t]{r} \neq \bot$; we use $\bleaves[\broker][t]{r} = \bot$ otherwise.
    
    \item We use $\bsignatures[\broker][t]{r}: \ids \rightarrowtail \signatures$ and $\bmultisignatures[\broker][t]{r}: \ids \rightarrowtail \multisignatures$ to respectively denote, if they exist, the values of $batches\qp{r}.signatures$ and $batches\qp{r}.reductions$ at $\broker$ at time $t$; we use $\bsignatures[\broker][t]{r} = \bmultisignatures[\broker][t]{r} = \bot$ otherwise.
    
    \item We use $\bcommitto[\broker][t]{r} \subseteq \servers$ and $\bcommittable[\broker][t]{r} \in \cp{\true, \false}$ to respectively denote, if they exist, the values of $batches\qp{r}.commit\_to$ and $batches\qp{r}.committable$ at $\broker$ at time $t$; we use $\bcommitto[\broker][t]{r} = \bcommittable[\broker][t]{r} = \bot$ otherwise.
    
    \item We use $\bwitnesses[\broker][t]{r}: \servers \rightarrowtail \multisignatures$ to denote, if it exists, the value of $batches\qp{r}.witnesses$ at $\broker$ at time $t$; we use $\bwitnesses[\broker][t]{r} = \bot$ otherwise. We underline that $batches\qp{r}.witnesses$ exists only if $batches\qp{r}$ exists and is \texttt{Witnessing}.
    
    \item We use $\bcommits[\broker][t]{r}: \servers \rightarrowtail \rp{\powerset{\ids} \times \multisignatures}$ to denote, if it exists, the value of $batches\qp{r}.commits$ at $\broker$ at time $t$; we use $\bcommits[\broker][t]{r} = \bot$ otherwise. We underline that $batches\qp{r}.commits$ exists only if $batches\qp{r}$ exists and is \texttt{Committing}.
    
    \item We use $\bexclusions[\broker][t]{r} \in \powerset{\ids}$ and $\bcompletions[\broker][t]{r}: \servers \rightarrowtail \multisignatures$ to respectively denote, if they exist, the values of $batches\qp{r}.exclusions$ and $batches\qp{r}.completions$ at $\broker$ at time $t$; we use $\bexclusions[\broker][t]{r} = \bcompletions[\broker][t]{r} = \bot$ otherwise. We underline that $batches\qp{r}.exclusions$ and $batches\qp{r}.completions$ exist only if $batches\qp{r}$ exists and is \texttt{Completing}.
\end{itemize}
\end{definition}

\begin{notation}[Broker variables]
Let $r$ be a root. Wherever it can be unequivocally inferred from context, we omit the correct broker and the time from $\bpayloads{r}$, $\bcontexts{r}$, $\bmessages{r}$, $\bleaves{r}$, $\bsignatures{r}$, $\bmultisignatures{r}$, $\bcommitto{r}$, $\bcommittable{r}$, $\bwitnesses{r}$, $\bcommits{r}$, $\bexclusions{r}$ and $\bcompletions{r}$.
\end{notation}

\begin{lemma}
\label{lemma:brokervariablesstructure}
Let $\broker$ be a correct broker, let $r$ be a root. We have
\begin{align*}
    &\rp{\bpayloads{r} = \bot} \Longleftrightarrow \rp{\bcontexts{r} = \bot} \Longleftrightarrow \rp{\bmessages{r} = \bot} \Longleftrightarrow \\
    \Longleftrightarrow \;&\rp{\bcommitto{r} = \bot} \Longleftrightarrow \rp{\bcommittable{r} = \bot} \Longleftrightarrow \rp{\bleaves{r} = \bot} \Longleftrightarrow \\
    \Longleftrightarrow \;&\rp{\bsignatures{r} = \bot} \Longleftrightarrow \rp{\bmultisignatures{r} = \bot}
\end{align*}
and
\begin{equation*}
    \rp{\bpayloads{r} = \bot} \implies \rp{\bwitnesses{r} = \bcommits{r} = \bexclusions{r} = \bcompletions{r} = \bot}
\end{equation*}
\begin{proof}
It follows immediately from Definition \ref{definition:brokervariables}.
\end{proof}
\end{lemma}

\begin{lemma}
\label{lemma:poolinbatches}
Let $\broker$ be a correct broker, let $i$ be an id, let $c$ be a context, let $m$ be a message such that
\begin{equation*}
    pool\qp{i} = Submission\cp{context: c, message: m, \tail}
\end{equation*}
at $\broker$. Eventually some root $r$ exists such that $\bpayloads{r}[i] = \rp{c, m}$.
\begin{proof}
By Lemma \ref{lemma:poolalwaysflushes}, eventually $pool = \cp{}$ at $\broker$. Moreover, $\broker$ resets $pool = \cp{}$ only by executing line \ref{line:bemptypool}. Upon first doing so, $\broker$ stores the original value of $pool$ in a variable $u$ (line \ref{line:bsubmissionsequalpool}), which it uses to define a variable $p$ by
\begin{equation*}
    \rp{\rp{i, \rp{c, m}} \in p} \Longleftrightarrow \rp{\rp{i, Submission\cp{c, m, \tail}} \in u}
\end{equation*}
(line \ref{line:bpayloaddefine}). We immediately have $p\qp{i} = \rp{c, m}$. Finally, for some $r$, $\broker$ adds $\rp{r, b}$ to $batches$, with $b.payloads = p$ (line \ref{line:bbatchesupdate}). The lemma follows immediately from Definition \ref{definition:brokervariables}.
\end{proof}
\end{lemma}

\begin{lemma}
\label{lemma:broadcastendsincorrectbatch}
Let $\client$ be a correct client, let $i = \directorymap{\client}$. Let $c$ be a context, let $m$ be a message such that $\client$ broadcast $m$ for $c$. Eventually, either a correct server delivers a message from $\client$ for $c$, or some correct broker $\broker$ and some root $r$ exist such that eventually $\bpayloads{r}[i] = \rp{c, m}$.
\begin{proof}
The lemma immediately follows from Lemmas \ref{lemma:brodcastissubmittedtocorrect}, \ref{lemma:submissionsaddedtopool} and \ref{lemma:poolinbatches}.
\end{proof}
\end{lemma}

\begin{lemma}
\label{lemma:brokerknowspool}
Let $\broker$ be a correct broker, let $i$ be an id. If $i \in pool$ at $\broker$, then $\broker$ knows $i$.
\begin{proof}
We start by noting that, upon initialization, $pool$ is empty at $\broker$ (line \ref{line:bpoolinit}). Moreover, $\broker$ adds $\rp{i, \any}$ to $pool$ only by executing line \ref{line:bpoolupdate}. $\broker$ does so only if $pending\qp{i}$ is not empty (line \ref{line:bexistsinpanding}). Upon initialization, all values of $pending$ are also empty at $\broker$ (line \ref{line:bpandingdefinition}). Finally, $\broker$ adds elements to $pending\qp{i}$ only by executing line \ref{line:bpushback}. Because $\broker$ does so only if $\broker$ knows $i$ (line \ref{line:bverifyclientsignature}), the lemma is proved.
\end{proof}
\end{lemma}

\begin{lemma}
\label{lemma:brokerknowsbatch}
Let $\broker$ be a correct broker, let $r$ be a root such that $\bpayloads{r} \neq \bot$. For all $i \in \mapdomain{\bpayloads{r}}$, $\broker$ knows $i$.
\begin{proof}
We start by noting that, upon initialization, $batches$ is empty at $\broker$ (line \ref{line:bbatchesinit}). Moreover, $\broker$ adds $\rp{r, b}$ to $batches$ only by executing line \ref{line:bbatchesupdate}. Upon doing so, $\broker$ satisfies $\mapdomain{b.payloads} = \mapdomain{u}$ (line \ref{line:bpayloaddefine}) for some $u$ initialized to $pool$ (line \ref{line:bsubmissionsequalpool}). The lemma immediately follows from Lemma \ref{lemma:brokerknowspool}.
\end{proof}
\end{lemma}

\begin{lemma}
\label{lemma:signaturepartition}
Let $\broker$ be a correct broker, let $r$ be a root such that $\rp{\bpayloads{r} \neq \bot} \Longleftrightarrow \rp{\bsignatures{r} \neq \bot}$. We have $\mapdomain{\bpayloads{r}} = \mapdomain{\bsignatures{r}} \cup \mapdomain{\bmultisignatures{r}} $ and $\mapdomain{\bsignatures{r}} \cap \mapdomain{\bmultisignatures{r}} = \emptyset$
\begin{proof}
We start by noting that, upon initialization, $batches$ is empty at $\broker$ (line \ref{line:bbatchesinit}). Moreover, $\broker$ adds $\rp{r, b}$ to $batches$ only by executing line \ref{line:bbatchesupdate}. Upon doing so, $\broker$ satisfies $\mapdomain{b.payloads} = \mapdomain{b.signatures}$ (lines \ref{line:bpayloaddefine} and \ref{line:bsignaturedefine}). Subsequently, $\broker$ updates $batches\qp{r}.signatures$ and $batches\qp{r}.reductions$ only concurrently, by executing lines \ref{line:bupdatereduction} and \ref{line:bupdateremoveid}. Upon doing so, $\broker$ shifts an id $i$ in $batches\qp{r}.payloads$ (line \ref{line:bcheckreducingid}) from $\mapdomain{batches\qp{r}.signatures}$ to $\mapdomain{batches\qp{r}.reductions}$.
\end{proof}
\end{lemma}

\begin{lemma}
\label{lemma:batchesarecorrectlysigned}
Let $\broker$ be a correct broker, let $r$ be a root such that $\bsignatures{r} \neq \bot$. Let $i \in \mapdomain{\bsignatures{r}}$, let $\client = \directorymap{i}$, let $\rp{c, m} = \bpayloads{r}[i]$, let $s = \bsignatures{r}[i]$. We have that $s$ is $\client$'s signature for $\qp{\texttt{Message}, c, m}$.
\begin{proof}
We underline that $c$ and $m$ are well-defined by Lemma \ref{lemma:signaturepartition}, and $i$ is well-defined by Lemmas \ref{lemma:signaturepartition} and \ref{lemma:brokerknowsbatch}. We start by noting that, upon initialization, $batches$ is empty at $\broker$ (line \ref{line:bbatchesinit}). Moreover, $\broker$ adds $\rp{r, b}$ to $batches$ only by executing line \ref{line:bbatchesupdate}. Upon last doing so, $\broker$ satisfies 
\begin{align*}
    \rp{c, m} = b.payloads\qp{i} &= \rp{w.context, w.message} \\
    s = b.signatures\qp{i} &= w.signature 
\end{align*}
with $w = u\qp{i}$ for some copy $u$ of $pool$ (lines \ref{line:bsubmissionsequalpool}, \ref{line:bpayloaddefine} and \ref{line:bsignaturedefine}). 

\medskip

Upon initialization, $pool$ is empty at $\broker$ as well (line \ref{line:bpoolinit}). Moreover, $\broker$ sets $pool\qp{i} = w$ only by executing line \ref{line:bpoolupdate}. It does so only if $w$ was in $pending\qp{i}$ (lines \ref{line:bexistsinpanding} and \ref{line:bpandingupdate}). Finally, $pending$ is initially empty at $\broker$ (line \ref{line:bpandingdefinition}), and $\broker$ adds $w$ to $pending\qp{i}$ only by executing line \ref{line:bpushback}. $\broker$ does so only if $w.signature$ is $\directorymap{i}$'s signature for $\qp{\texttt{Message}, w.context, w.message}$ (line \ref{line:bverifyclientsignature}). This proves that $s$ is $\client$'s signature for $\qp{\texttt{Message}, c, m}$ and concludes the lemma.
\end{proof}
\end{lemma}

\begin{lemma}
\label{lemma:batchesarecorrectlyreduced}
Let $\broker$ be a correct broker, let $r$ be a root such that $\bmultisignatures{r} \neq \bot$. Let $i \in \mapdomain{\bmultisignatures{r}}$, let $\client = \directorymap{i}$, let $q = \bmultisignatures{r}[i]$. We have that $q$ is $\client$'s multisignature for $\qp{\texttt{Reduction}, r}$.
\begin{proof}
We start by noting that, upon initialization, $batches$ is empty at $\broker$ (line \ref{line:bbatchesinit}). Moreover, $\broker$ adds $\rp{r, b}$ to $batches$ only by executing line \ref{line:bbatchesupdate}. Upon doing so, $\broker$ satisfies $b.reductions = \cp{}$. Finally, $\broker$ adds $\rp{i, q}$ to $batches\qp{r}.reductions$ only by executing line \ref{line:bupdatereduction}. $\broker$ does so only if $q$ is $\rp{\client = \directorymap{i}}$'s multisignature for $\qp{\texttt{Reduction}, r}$ (lines \ref{line:bcheckreducingid} and \ref{line:bcheckmultisignereduction}). 
\end{proof}
\end{lemma}

\begin{lemma}
\label{lemma:brokercomputesmerkletree}
Let $\broker$ be a correct broker, let $r$ be a root such that $\bpayloads{r} \neq \bot$. We have $r = \aroot{\bleaves{r}}$.
\begin{proof}
We start by noting that $\broker$ adds elements to $batches$ only by executing line \ref{line:bbatchesupdate}, and $\broker$ never updates $batches\qp{r}.payloads$. Immediately before adding $\rp{r, b}$ to $batches$ (line \ref{line:bbatchesupdate}), $\broker$ computes $r = \aroot{\sort{l}}$ (lines \ref{line:btreecostruction} and \ref{line:brootcostruction}), with
\begin{align*}
    \rp{\rp{i, c, m} \in l} &\Longleftrightarrow \rp{\rp{i, Submission\cp{c, m, \tail}} \in u} \\
    &\Longleftrightarrow \rp{\rp{i, \rp{c, m}} \in b.payloads}
\end{align*}
for some $u$ (lines \ref{line:bleavescostruction} and line \ref{line:bpayloaddefine}). The lemma immediately follows from Definition \ref{definition:brokervariables}.
\end{proof}
\end{lemma}

\begin{lemma}
\label{lemma:payloadsarestable}
Let $\broker$ be a correct broker, let $r$ be a root, let $t, t' \in \mathbb{R}$ such that $\bpayloads[\broker][t]{r} \neq \bot$ and $\bpayloads[\broker][t']{r} \neq \bot$. We have $\bpayloads[\broker][t]{r} = \bpayloads[\broker][t']{r}$.
\begin{proof}
It follows immediately from Lemmas \ref{lemma:brokercomputesmerkletree} and \ref{lemma:rootisinjective} and Definition \ref{definition:brokervariables}.
\end{proof}
\end{lemma}

\begin{lemma}
\label{lemma:payloadsmatchbetweenbrokerandserver}
Let $\broker$ be a correct broker, let $\server$ be a correct server, let $r$ be a root such that $\bpayloads{r} \neq \bot$ and $r \in batches$ at $\server$. Let $b = batches\qp{r}$ at $\server$, let
\begin{align*}
    i &= b.ids \\
    p &= b.payloads
\end{align*}
We have
\begin{align*}
    i &= \sort{\mapdomain{\bpayloads{r}}} \\
    p &= \sort{\mapcodomain{\bpayloads{r}}}
\end{align*}
\begin{proof}
It follows immediately from Definition \ref{definition:brokervariables}, Lemmas \ref{lemma:brokercomputesmerkletree} and \ref{lemma:servercomputesmerkletree}, and Lemma \ref{lemma:rootisinjective}.
\end{proof}
\end{lemma}

\begin{lemma}
\label{lemma:batchbeyondwitnessingwaswitnessed}
Let $\broker$ be a correct broker, let $r$ be a root, let $t, t'' \in \mathbb{R}$ such that $t'' > t$, $\bwitnesses[\broker][t]{r} \neq \bot$ and $\bwitnesses[\broker][t'']{r} = \bot$. For some $t' \in \qp{t, t''}$ we have $\abs{\bwitnesses[\broker][t']{r}} \geq f + 1$.
\begin{proof}
We start by noting that, by Definition \ref{definition:brokervariables}, $batches\qp{r}$ is $\texttt{Witnessing}$ at $\broker$ at time $t$. Let $t'$ identify the first moment after $t$ when $\broker$ updates $batches\qp{r}$'s variant. $\broker$ does so only by executing line \ref{line:bcommiting}, and only if $\abs{batches\qp{r}.witnesses} \geq f + 1$ (line \ref{line:binwitnessing}). We then have $\abs{\bwitnesses[\broker][t']{r}} \geq f + 1$, and the lemma is proved.
\end{proof}
\end{lemma}

\begin{notation}[Sequence elements]
Let $X$ be a set, let $z \in X^{<\infty}$. We use
\begin{equation}
    \cp{z} = \cp{z_n \suchthat n \leq \abs{z}}
\end{equation}
to denote the \textbf{elements} of $z$.
\end{notation}

\begin{notation}[Sequence indexing]
Let $X$ be a set, let $z \in X^{<\infty}$ such that all elements of $z$ are distinct. Let $x \in \cp{z}$. We use
\begin{equation*}
    \rp{z_x} = \rp{n \Longleftrightarrow z_n = x}
\end{equation*}
to identify the \textbf{index} of $x$ in $z$.
\end{notation}

\begin{lemma}
\label{lemma:sentbatchbeforewitnessing}
Let $\broker$ be a correct broker, let $r$ be a root such that $\bwitnesses{r} \neq \bot$. $\broker$ has sent a $\qp{\texttt{Batch}, \tilde i, p}$ message to all servers, with
\begin{align*}
    \tilde{i} &= \compress{i} \\
    i &= \sort{\mapdomain{\bpayloads{r}}} \\
    p &= \sort{\mapcodomain{\bpayloads{r}}}
\end{align*}
\begin{proof}
$\broker$ sets $batches\qp{r}$'s variant to $\texttt{Witnessing}$ only by executing line \ref{line:bbatchinwitnessing}. The lemma immediately follows from lines \ref{line:breducefetchbatch}, \ref{line:bcompressids}, \ref{line:bpayloads}, \ref{line:bserverinserver} and \ref{line:bsendbacth} and Definition \ref{definition:brokervariables}.
\end{proof}
\end{lemma}

\begin{lemma}
\label{lemma:serverreceivesbatch}
Let $\broker$ be a correct broker, let $r$ be a root such that $\bpayloads{r} \neq \bot$, let
\begin{align*}
    \tilde{i} &= \compress{i} \\
    i &= \sort{\mapdomain{\bpayloads{r}}} \\
    p &= \sort{\mapcodomain{\bpayloads{r}}}
\end{align*}
Let $\server$ be a correct server. Upon delivering a $\qp{\texttt{Batch}, \tilde i, p}$ message from $\broker$, $\server$ sets
\begin{equation*}
    batches\qp{r} = Batch\cp{ids: i, payloads: p, \tail}
\end{equation*}
and sends a $\qp{\texttt{BatchAcquired}, r, u}$ message back to $\broker$, with $u \subseteq \mapdomain{\bpayloads{r}}$ such that $\server$ knows all elements of $\rp{\mapdomain{\bpayloads{r}} \setminus u}$.
\begin{proof}
Noting that $i$ and $p$ respectively list the domain and codomain of the same map, we have that $i$ has no duplicates and $\abs{i} = \abs{p}$. Moreover, by Definitions \ref{definition:joinfunction} and \ref{definition:brokervariables}, we have $\join{i}{p} = \bleaves{r}$. By Lemma \ref{lemma:brokercomputesmerkletree}, this proves $r = \aroot{\join{i}{p}}$.

\medskip

Upon delivering $\qp{\texttt{Batch}, \tilde i, p}$ (line \ref{line:sbatchrecive}), $\server$ expands $\tilde{i}$ back into $i = \expand{\tilde{i}}$ (line \ref{line:sexpandprocedure}) and verifies that $i$ has no duplicates and $\abs{i} = \abs{p}$ (line \ref{line:sbatchcheck}). $\server$ then computes the set $u$ of ids in $\cp{i}$ that $\server_n$ does not know (line \ref{line:bcomputeunknowns}). We immediately have $u \subseteq \mapdomain{\bpayloads{r}}$, and $\server$ knows all elements of $\rp{\mapdomain{\bpayloads{r}} \setminus u}$. Next, $\server_n$ computes $r = \aroot{\join{i}{p}}$ (lines \ref{line:scomputeleaves}, \ref{line:scomputetree} and \ref{line:scomputeroot}). Finally, $\server_n$ sets $batches\qp{r} = Batch\cp{ids: i, payloads: p, \tail}$ (line \ref{line:sbatchesupdate}), and sends a $\qp{\texttt{BatchAcquired}, r, u}$ message back to $\broker$ (lines \ref{line:sbatchacquired}, \ref{line:sbatchreceivecheckresponse} and \ref{line:sbatchreceivesendresponse}).
\end{proof}
\end{lemma}

\begin{lemma}
\label{lemma:batchesarewitnessed}
Let $\broker$ be a correct broker, let $r$ be a root such that, at some point in time, we have $\bpayloads{r} \neq \bot$. At some point in time we have $\abs{\bwitnesses{r}} \geq f + 1$.
\begin{proof}
Upon initialization, $batches$ is empty at $\broker$ (line \ref{line:bbatchesinit}). Moreover, upon adding $\rp{r, \any}$ to $batches$ (line \ref{line:bbatchesupdate} only), $\broker$ sets a $\qp{\texttt{Reduce}, r}$ timer (line \ref{line:btimerset}). When $\qp{\texttt{Reduce}, r}$ eventually rings at $\broker$ (line \ref{line:breducering}), $\broker$ updates $batches\qp{r}$'s variant to $\texttt{Witnessing}$. By Lemma \ref{lemma:sentbatchbeforewitnessing}, before doing so $\broker$ sends a $\qp{\texttt{Batch}, \tilde i, p}$ message to all servers, with
\begin{align*}
    \tilde{i} &= \compress{i} \\
    i &= \sort{\mapdomain{\bpayloads{r}}} \\
    p &= \sort{\mapcodomain{\bpayloads{r}}}
\end{align*}

\bigskip

Let $\server_1, \ldots, \server_{f + 1}$ be distinct correct servers (noting that at most $f$ servers are Byzantine, $\server_1, \ldots, \server_{f + 1}$ are guaranteed to exist). Let $n \leq f + 1$. By Lemma \ref{lemma:serverreceivesbatch}, upon delivering $\qp{\texttt{Batch}, \tilde i, p}$ $\server_n$ sets $batches\qp{r} = Batch\cp{ids: i, payloads: p, \tail}$, and sends a $\qp{\texttt{BatchAcquired}, r, u}$ message back to $\broker$, with $u \subseteq \mapdomain{\bpayloads{r}}$ such that $\server$ knows all elements of $\rp{\mapdomain{\bpayloads{r}} \setminus u}$.

\bigskip

Let us assume that, upon delivering $\qp{\texttt{BatchAcquired}, r, u}$ from $\server_n$, we have $r \notin batches$ at $\broker$. By Definition \ref{definition:brokervariables} we have $\bpayloads{r} = \bot$ which, by Lemma \ref{lemma:brokervariablesstructure}, implies $\bwitnesses{r} = \bot$. As a result, by Lemma \ref{lemma:batchbeyondwitnessingwaswitnessed}, at some point in time we must have had $\abs{\bwitnesses{r}} \geq f + 1$. Throughout the remainder of this proof we assume that, upon delivering $\qp{\texttt{BatchAcquired}, r, u}$ from $\server_n$, we have $r \in batches$ at $\broker$.

\medskip

Upon delivering $\qp{\texttt{BatchAcquired}, r, u}$ from $\server_n$ (line \ref{line:bdeliverbatchacquired}), $\broker$ verifies that $b \in batches$ (line \ref{line:bbatchacquiredgetbatch}) and, because $\broker$ knows all elements of $u$ (line \ref{line:bbatchacquiredcheckids}), $\broker$ maps $u$ into a corresponding set of assignments $a$ (line \ref{line:bbatchacquiredexportassignments}). Next, $\broker$ aggregates all elements of $\mapcodomain{\bmultisignatures{r}}$ into a multisignature $q$ (line \ref{line:bbatchacquiredaggregate}) and copies $\bsignatures{r}$ into a map $s$ (line \ref{line:bbatchacquiredcopysignatures}). 

\medskip

By Lemmas \ref{lemma:payloadsarestable} and \ref{lemma:signaturepartition} we have
\begin{equation}
\label{equation:batchesarewitnessed.signaturepartition}
    \cp{i} \setminus \mapdomain{s} = \mapdomain{\bpayloads{r}} \setminus \mapdomain{\bsignatures{r}} = \mapdomain{\bmultisignatures{r}}
\end{equation}
By Lemma \ref{lemma:batchesarecorrectlysigned}, for all $\rp{\hat i, \hat s} \in s$, $\rp{\hat s = s\qp{\hat i} = \bsignatures{r}[i]}$ is $\directorymap{\hat i}$'s signature for $\qp{\texttt{Message}, c, m}$, with 
\begin{equation*}
    \rp{c, m} = p_\rp{i_{\hat i}} = \bpayloads{r}[\hat i]
\end{equation*}
By Lemma \ref{lemma:batchesarecorrectlyreduced} and Equation \ref{equation:batchesarewitnessed.signaturepartition}, $q$ is $\rp{\rp{\cp{i} \setminus \mapdomain{s}} = \mapdomain{\bmultisignatures{r}}}$'s multisignature for $\qp{\texttt{Reduction}, r}$. 

\medskip

Having computed $a$, $q$ and $s$, $\broker$ sends a $\qp{\texttt{Signatures}, r, a, q, s}$ message back to $\server_n$ (line \ref{line:bsendsignatures}).

\bigskip

Upon delivering $\qp{\texttt{Signatures}, r, a, q, s}$ from $\broker$ (line \ref{line:sdeliversignatures}), $\server_n$ imports all elements of $a$ (line \ref{line:shandlesignaturesimportassignments}). Noting that $a$ contains assignments for all elements of $u$, and any element of $\rp{\cp{i} \setminus u}$ was known to $\server_n$ upon delivering $\qp{\texttt{Batch}, \ldots}$, $\server_n$ knows all elements of $i$. Noting that $\server_n$ never modifies or removes elements of $batches$, $\server_n$ successfully retrieves $i$ and $p$ from $batches\qp{r}$ (lines \ref{line:scheckbatcheshandlesignature} and \ref{line:sbatchload}). $\server_n$ then verifies to know all elements of $i$ (line \ref{line:sdircheck}). 

\medskip

Next, $\server_n$ loops through all elements of $s$. For each $\rp{\hat i, \hat s}$ in $s$, $\server_n$ verifies that $\hat i \in i$ (line \ref{line:ssignatureclientmissingcheck}, see Lemma \ref{lemma:signaturepartition}), then verifies that $\hat s$ is $\directorymap{\hat i}$'s signature for $\qp{\texttt{Message}, c, m}$, with $\rp{c, m} = p_\rp{i_{\hat i}}$ (lines \ref{line:stakepayloads} and \ref{line:ssignatureclientverification}). Subsequently, $\server_n$ verifies that $q$ is $\rp{i \setminus \mapdomain{s}}$'s multisignature for $\qp{\texttt{Reduction}, r}$ (lines \ref{line:sdeterminemultisigners} and \ref{line:smultiverifyreduction}). Finally, $\server$ produces a multisignature $w$ for $\qp{\texttt{Witness, r}}$ (line \ref{line:ssignwitness}) and sends a $\qp{\texttt{WitnessShard, r, w}}$ message back to $\broker$ (lines \ref{line:sreturnwitnessshard}, \ref{line:shandlesignaturescheckresponse} and \ref{line:shandlesignaturessendresponse}).

\bigskip

Let us assume that, upon delivering $\qp{\texttt{WitnessShard}, r, w}$ from $\server_n$, we have $r \notin batches$ or $batches\qp{r}$ not $\texttt{Witnessing}$ at $\broker$. By Definition \ref{definition:brokervariables} we have $\bwitnesses{r} = \bot$. As a result, by Lemma \ref{lemma:batchbeyondwitnessingwaswitnessed}, at some point in time we must have had $\abs{\bwitnesses{r}} \geq f + 1$. Throughout the remainder of this proof we assume that, upon delivering $\qp{\texttt{WitnessShard}, r, w}$, we have $r \in batches$ and $batches\qp{r}$ $\texttt{Witnessing}$ at $\broker$.

\medskip

Upon delivering $\qp{\texttt{WitnessShard}, r, w}$ from $\server_n$ (line \ref{line:bdeliverwitnessshard}), $\broker$ verifies that $r \in batches$ and $batches\qp{r}$ is $\texttt{Witnessing}$ (lines \ref{line:brootandcommitable} and \ref{line:bwitnessshardcheckbatch}), then verifies that $w$ is $\server_n$'s multisignature for $\qp{\texttt{Witness}, r}$ (line \ref{line:bwitnessshardchecksignature}) and finally adds $\rp{\server_n, w}$ to \\ $batches\qp{r}.witnesses$ (line \ref{line:bstorewitnessshard}).

\medskip

In summary, for all $n \leq f + 1$, either $\abs{\bwitnesses{r}} \geq f + 1$, or $\broker$ adds a distinct element to $batches\qp{r}.witnesses$. Noting that $\broker$ never removes elements from $batches\qp{r}.witnesses$, this trivially reduces to $\broker$ eventually satisfying $\abs{\bwitnesses{r}} \geq f + 1$, and the lemma is proved.
\end{proof}
\end{lemma}

\begin{lemma}
\label{lemma:serverwitnessesarecorrect}
Let $\server$ be a correct server, let $\rp{r, w} \in witnesses$ at $\server$. We have that $w$ is a plurality certificate for $\qp{\texttt{Witness}, r}$.
\begin{proof}
Upon initialization, $witnesses$ is empty at $\server$ (line \ref{line:sinitwitnesses}). Moreover, $\server$ adds $\rp{r, w}$ to $witnesses$ only by executing line \ref{line:sstorewitness}. Immediately before doing so, $\server$ verifies that $w$ is a plurality certificate for $\qp{\texttt{Witness}, r}$ (lines \ref{line:switnesssignaturecheck} and \ref{line:switnesssignaturecheckreturn}).
\end{proof}
\end{lemma}

\begin{lemma}
\label{lemma:serverkeepsbatches}
Let $\server$ be a correct server, let $t \in \mathbb{R}$, let $\rp{r, b} \in batches$ at $\server$ at time $t$. For all $t' \geq t$, $\rp{r, b} \in batches$ at $\server$ at time $t$.
\begin{proof}
Let $t' \geq t$. Because $\server$ never removes elements from $batches$, for some $b'$ we have $\rp{r, b'} \in batches$ at $\server$ at time $t'$. By Lemmas \ref{lemma:servercomputesmerkletree} and \ref{lemma:rootisinjective} we have $b' = b$, and the lemma is proved.
\end{proof}
\end{lemma}

\begin{lemma}
\label{lemma:messagesarenotdangling}
Let $\server$ be a correct server, let $\rp{\rp{i, c}, \rp{\any, r}} \in messages$ at $\server$. We have $r \in batches$ at $\server$.
\begin{proof}
We start by noting that, upon initialization, $messages$ is empty at $\server$ (line \ref{line:smessagesinit}). Moreover, $\server$ adds $\rp{\rp{i, c}, \rp{\any, r}}$ to $messages$ only by executing line \ref{line:updatevaraiablemessage}. $\server$ does so only if $r \in batches$ (lines \ref{line:srootinbatchesw} and \ref{line:srootinbatchesreturnw}). The lemma immediately follows from Lemma \ref{lemma:serverkeepsbatches}.
\end{proof}
\end{lemma}

\begin{lemma}
\label{lemma:checkleafinmessages}
Let $\server$ be a correct server, let $\rp{\rp{i, c}, \rp{m, r}} \in messages$ at $\server$. Let $b = batches\qp{r}$ at $\server$, let
\begin{align*}
    i &= b.ids \\
    p &= b.payloads
\end{align*}
let $l = \join{i}{p}$. We have $\rp{i, c, m} \in \cp{l}$.
\begin{proof}
We underline that, by Lemma \ref{lemma:messagesarenotdangling}, $b$ is well-defined. We start by noting that, upon initialization, $messages$ is empty at $\server$ (line \ref{line:smessagesinit}). Moreover, $\server$ adds $\rp{\rp{i, c}, \rp{m, r}}$ to $messages$ only by executing line \ref{line:updatevaraiablemessage}. $\server$ does so only if (see Lemma \ref{lemma:serverkeepsbatches}) $\rp{i, c, m} \in \cp{l}$ (lines \ref{line:sbatchesnamew}, \ref{line:scheckidsummision}).
\end{proof}
\end{lemma}

\begin{lemma}
\label{lemma:batchbeyondcommittingiscommitted}
Let $\broker$ be a correct broker, let $r$ be a root, let $t, t'' \in \mathbb{R}$ such that $t'' > t$, $\bcommits[\broker][t]{r} \neq \bot$ and $\bcommits[\broker][t'']{r} = \bot$. For some $t' \in \qp{t, t''}$ we have $\abs{\bcommits[\broker][t']{r}} \geq 2f + 1$.
\begin{proof}
We start by noting that, by Definition \ref{definition:brokervariables}, $batches\qp{r}$ is $\texttt{Committing}$ at $\broker$ at time $t$. Let $t'$ identify the first moment after $t$ when $\broker$ updates $batches\qp{r}$'s variant. $\broker$ does so only by executing line \ref{line:bcompleting}, and only if $\abs{batches\qp{r}.commits} \geq 2f + 1$ (line \ref{line:bpluralitycommitingcheck}). We then have $\abs{\bcommits[\broker][t']{r}} \geq 2f + 1$, and the lemma is proved.
\end{proof}
\end{lemma}

\begin{lemma}
\label{lemma:batchesarecommitted}
Let $\broker$ be a correct broker, let $r$ be a root such that $\abs{\bwitnesses{r}} \geq f + 1$. At some point in time we have $\abs{\bcommits{r}} \geq 2f + 1$.
\begin{proof}
We start by noting that, if $batches\qp{r}$ is $\texttt{Witnessing}$ at broker, $\broker$ never removes elements from $batches\qp{r}.witnesses$, and $\broker$ updates $batches\qp{r}$'s variant only by executing line \ref{line:bcommiting}. As a result, $\broker$ is eventually guaranteed to detect that $batches\qp{r}$ is $\texttt{Witnessing}$ and that $\abs{batches\qp{r}.witnesses} \geq f + 1$ (line \ref{line:binwitnessing}). Upon doing so, $\broker$ aggregates $batch\qp{r}.witnesses$ into a certificate $c$ (lines \ref{line:binwitnessingfetchbatch} and \ref{line:baggregatewitnesses}) and sends a $\qp{\texttt{Witness}, r, c}$ message to every server (lines \ref{line:bbroadcastwitnessesloop} and \ref{line:bbroadcastwitnesses}). 

\bigskip

Let $\server_1, \ldots, \server_{2f + 1}$ be distinct correct servers (noting that at most $f$ servers are Byzantine, $\server_1, \ldots, \server_{2f + 1}$ are guaranteed to exist). Let $n \leq 2f + 1$. Noting that $\bwitnesses{r} \neq \bot$, by Lemma \ref{lemma:sentbatchbeforewitnessing}, the source-order delivery of perfect links and Lemmas \ref{lemma:serverreceivesbatch} and \ref{lemma:serverkeepsbatches}, by the time $\server_n$ handles the delivery of $\qp{\texttt{Witness}, r, c}$, $\server_n$ satisfies
\begin{equation*}
    batches\qp{r} = Batch \cp{ids: i, payloads: p, \tail}
\end{equation*}
with
\begin{align*}
    i &= \sort{\mapdomain{\bpayloads{r}}} \\
    p &= \sort{\mapcodomain{\bpayloads{r}}}
\end{align*}

\medskip

Upon delivering $\qp{\texttt{Witness}, r, c}$ (line \ref{line:sdeliverwitness}), $\server_n$ verifies that $r \in batches$ (line \ref{line:srootinbatchesw}) and retrieves $i$ and $p$ from $batches\qp{r}$ (line \ref{line:sbatchesnamew}). $\server_n$ then initializes an empty map $f: \ids \rightarrowtail \rp{\aroots, \multisignatures, \aproofs, \messages}$ (line \ref{line:sinitializeconflicts}), and loops through all elements of $\join{i}{p}$ (line \ref{line:scheckidsummision}).

\medskip

For all $j \leq \rp{\abs{i} = \abs{p}}$, let $\rp{c_j, m_j} = p_j$. Noting that $\server_n$ adds to $f$ only keys that belong to $i$ (line \ref{line:sconflictsupdate}), let $k \leq \rp{\abs{i} = \abs{p}}$ such that $\server_n$ sets
\begin{equation*}
    f\qp{i_k} = \rp{r'_k, w'_k, p'_k, m'_k}
\end{equation*}
By line \ref{line:sconflictcheck} we immediately have $m'_k \neq m_k$. Additionally, because $w'_k = witnesses\qp{r'_k}$ at $\server$ (line \ref{line:sfetchconflictwitness}), by Lemma \ref{lemma:serverwitnessesarecorrect} $w'_k$ is a plurality certificate for $\qp{\texttt{Witness}, r'_k}$. Finally, by line \ref{line:sconflictcheck} we have $\rp{\rp{i_k, c_k}, \rp{m'_k, r'_k}} \in messages$ at $\server_n$. Let $b'_k = batches\qp{r'_k}$ at $\server_n$, let
\begin{equation*}
    l'_k = \join{b'_k.ids}{b'_k.payloads}
\end{equation*}
By Lemma \ref{lemma:checkleafinmessages} we have $\rp{i_k, c_k, m'_k} \in l'_k$. As a result, by lines \ref{line:soriginalbatch}, \ref{line:sextraxtoriginalbatch}, \ref{line:soriginalleaf} and \ref{line:sconflictsupdate}, we have that $p'_k$ is a proof for $\rp{i_k, c_k, m'_k}$ from $r'_k$.

\medskip

Having looped over all elements of $\join{i}{p}$, $\server_n$ produces a signature $s$ for $\qp{\texttt{Commit}, r, \mapdomain{f}}$ (line \ref{line:ssigncommit}) and sends a $\qp{\texttt{CommitShard}, r, f, s}$ message back to $\broker$ (lines \ref{line:sreturncommitshard}, \ref{line:sresponseempty} and \ref{line:striggertobroker}).

\bigskip

Let us assume that, upon delivering $\qp{\texttt{CommitShard}, r, f, s}$ from $\server_n$, we have $r \notin batches$ or $batches\qp{r}$ not $\texttt{Witnessing}$ at $\broker$. By Definition \ref{definition:brokervariables} we have $\bcommits{r} = \bot$. As a result, by Lemma \ref{lemma:batchbeyondcommittingiscommitted}, at some point in time we must have had $\abs{\bcommits{r}} \geq 2f + 1$. Throughout the remainder of this proof we assume that, upon delivering $\qp{\texttt{CommitShard}, r, f, s}$, we have $r \in batches$ and $batches\qp{r}$ $\texttt{Witnessing}$ at $\broker$.

\medskip

Upon delivering $\qp{\texttt{CommitShard}, r, f, s}$ from $\server_n$ (line \ref{line:bdelivercommitshard}), $\broker$ verifies that $r \in batches$ and $batches\qp{r}$ is $\texttt{Witnessing}$ (line \ref{line:bcommitshardfetchbatch}). $\broker$ then verifies that $s$ is $\server_n$'s multisignature for $\qp{\texttt{Commit}, r, \mapdomain{f}}$ (line \ref{line:bmultiverifycommit}). Recalling that $\mapdomain{f} \subseteq \cp{i}$ and, by Lemma \ref{lemma:payloadsarestable}, we still have
\begin{align*}
    i &= \sort{\mapdomain{\bpayloads{r}}} \\
    p &= \sort{\mapcodomain{\bpayloads{r}}}
\end{align*}
for all $k \leq \abs{i}$ such that $\rp{i_k, \rp{r'_k, w'_k, p'_k, m'_k}} \in f$ (line \ref{line:bcommitshardloop})), $\broker$ successfully verifies that: $i_k \in \cp{i}$ (line \ref{line:bcommitshardcheckid}); $w'_k$  is a plurality certificate for $\qp{\texttt{Witness}, r'_k}$ (line \ref{line:bverifyconflictroot}); $p'_k$ is a proof for $\rp{i_k, c_k, m'_k}$ from $r'_k$ (lines \ref{line:bextraxtxontexmessage} and \ref{line:bverifyconflictproof}); and $m'_k \neq m_k$ (lines \ref{line:bextraxtxontexmessage} and \ref{line:bverifyconflictmessage}). Having done so, $\broker$ adds $\rp{\server_n, \any}$ to $batches\qp{r}.commits$ (lines \ref{line:bcommitshardfetchbatch} and \ref{line:bstorecommit}).

\medskip

In summary, for all $n \leq 2f + 1$, either $\abs{\bcommits{r}} \geq 2f + 1$, or $\broker$ adds a distinct element to $batches\qp{r}.commits$. Noting that $\broker$ never removes elements from $batches\qp{r}.commits$, this trivially reduces to $\broker$ eventually satisfying $\abs{\bcommits{r}} \geq 2f + 1$, and the lemma is proved.
\end{proof}
\end{lemma}


\begin{lemma}
\label{lemma:storedcommitshavenoforeignids}
Let $\broker$ be a correct broker, let $r$ be a root such that $\bcommits{r} \neq \bot$. Let $\rp{\server, \rp{i, \any}} \in \bcommits{r}$. We have $i \subseteq \mapdomain{\bpayloads{r}}$.
\begin{proof}
Upon first setting $batches\qp{r}$'s variant to $\texttt{Committing}$ (line \ref{line:bcommiting} only), $\broker$ sets $batches\qp{r}.commits = \cp{}$. Moreover, $\broker$ adds $\rp{\server, \rp{i, \any}}$ to $batches\qp{r}.commits$ only by executing line \ref{line:bstorecommit}. Before doing so, $\broker$ retrieves $b = batches\qp{r}$ (line \ref{line:bcommitshardfetchbatch}), loops through every element $\hat i$ of $i$ (line \ref{line:bcommitshardloop}) and verifies $\hat i \in b.payloads$ (line \ref{line:bcommitshardcheckid}). The lemma immediately follows from Definition \ref{definition:brokervariables}.
\end{proof}
\end{lemma}

\begin{lemma}
\label{lemma:payloadsarebroadcast}
Let $\broker$ be a correct broker, let $r$ be a root such that $\bpayloads{r} \neq \bot$, let $\rp{i, \rp{c, m}} \in \bpayloads{r}$ such that $\client = \directorymap{i}$ is correct. We have that $\client$ broadcast $m$ for $c$.
\begin{proof}
Upon initialization, $batches$ is empty at $\broker$ (line \ref{line:bbatchesinit}). Moreover, $\broker$ adds $\rp{r, b}$ to $batches$ only by executing line \ref{line:bbatchesupdate}. Upon doing so, $\broker$ sets $b.reductions = \cp{}$. By Definition \ref{definition:brokervariables} and Lemma \ref{lemma:signaturepartition} we then have $\mapdomain{\bsignatures{r}} = \mapdomain{\bpayloads{r}}$. Therefore, by Lemma \ref{lemma:batchesarecorrectlysigned}, $\client$ signed $\qp{\texttt{Message}, c, m}$. Because $\client$ does so (line \ref{line:csignaturemessage} only) only upon broadcasting $m$ for $c$ (line \ref{line:cbrodcastmessage}), the lemma is proved.
\end{proof}
\end{lemma}

\begin{lemma}
\label{lemma:correctserveronlywitnessestrees}
Let $\server$ be a correct server, let $r$ be a root such that $\server$ signs $\qp{\texttt{Witness}, r}$. Some $l \in \sorted{\rp{\ids \times \contexts \times \messages}}$ exists such that $r = \aroot{l}$.
\begin{proof}
$\server$ signs $\qp{\texttt{Witness}, r}$ only by executing line \ref{line:ssignwitness}. $\server$ does so only if $r \in batches$ (lines \ref{line:scheckbatcheshandlesignature} and \ref{line:scheckbatcheshandlesignaturereturn}). Upon initialization, $batches$ is empty at $\server$ (line \ref{line:sinitbatches}. Moreover, $\server$ adds $\rp{r, \any}$ to $batches$ (line \ref{line:sbatchesupdate} only) only if, for some sorted $l$ (lines \ref{line:sexpandprocedure} and \ref{line:scomputeleaves}), we have $r = \aroot{l}$ (lines \ref{line:scomputetree} and \ref{line:scomputeroot}).
\end{proof}
\end{lemma}

\begin{lemma}
\label{lemma:exceptionsarebyzantine}
Let $\broker$ be a correct broker, let $r$ be a root such that $\bcommits{r} \neq \bot$. Let $\rp{\any, \rp{i, \any}} \in \bcommits{r}$, let $\hat i \in i$, let $\client = \directorymap{\hat i}$. We have that $\client$ is Byzantine.
\begin{proof}
We underline that, by Lemmas \ref{lemma:storedcommitshavenoforeignids} and \ref{lemma:brokerknowsbatch}, $\client$ is well-defined. Upon first setting $batches\qp{r}$'s variant to $\texttt{Committing}$ (line \ref{line:bcommiting} only), $\broker$ sets $batches\qp{r}.commits = \cp{}$. Moreover, $\broker$ adds $\rp{\any, \rp{i, \any}}$ to $batches\qp{r}.commits$ only by executing line \ref{line:bstorecommit}. Before doing so, $\broker$ retrieves $b = batches\qp{r}$ (line \ref{line:bcommitshardfetchbatch}), then loops through all elements of $i$ (line \ref{line:bcommitshardloop}). Upon looping over $\hat i$, $\broker$ retrieves $\rp{c, m} = b.payloads\qp{r}$. $\broker$ then verifies that, for some root $r'$, at least $f + 1$ servers signed $\qp{\texttt{Witness}, r'}$ (line \ref{line:bverifyconflictroot}). Next, $\broker$ verifies, that for some proof $p$, index $n$ and message $m'$, we have
\begin{equation}
\label{equation:exceptionsarebyzantine.averify}
    \averify{r'}{p}{n}{\rp{\hat i, c, m'}} = \true
\end{equation}
(line \ref{line:bverifyconflictproof}). Finally, $\broker$ verifies that $m \neq m'$ (line \ref{line:bverifyconflictmessage}). 

\medskip

Let us assume by contradiction that $\client$ is correct. By Lemma \ref{lemma:payloadsarebroadcast} we immediately have that $\client$ broadcast $m$ for $c$. Moreover, noting that at most $f$ servers are Byzantine, at least one correct server signed $\qp{\texttt{Witness}, r'}$. As a result, by Lemma \ref{lemma:correctserveronlywitnessestrees}, some $l' \in \sorted{\ids \times \contexts \times \messages}$ exists such that $r' = \aroot{l'}$. Moreover, by Equation \ref{equation:exceptionsarebyzantine.averify} and Definition \ref{definition:idealaccumulator}, we have $l'_n = \rp{i, c, m'}$. By Lemma \ref{lemma:witnessedbatchissigned}, this proves that $\client$ broadcast $m'$ for $c$. In summary, $\client$ broadcast $m$ and $m' \neq m$ for $c$. This contradicts $\client$ being correct and proves the lemma.
\end{proof}
\end{lemma}

\begin{lemma}
\label{lemma:commitsarecorrectlysigned}
Let $\broker$ be a correct broker, let $r$ be a root such that $\bcommits{r} \neq \bot$. Let $\rp{\server, \rp{\epsilon, s}} \in \bcommits{r}$. We have that $s$ is $\server$'s multisignature for $\qp{\texttt{Commit}, r, \epsilon}$.
\begin{proof}
Upon setting $batches\qp{r}$'s variant to $\texttt{Committing}$ (line \ref{line:bcommiting} only), $\broker$ initializes $batches\qp{r}.commits$ to an empty map. Moreover, $\broker$ adds $\rp{\server, \rp{\epsilon, s}}$ to $batches\qp{r}.commits$ only by executing line \ref{line:bstorecommit}. $\broker$ does so only if $s$ is $\server$'s multisignature for $\qp{\texttt{Commit}, r, \epsilon}$ (lines \ref{line:bmultiverifycommit}, \ref{line:bmultiverifycommitreturn}).
\end{proof}
\end{lemma}

\begin{lemma}
\label{lemma:committohasids}Let $\broker$ be a correct broker, let $r$ be a root, let $\server \in \bcommitto{r}$ be a correct server. We have that $r \in batches$ at server, and $\server$ knows all elements of $\mapdomain{\bpayloads{r}}$.
\begin{proof}
Upon initialization of $batches\qp{r}$, we have $batches\qp{r}.commit\_to = \emptyset$ at $\broker$ (line \ref{line:bbatchesupdate} only). Moreover, $\broker$ adds $\server$ to $batches\qp{r}.commit\_to$ only by executing line \ref{line:baddtocommitto}. $\broker$ does so only upon receiving a $\qp{\texttt{WitnessShard}, r}$ message from $\server$ (line \ref{line:bdeliverwitnessshard}). In turn, $\server$ sends a $\qp{\texttt{WitnessShard}, r}$ message to $\broker$ only by executing line \ref{line:shandlesignaturessendresponse}. $\server$ does so only if $r \in batches$ at $\server$ (lines \ref{line:scheckbatcheshandlesignature} and \ref{line:scheckbatcheshandlesignaturereturn}) and $\broker$ knows all elements of $batches\qp{r}.ids$ (lines \ref{line:ssignatureclientverification} and \ref{line:ssignatureclientverificationreturn}). By Lemma \ref{lemma:payloadsmatchbetweenbrokerandserver}, however, we have $batches\qp{r}.ids = \sort{\mapdomain{\bpayloads{r}}}$, and the lemma is proved.
\end{proof}
\end{lemma}

\begin{lemma}
\label{lemma:committoserverbecameplurality}
Let $\broker$ be a correct broker, let $r$ be a root such that $\bcommits{r} \neq \bot$. We have $\abs{\bcommitto{r}} \geq f + 1$.
\begin{proof}
We start by noting that $\broker$ updates $batches\qp{r}$'s variant to $\mathtt{Committing}$ only by executing line \ref{line:bcommiting}. $\broker$ does so only if $batches\qp{r}$ is $\texttt{Witnessing}$ and $\abs{batches\qp{r}.witnesses} >= f+1$ (line \ref{line:binwitnessing}). Upon setting $batches\qp{r}$'s variant to $\texttt{Witnessing}$ (line \ref{line:bbatchinwitnessing} only), $\broker$ initializes $batches\qp{r}.witnesses$ to an empty map. Finally, whenever $\broker$ adds $\rp{\server, \any}$ to $batches\qp{r}.witnesses$ (line \ref{line:bstorewitnessshard} only), $\broker$ also adds $\server$ to $batches\qp{r}.commit\_to$ (line \ref{line:baddtocommitto}). The lemma immediately follows from the observation that $\broker$ never removes elements from $batches\qp{r}.commit\_to$.
\end{proof}
\end{lemma}

\begin{lemma}
\label{lemma:eventuallycommittable}
Let $\broker$ be a correct broker, let $r$ be a root such that $\bcommits{r} \neq \bot$. We eventually have $\bcommittable{r} = \true$.
\begin{proof}
We start by noting that $\broker$ removes $r$ from $batches$ (line \ref{line:sremoveroot} only) only if $batches\qp{r}$ is $\texttt{Completing}$ (line \ref{line:bbatchcompleting}). Moreover, $\broker$ sets $batches\qp{r}$'s variant to $\texttt{Completing}$ (line  \ref{line:bcompleting} only) only if $batches\qp{r}.committable = \true$. In other words, if $\bcommits{r} \neq \bot$, then $\broker$ removes $r$ from $batches$ only if $\bcommittable{r} = \true$.

\medskip

$\broker$ updates $batches\qp{r}$'s variant to $\texttt{Committing}$ (line \ref{line:bcommiting} only) only if $batches\qp{r}$'s variant is $\texttt{Witnessing}$. Immediately after setting $batches\qp{r}$'s variant to $\texttt{Witnessing}$ (line \ref{line:bbatchinwitnessing} only), $\broker$ sets a $\qp{\texttt{Committable}, r}$ timer (line \ref{line:bcommitabletimer}). When $\qp{\texttt{Committable}, r}$ eventually rings (line \ref{line:bcommitabletimerring}), $\broker$ checks if $r \in batches$ (line \ref{line:bhavebacthring}). If so, $\broker$ sets $batches\qp{r}.committable$ to $\true$. Otherwise, as we proved above, we previously had $batches\qp{r}.committable = \true$ at $\broker$, and the lemma is proved.
\end{proof}
\end{lemma}

\begin{lemma}
\label{lemma:committedbatchisdelivered}
Let $\broker$ be a correct broker, let $r$ be a root such that $\abs{\bcommits{r}} \geq 2f + 1$. Let
\begin{equation*}
    E = \bigcup_{\rp{\any, \rp{\epsilon, \any}} \in \bcommits{r}} \epsilon
\end{equation*}
Let $\hat i \in \rp{\mapdomain{\bpayloads{r}} \setminus E}$, let $\hat \client = \directorymap{\hat i}$, let $\rp{\hat c, \any} = \bpayloads{r}\qp{\hat i}$. Some correct server $\server$  exists such that $\server$ eventually delivers a message from $\hat \client$ for $\hat c$.
\begin{proof}
We start by noting that, if $batches\qp{r}$ is $\texttt{Committing}$ at $\broker$, $\broker$ never removes elements from $batches\qp{r}.commits$, and $\broker$ updates $batches\qp{r}$'s variant only by executing line \ref{line:bcompleting}. As a result, by Lemma \ref{lemma:eventuallycommittable}, $\broker$ is eventually guaranteed to detect that: $batches\qp{r}$ is $\texttt{Committing}$; $batches\qp{r}.commitable = true$; and $\abs{batches\qp{r}.commits} \geq 2f + 1$ (line \ref{line:bpluralitycommitingcheck}). Upon doing so, $\broker$ builds a map $p: \powerset{\ids} \rightarrowtail \powerset{\multisignatures}$ such that
\begin{equation}
\label{equation:committedbatchisdelivered.patcheskeys}
    \rp{\rp{\epsilon, \any} \in p} \Longleftrightarrow \rp{\rp{\any, \rp{\epsilon, \any}} \in \bcommits{r}}
\end{equation}
and
\begin{equation}
\label{equation:committedbatchisdelivered.patchesvalues}
    p\qp{\epsilon} = \bigcup_{\rp{\any, \rp{\epsilon, s}} \in \bcommits{r}} s
\end{equation}
(lines \ref{line:bpatchesinit}, \ref{line:blooponthecommits}, \ref{line:bpatchesupdate}). From $p$, $\broker$ builds a map $q: \powerset{\ids} \rightarrowtail \multisignatures$ that to each $\epsilon$ in $p$ associates the aggregation of $p\qp{\epsilon}$ (line \ref{line:patchesaggregation}). By Equation \ref{equation:committedbatchisdelivered.patcheskeys} we immediately have
\begin{equation*}
    \bigcup_{\rp{\epsilon, \any} \in q} \epsilon = E
\end{equation*}
For all $\rp{\any, c} \in q$, let $\abs{c}$ denote the number of signers of $c$. 
By Equations \ref{equation:committedbatchisdelivered.patcheskeys} and \ref{equation:committedbatchisdelivered.patchesvalues} we have
\begin{equation*}
    \sum_{\rp{\any, c} \in q} \abs{c} = \sum_{\rp{\any, s} \in p} \abs{s} = \abs{\bcommits{r}} \geq 2f + 1
\end{equation*}
Finally, by Lemma \ref{lemma:commitsarecorrectlysigned}, for every $\rp{\epsilon, c} \in q$, $c$ certifies $\qp{\texttt{Commit}, r, \epsilon}$. Having computed $q$, $\broker$ sends a $\qp{\texttt{Commit}, r, q}$ message to all servers in $\bcommitto{r}$ (lines \ref{line:bselectcommittoserver} and \ref{line:bsendtocommittoserver}).
    
\medskip
    
By Lemma \ref{lemma:committoserverbecameplurality}, we have $\abs{\bcommitto{r}} \geq f + 1$. Let $\server \in \bcommits{r}$ be a correct server. Noting that at most $f$ servers are Byzantine, $\server$ is guaranteed to exist. By Lemma \ref{lemma:committohasids}, upon receiving a $\qp{\texttt{Commit}, r, q}$ message from $\broker$ (line \ref{line:sdelivercommitpatches}), $\server$ verifies that $r \in batches$ (line \ref{line:srootinbatches}). By Lemma \ref{lemma:payloadsmatchbetweenbrokerandserver}, $\server$ then verifies to know all elements $\rp{\cp{batches\qp{r}.ids} = \mapdomain{\bpayloads{r}}}$ (line \ref{line:sdircheckcommit}). Next, $\server$ verifies that, for every $(\epsilon, c) \in q$, $c$ certifies $\qp{\texttt{Commit}, r, \epsilon}$ (lines \ref{line:spatches} and \ref{line:svaliditysignatureconflicts}). Finally, $\server$ verifies that
\begin{equation*}
    \sum_{\rp{\any, c}} \abs{c} \geq 2f + 1
\end{equation*}
(lines \ref{line:ssignerinit}, \ref{line:spatches}, \ref{line:ssignerextend} and \ref{line:squorumcsignatureconflicts}). Having done so, $\server$ computes $E$ (line \ref{line:sunionexclusion}). Next, by Lemma \ref{lemma:payloadsmatchbetweenbrokerandserver} $\server$ loops through all elements of $\bpayloads{r}$ (line \ref{line:sdeliverfrombrokerloop}). For every $\rp{i, \rp{c, m}} \in \bpayloads{r}$ such that $i \notin E$ (line \ref{line:sdelivercheck}), $\server$ either delivers $m$ from $\directorymap{i}$ for $c$ (lines \ref{line:stakekeycard} and \ref{line:sdeliver}), or $\server$ has already delivered a message from $\directorymap{i}$ for $c$ (line \ref{line:sdelivercheck}, see Lemma \ref{lemma:deliveredtracksdelivered}). This proves in particular that $\server$ delivers a message from $\hat \client$ for $\hat c$, and concludes the lemma.
\end{proof}
\end{lemma}

\begin{theorem}
\csbal satisfies validity.
\begin{proof}
Let $\client$ be a correct client, let $c$ be a context, let $m$ be a message such that $\client$ broadcasts $m$ for $c$. By Lemma \ref{lemma:broadcastendsincorrectbatch}, either a correct server delivers message from $\client$ for $c$, or some correct broker $\broker$ and some root $r$ exist such that eventually $\bpayloads{r}\qp{i} = \rp{c, m}$. Throughout the remainder of this proof, we assume the existence of $\broker$ and $r$. By Lemma \ref{lemma:batchesarewitnessed}, at some point in time we have $\abs{\bwitnesses{r}} \geq f + 1$. Therefore, by Lemma \ref{lemma:batchesarecommitted}, at some point in time we have $\abs{\bcommits{r}} \geq 2f + 1$. Moreover, by Lemma \ref{lemma:exceptionsarebyzantine}, for all $\rp{\any, \rp{\epsilon, \any}} \in \bcommits{r}$, we have $\directorymap{\client} \notin \epsilon$. As a result, by Lemma \ref{lemma:committedbatchisdelivered}, some correct server delivers a message from $\client$ for $c$, and the theorem is proved.
\end{proof}
\end{theorem}
\subsection{Complexity}
\label{subsection:csbal.complexity}

In this section, we prove to the fullest extent of formal detail the good-case signature and communication complexity of \csbal.

\subsubsection{Auxiliary results}

In this section we gather definitions and lemmas that we will use to prove, in the next sections, the good-case signature and communication complexity of \csbal. The results presented in this section hold independently of \csbal itself, and could be applicable to a broader spectrum of analyses.

\begin{notation}[Bit strings]
We use $\bitstrings = \cp{0, 1}^{<\infty}$ to denote all finite strings of bits. We use $\emptyseq$ to denote the empty sequence of bits. We use $\bitstrings^{<b} = \cp{0, 1}^{<b}$,  $\bitstrings^{\leq b} = \cp{0, 1}^{\leq b}$, $\bitstrings^b = \cp{0, 1}^b$, $\bitstrings^{\geq b} = \cp{0, 1}^{\geq b}$ and $\bitstrings^{>b} = \cp{0, 1}^{>b}$. We use $\bitstrings^{even}$ and $\bitstrings^{odd}$ to denote the sets of strings with an even and odd number of bits, respectively. We use programming notation when indexing a string of bits: for all $s \in \bitstrings$ we use $s = \rp{s_0, s_1, \ldots}$.
\end{notation}

\begin{notation}[Cropping]
Let $n \in \mathbb{N}$, let $s \in \bitstrings$. We use the following \textbf{cropping} notation:
\begin{align*}
    \crop{s}{\leq n} &= \rp{s_0, \ldots, s_n} \\
    \crop{s}{\geq n} &= \rp{s_n, \ldots, s_{\abs{s} - 1}}
\end{align*}
We also use $\crop{s}{<n} = \crop{s}{\leq n - 1}$ and $\crop{s}{>n} = \crop{s}{\geq n + 1}$.
\end{notation}

\begin{definition}[Integer encoding]
\label{definition:integerencoding}
For all $b \in \mathbb{N}$, the \textbf{$b$-bits integer representation} $\intreprsym{b}: \rp{0..\rp{2^b-1}} \leftrightarrow \bitstrings^b$ is defined by
\begin{align*}
    \intrepr{b}{n}_i &= \floor{\frac{n}{2^i}} \bmod 2 \\
    \intread{b}{s} &= \sum_{i = 0}^b 2^i s_i
\end{align*}
The \textbf{$b$-bits integer encoding} $\intencsym{b}: \bitstrings \times \rp{0..\rp{2^b - 1}} \leftrightarrow \bitstrings^{\geq b}$ is defined by
\begin{align*}
    \intenc{b}{s}{n} &= \intrepr{b}{n} \append s \\
    \intdec{b}{s} &= \rp{\crop{s}{\geq b}, \intread{b}{\crop{s}{<b}}}
\end{align*}
\end{definition}

\begin{lemma}
\label{lemma:intreadisinjective}
Let $b \in \mathbb{N}$. $\intreadsym{b}$ is injective.
\begin{proof}
Let $s, s' \in \bitstrings$ such that $s \neq s'$. Let
\begin{equation*}
    k = \max i \suchthat s_i \neq s'_i
\end{equation*}
Noting that $s \neq s'$, $k$ is guaranteed to exist. Without loss of generality, let us assume $s_k = 1$, $s'_k = 0$. We by Definition \ref{definition:integerencoding} have
\begin{align*}
    \intread{b}{s} - \intread{b}{s'} &= \sum_{i = 0} 2^i \rp{s_i - s'_i} = \\
    &= \sum_{i = 0}^{k - 1} 2^i \rp{s_i - s'_i} + 2^k \geq \\
    &\geq \sum_{i = 0}^{k - 1} -2^i + 2^k \geq 1 
\end{align*}
which proves $\intread{b}{s} \neq \intread{b}{s'}$ and concludes the lemma
\end{proof}
\end{lemma}

\begin{lemma}
\label{lemma:intreprinvertsintread}
Let $b \in \mathbb{N}$, let $s \in \bitstrings^b$. We have $\intrepr{b}{\intread{b}{s}} = s$.
\begin{proof}
By Definition \ref{definition:integerencoding} we have
\begin{align*}
    \intrepr{b}{\intread{b}{s}}_i &= \floor{\frac{\sum_{j = 0}^b 2^j s_j}{2^i}} \bmod 2 = \\
    &= \floor{\frac{\sum_{j = 0}^{i - 1} 2^j s_j}{2^i} + s_i + \frac{\sum_{j = i + 1}^b 2^j s_j}{2^i}} \bmod 2 = \\
    &= \floor{\underbrace{\frac{\sum_{j = 0}^{i - 1} 2^j s_j}{2^i}}_{< 1} + \underbrace{s_i + 2\sum_{j = 0}^{b - i - 1} 2^j s_{j + i + 1}}_{\in \mathbb{N}}} \bmod 2 = \\
    &= \rp{s_i + 2\underbrace{\sum_{j = 0}^{b - i - 1} 2^j s_{j + i + 1}}_{\in \mathbb{N}}} \bmod 2 = s_i
\end{align*}
\end{proof}
\end{lemma}

\begin{lemma}
\label{lemma:intreprisbijection}
Let $b \in \mathbb{N}$. $\intreprsym{b}$ is is a bijection.
\begin{proof}
It follows immediately from Lemmas \ref{lemma:intreadisinjective} and \ref{lemma:intreprinvertsintread}.
\end{proof}
\end{lemma}

\begin{lemma}
\label{lemma:intencisbijection}
Let $b \in \mathbb{N}$. $\intencsym{b}$ is a bijection.
\begin{proof}
It follows immediately from Definition \ref{definition:integerencoding} and Lemma \ref{lemma:intreprisbijection}.
\end{proof}
\end{lemma}

\begin{definition}[Varint encoding]
\label{definition:varintencoding}
The \textbf{varint representation} $\varreprsym: \mathbb{N}^+ \leftrightarrow \bitstrings^{even}$ is defined by
\begin{align*}
    \abs{\varrepr{n}} &= 2 \ceil{\log_2\rp{n + 1}} \\
    \varrepr{n}_i &= \begin{cases}
        1 \caseiff i \bmod 2 = 0, i < 2 \ceil{\log_2\rp{n + 1}} - 2\\
        0 \caseiff i = 2 \ceil{\log_2\rp{n + 1}} - 2\\
        \intrepr{\ceil{\log_2\rp{n + 1}}}{n}_{\frac{i - 1}{2}} \caseotherwise
    \end{cases} \\
    \varread{s} &= \intread{\frac{\abs{s}}{2}}{s_1, s_3, s_5, \ldots, s_{\abs{s} - 1}}
\end{align*}
The set of \textbf{varint-parseable strings} is the set
\begin{equation*}
    \varparseable = \cp{s \in \bitstrings \suchthat \rp{\exists k \in \mathbb{N} \suchthat s_{2k} = 0}}
\end{equation*}
The \textbf{varint encoding} $\varencsym: \rp{\bitstrings \times \mathbb{N}^+ \leftrightarrow \varparseable}$ is defined by
\begin{align*}
    \varenc{s}{n} &= \varrepr{n} \append s \\
    \vardec{s} &= \rp{\crop{s}{\geq \varlen{s}}, \varread{\crop{s}{< \varlen{s}}}}
\end{align*}
with $\varlensym: \bitstrings \rightarrow \mathbb{N}$ defined by
\begin{equation*}
    \varlen{s} = 2 \rp{\min k \suchthat s_{2k} = 0} + 2
\end{equation*}
\end{definition}

Definition \ref{definition:varintencoding} contains a slight abuse of notation: we prove below that $\varreadsym$ inverts $\varreprsym$, but we do not prove that $\varreprsym$ inverts $\varreadsym$ (it does not). Similarly we prove that  $\vardecsym$ inverts $\varencsym$, but not that $\varencsym$ inverts $\vardecsym$.

\begin{lemma}
\label{lemma:varreadinvertsvarrepr}
Let $n \in \mathbb{N}^+$. We have $\varread{\varrepr{n}} = n$.
\begin{proof}
By Definition \ref{definition:varintencoding} and Lemma \ref{lemma:intencisbijection} we have
\begin{align*}
    \varread{\varrepr{n}} &= \intread{\frac{2 \ceil{\log_2\rp{n + 1}}}{2}}{\intrepr{\ceil{\log_2\rp{n + 1}}}{n}_0, \intrepr{\ceil{\log_2\rp{n + 1}}}{n}_1, \ldots, \intrepr{\ceil{\log_2\rp{n + 1}}}{n}_{\ceil{\log_2\rp{n + 1}} - 1}} = \\
    &= \intread{\log_2\rp{n + 1}}{\intrepr{\log_2\rp{n + 1}}{n}} = n
\end{align*}
\end{proof}
\end{lemma}

\begin{lemma}
\label{lemma:vardecinvertsvarenc}
Let let $s \in \bitstrings$, let $n \in \mathbb{N}^+$. We have $\vardec{\varenc{s}{n}} = \rp{s, n}$.
\begin{proof}
All derivations in this lemma follow directly from Definition \ref{definition:varintencoding}. We have
\begin{equation*}
    \varenc{s}{n} = \varrepr{n} \append s
\end{equation*}
Moreover, we have
\begin{equation*}
    \forall k < \ceil{\log_2{n + 1}} - 1, \varenc{s}{n}_{2k} = \varrepr{n}_{2k} = 1
\end{equation*}
and
\begin{equation*}
    \varenc{s}{n}_{2 \rp{\ceil{\log_2\rp{n + 1}} - 1}} = \varread{n}_{2 \rp{\ceil{\log_2\rp{n + 1}} - 1}} = 0
\end{equation*}
which proves
\begin{equation*}
    \varlen{\varenc{s}{n}} = 2\ceil{\log_2\rp{n + 1}}
\end{equation*}
This implies
\begin{align*}
    \crop{\varenc{s}{n}}{<\varlen{\varenc{s}{n}}} = \crop{\varrepr{n} \append s}{< 2\ceil{\log_2\rp{n + 1}}} = \varrepr{n}
\end{align*}
and similarly
\begin{equation*}
    \crop{\varenc{s}{n}}{\geq \varlen{\varenc{s}{n}}} = s
\end{equation*}
By Lemma \ref{lemma:varreadinvertsvarrepr} we then have
\begin{equation*}
    \vardec{\varenc{s}{n}} = \rp{s, \varread{\varrepr{n}}} = \rp{s, n}
\end{equation*}
and the lemma is proved.
\end{proof}
\end{lemma}

\begin{notation}[Integer partition]
\label{notation:integerpartition}
Let $X$ be a finite set, let $\mu: X \rightarrow \powerset[<\infty]{\mathbb{N}}$. We call $\mu$ a \textbf{integer partition} on $X$. We use
\begin{gather*}
    \abs{\mu} = \sum_{x \in X} \abs{\mu\rp{x}} \\
    \max \mu = \max_{x \in X} \max \mu\rp{x}
\end{gather*}
\end{notation}

\begin{notation}[Enumeration]
Let $X = \cp{x_1, \ldots, x_{\abs{X}}}$ be a finite set. We use $\enumeration{X} = \rp{x_1, \ldots, x_{\abs{X}}}$ to denote any specific \textbf{enumeration} of $X$.
\end{notation}

\begin{definition}[Partition representation]
\label{definition:partitionrepresentation}
Let $X$ be a finite set. The \textbf{partition representation} on $X$ is the function $\partitionreprsym: \rp{X \rightarrow \powerset[<\infty]{\mathbb{N}}} \leftrightarrow \bitstrings$ defined, with
\begin{equation*}
    \rp{x_1, \ldots, x_{\abs{X}}} = \enumeration{X}
\end{equation*}
by
\begin{align*}
    w\rp{\mu} &= \ceil{\log_2\rp{\max \mu + 1}} \\
    d\rp{\mu} &= \ceil{\log_2\rp{\abs{\mu} + 1}} \\
    y_n\rp{\mu} &= \abs{\mu\rp{x_n}} \\
    \rp{z_1\rp{\mu}, \ldots, z_{\abs{\mu}}\rp{\mu}} &= \enumeration{\mu\rp{x_1}} \append \ldots \append \enumeration{\mu\rp{x_{\abs{X}}}} \\
    \mathtt{v}_0\rp{\mu} &= \emptyseq \\
    \mathtt{v}_n\rp{\mu} &= \intenc{w\rp{\mu}}{\mathtt{v}_{n - 1}\rp{\mu}}{z_n\rp{\mu}} \\
    \mathtt{v}\rp{\mu} &= \mathtt{v}_{\abs{\mu}}\rp{\mu} \\
    \mathtt{k}_0\rp{\mu} &= \mathtt{v}\rp{\mu} \\
    \mathtt{k}_n\rp{\mu} &= \intenc{d\rp{\mu}}{\mathtt{k}_{n - 1}\rp{\mu}}{y_n\rp{\mu}} \\
    \mathtt{k}\rp{\mu} &= \mathtt{k}_{\abs{X}}\rp{\mu} \\
    \mathtt{g}\rp{\mu} &= \varenc{\mathtt{k}\rp{\mu}}{w\rp{\mu}} \\
    \mathtt{h}\rp{\mu} &= \varenc{\mathtt{g}\rp{\mu}}{d\rp{\mu}} \\
    \partitionrepr{\mu} &= \mathtt{h}\rp{\mu}
\end{align*}
and
\begin{align*}
    \mathtt{h}\rp{s} &= s \\
    \rp{\mathtt{g}\rp{s}, d\rp{s}} &= \vardec{\mathtt{h}\rp{s}} \\
    \rp{\mathtt{k}\rp{s}, w\rp{s}} &= \vardec{\mathtt{g}\rp{s}} \\
    \mathtt{k}_{\abs{X}}\rp{s} &= \mathtt{k}\rp{s} \\
    \rp{\mathtt{k}_{n - 1}\rp{s}, y_n\rp{s}} &= \intdec{d\rp{s}}{\mathtt{k}_n\rp{s}} \\
    t\rp{s} &= \sum_{n = 1}^{\abs{X}} y_n\rp{s} \\
    \mathtt{v}\rp{s} &= \mathtt{k}_0\rp{s} \\
    \mathtt{v}_{t\rp{s}}\rp{s} &= \mathtt{v}\rp{s} \\
    \rp{\mathtt{v}_{n - 1}\rp{s}, z_n\rp{s}} &= \intdec{w\rp{s}}{\mathtt{v}_n\rp{s}} \\
    p_n\rp{s} &= \sum_{i = 1}^n y_n\rp{s} \\
    \partitionread{s}\rp{x_n} &= \cp{z_{p_{n - 1}\rp{s} + 1}\rp{s}, \ldots, z_{p_n\rp{s}}\rp{s}}
\end{align*}
\end{definition}

Definition \ref{definition:partitionrepresentation} contains a slight abuse of notation: we prove below that $\partitionreadsym$ inverts $\partitionreprsym$, but $\partitionreprsym$ is obviously not defined on all of $\bitstrings$ ($\emptyseq$ being a trivial counterexample).

\begin{lemma}
\label{lemma:partitionreadinvertspartitionrepr}
Let $X$ be a finite set, let $\mu: X \rightarrow \powerset[<\infty]{\mathbb{N}}$. We have $\partitionread{\partitionrepr{\mu}} = \mu$.
\begin{proof}
All derivations in this lemma follow from Notation \ref{notation:integerpartition}, Definition \ref{definition:partitionrepresentation} and Lemmas \ref{lemma:intencisbijection} and \ref{lemma:vardecinvertsvarenc}. Let $s = \partitionrepr{\mu}$. We trivially have
\begin{equation*}
    \mathtt{h}\rp{s} = s = \rho\rp{\mu} = \mathtt{h}\rp{\mu}
\end{equation*}
In steps, we can derive
\begin{equation*}
    \rp{\mathtt{g}\rp{s}, d\rp{s}} = \vardec{\mathtt{h}\rp{s}} = \vardec{\mathtt{h}\rp{\mu}} = \vardec{\varenc{\mathtt{g}\rp{\mu}}{d\rp{\mu}}} = \rp{\mathtt{g}\rp{\mu}, d\rp{\mu}}
\end{equation*}
\begin{equation*}
    \rp{\mathtt{k}\rp{s}, w\rp{s}} = \vardec{\mathtt{g}\rp{s}} = \vardec{\mathtt{g}\rp{\mu}} = \vardec{\varenc{\mathtt{k}\rp{\mu}}{w\rp{\mu}}} = \rp{\mathtt{k}\rp{\mu}, w\rp{\mu}}
\end{equation*}
and
\begin{equation*}
    \mathtt{k}_{\abs{X}}\rp{s} = \mathtt{k}\rp{s} = \mathtt{k}\rp{\mu} = \mathtt{k}_{\abs{X}}\rp{\mu}
\end{equation*}
By backwards induction, for all $n \in 1..\abs{X}$ we then have
\begin{align*}
    \rp{\mathtt{k}_{n - 1}\rp{s}, y_n\rp{s}} &= \intdec{d\rp{s}}{\mathtt{k}_n\rp{s}} = \intdec{d\rp{\mu}}{\mathtt{k}_n\rp{\mu}} = \\
    &= \intdec{d\rp{\mu}}{\intenc{d\rp{\mu}}{\mathtt{k}_{n - 1}\rp{\mu}}{y_n\rp{\mu}}} = \rp{\mathtt{k}_{n - 1}\rp{\mu}, y_n\rp{\mu}}
\end{align*}
from which follows
\begin{equation*}
    t\rp{s} = \sum_{n = 1}^{\abs{X}} y_n\rp{s} = \sum_{n = 1}^{\abs{X}} y_n\rp{\mu} = \sum_{n = 1}^{\abs{X}} \abs{\mu\rp{x_n}} = \abs{\mu}
\end{equation*}
In steps, we can then derive
\begin{equation*}
    \mathtt{v}\rp{s} = \mathtt{k}_0\rp{s} = \mathtt{k}_0\rp{\mu} = \mathtt{v}\rp{\mu}
\end{equation*}
\begin{equation*}
    \mathtt{v}_{\abs{\mu}}\rp{s} = \mathtt{v}_{t\rp{s}}\rp{s} = \mathtt{v}\rp{s} = \mathtt{v}\rp{\mu} = \mathtt{v}_{\abs{\mu}}\rp{\mu}
\end{equation*}
Again by backwards induction, for all $n \in 1..\abs{\mu}$ we then have
\begin{align*}
    \rp{\mathtt{v}_{n - 1}\rp{s}, z_n\rp{s}} &= \intdec{w\rp{s}}{\mathtt{v}_n\rp{s}} = \intdec{w\rp{\mu}}{\mathtt{v}_n\rp{\mu}} \\
    &= \intdec{w\rp{\mu}}{\intenc{w\rp{\mu}}{\mathtt{v}_{n - 1}\rp{\mu}}{z_n\rp{\mu}}} = \rp{\mathtt{v}_{n - 1}\rp{\mu}, z_n\rp{\mu}}
\end{align*}
In summary we have
\begin{gather*}
    \forall n \in 1..\abs{X}, y_n\rp{s} = y_n\rp{\mu} \\
    \forall n \in 1..\abs{\mu}, z_n\rp{s} = z_n\rp{\mu}
\end{gather*}
from which finally follows
\begin{align*}
    \partitionread{s}\rp{x_n} &= \cp{z_{p_{n - 1}\rp{s} + 1}\rp{s}, \ldots, z_{p_n\rp{s}}\rp{s}} = \\
    &= \cp{z_{p_{n - 1}\rp{s} + 1}\rp{\mu}, \ldots, z_{p_n\rp{s}}\rp{\mu}} = \\
    &= \cp{z_{\rp{\sum_{i = 1}^{n - 1} y_i\rp{s} + 1}}\rp{\mu}, \ldots, z_{\rp{\sum_{i = 1}^n y_i\rp{s}}}\rp{\mu}} = \\
    &= \cp{z_{\rp{\sum_{i = 1}^{n - 1} y_i\rp{\mu} + 1}}\rp{\mu}, \ldots, z_{\rp{\sum_{i = 1}^n y_i\rp{\mu}}}\rp{\mu}} = \\
    &= \cp{z_{\rp{\sum_{i = 1}^{n - 1} \abs{\mu\rp{x_i}} + 1}}\rp{\mu}, \ldots, z_{\rp{\sum_{i = 1}^n \abs{\mu\rp{x_i}}}}\rp{\mu}} = \\
    &= \cp{z_{\rp{\sum_{i = 1}^{n - 1} \abs{\enumeration{\mu\rp{x_i}}} + 1}}\rp{\mu}, \ldots, z_{\rp{\sum_{i = 1}^n \abs{\enumeration{\mu\rp{x_i}}}}}\rp{\mu}} = \\
    &= \cp{\enumeration{\mu\rp{x_n}}_1, \ldots, \enumeration{\mu\rp{x_n}}_{\abs{\mu\rp{x_n}}}} = \mu\rp{x_n}
\end{align*}
which proves $\partitionread{\partitionrepr{\mu}} = \partitionread{s} = \mu$, and concludes the lemma.
\end{proof}
\end{lemma}

\begin{lemma}
\label{lemma:partitionreprisefficient}
Let $X$ be a finite set, let $\mu: \mathbb{N} \rightarrow \rp{X \rightarrow \powerset[<\infty]{\mathbb{N}}}$ be a sequence of partition represenations such that
\begin{align*}
    &\lim_{n \rightarrow \infty} \abs{\mu_n} = \infty \\
    &\lim_{n \rightarrow \infty} \frac{\log_2\rp{\log_2\rp{\max \mu_n}}}{\abs{\mu_n}} = 0
\end{align*}
We have
\begin{equation*}
    \lim_{n \rightarrow \infty} \frac{\abs{\partitionrepr{\mu_n}}}{\abs{\mu_n}} = \ceil{\log_2\rp{\max \mu_n + 1}}
\end{equation*}
\begin{proof}
All derivations in this proof follow from Definitions \ref{definition:partitionrepresentation}, \ref{definition:varintencoding} and \ref{definition:integerencoding}. Let $n \in \mathbb{N}$. We have
\begin{equation*}
    \abs{\mathtt{v}_0\rp{\mu_n}} = 0
\end{equation*}
and, for all $k \in 1..\abs{\mu}$,
\begin{equation*}
    \abs{\mathtt{v}_k\rp{\mu_n}} = \abs{\mathtt{v}_{k - 1}\rp{\mu_n}} + w\rp{\mu_n} = \abs{\mathtt{v}_{k - 1}\rp{\mu_n}} + \ceil{\log_2\rp{\max \mu_n + 1}}
\end{equation*}
from which, by induction, follows
\begin{equation*}
    \abs{\mathtt{v}\rp{\mu_n}} = \abs{\mu_n} \ceil{\log_2\rp{\max \mu_n + 1}}
\end{equation*}
Similarly we have
\begin{equation*}
    \abs{\mathtt{k}_0\rp{\mu_n}} = \abs{\mu_n} \ceil{\log_2\rp{\max \mu + 1}}
\end{equation*}
and, for all $k \in 1..\abs{X}$,
\begin{equation*}
    \abs{\mathtt{k}_k\rp{\mu_n}} = \abs{\mathtt{k}_{k - 1}\rp{\mu_n}} + d\rp{\mu_n} = \abs{\mathtt{k}_{k - 1}\rp{\mu_n}} + \ceil{\log_2\rp{\abs{\mu_n} + 1}}
\end{equation*}
from which, by induction, follows
\begin{equation*}
    \abs{\mathtt{k}\rp{\mu_n}} = \abs{\mu_n} \ceil{\log_2\rp{\max \mu_n + 1}} + \abs{X} \ceil{\log_2\rp{\abs{\mu_n} + 1}}
\end{equation*}
In steps we can then derive
\begin{align*}
    &\abs{\mathtt{g}\rp{\mu_n}} = \abs{\mathtt{k}\rp{\mu_n}} + 2 \ceil{\log_2\rp{w\rp{\mu_n} + 1}} = \\
    &= \abs{\mu_n} \ceil{\log_2\rp{\max \mu_n + 1}} + \abs{X} \ceil{\log_2\rp{\abs{\mu_n} + 1}} + 2 \ceil{\log_2\rp{\ceil{\log_2\rp{\max \mu_n + 1}} + 1}}
\end{align*}
and finally
\begin{align*}
    &\abs{\partitionrepr{\mu_n}} = \abs{\mathtt{h}\rp{\mu_n}} = \abs{\mathtt{g}\rp{\mu_n}} + 2 \ceil{\log_2\rp{d\rp{\mu_n}}} = \\
    &= \abs{\mu_n} \ceil{\log_2\rp{\max \mu_n + 1}} + \abs{X} \ceil{\log_2\rp{\abs{\mu_n} + 1}} + 2 \ceil{\log_2\rp{\ceil{\log_2\rp{\max \mu_n + 1}} + 1}} \\ 
    &\;\;\; + 2 \ceil{\log_2\rp{\ceil{\log_2\rp{\abs{\mu_n} + 1}} + 1}}
\end{align*}
The above holds for any $n \in \mathbb{N}$. Moreover, by hypothesis we have
\begin{align*}
    &\lim_{n \rightarrow \infty} \frac{\abs{X} \ceil{\log_2\rp{\abs{\mu_n} + 1}}}{\abs{\mu_n}} = 0 \\
    &\lim_{n \rightarrow \infty} \frac{2 \ceil{\log_2\rp{\ceil{\log_2\rp{\abs{\mu_n} + 1}} + 1}}}{\abs{\mu_n}} = 0
\end{align*}
and
\begin{align*}
    &\lim_{n \rightarrow \infty} \frac{2 \ceil{\log_2\rp{\ceil{\log_2\rp{\max \mu_n + 1}} + 1}}}{\abs{\mu_n}} = \\
    &\lim_{n \rightarrow \infty} \frac{2 \ceil{\log_2\rp{\ceil{\log_2\rp{\max \mu_n}}}}}{\abs{\mu_n}} \cancel{\frac{\ceil{\log_2\rp{\ceil{\log_2\rp{\max \mu_n + 1}} + 1}}}{\ceil{\log_2\rp{\ceil{\log_2\rp{\max \mu_n}}}}}} = 0
\end{align*}
This proves that
\begin{equation*}
    \lim_{n \rightarrow \infty} \frac{\abs{\partitionrepr{\mu_n}}}{\abs{\mu_n}} = \ceil{\log_2\rp{\max \mu_n + 1}}
\end{equation*}
and concludes the lemma.
\end{proof}
\end{lemma}

\subsubsection{Batching limit}
\label{subsubsection:csbal.complexity.batchinglimit}

As we discussed in Section \ref{section:introduction}, \csbal is designed to asymptotically match, in the good case, the signature and communication complexity of \textsf{Oracle-\csbshort}, a toy implementation of \csbshort that relies on an infallible oracle to uphold all \csbshort properties. As we discussed, \textsf{Oracle-\csbshort} achieves optimal signature complexity (as it requires no signature verification) and optimal communication complexity (if the frequency at which clients broadcast their payloads is uniform or unknown). Below we establish and discuss \csbal's \textbf{batching limit}, i.e., the collection of assumptions and limits at which \csbal matches \textsf{Oracle-\csbshort}'s complexity:
\begin{itemize}
    \item \textbf{Assumption: good-case execution}. As we discussed in Section \ref{section:model}, in the good case: \emph{links are synchronous} (messages are delivered at most one time unit after they are sent); \emph{all processes are correct}; and \emph{the set of brokers contains only one element}. \textbf{Discussion.} We assume only one broker for pedagogical reasons: as long as every broker is exposed to a high enough rate of submissions, all derivations in this section still hold true. In the real world, the assumptions of synchrony and correctness are not strict for clients: if a small fraction of broadcasting clients is slow or Byzantine, \csbal still achieves near-oracular efficiency, linearly degrading its performance as more client fail to engage with the broker to reduce its batch\footnote{Proving this result is beyond the scope of this paper, and left as an exercise to the interested reader.}.
    \item \textbf{Assumption: steady-state directory}. We assume that \emph{all servers know all broadcasting clients}. \textbf{Discussion.} This assumption is naturally satisfied if all broadcasting clients have already broadcast at least one message. In a real-world, long-lived system, most clients can safely be assumed to broadcast more than one message (this is especially true in the context of a cryptocurrency). Similarly to client synchrony and correctness, \csbal's performance degrades linearly in the number of unknown, broadcasting clients.
    \item \textbf{Assumption: concurrent broadcasts}. We assume that \emph{all clients broadcast their message within $\csbalbatchingwindow$ time units of each other}, where $\csbalbatchingwindow$ is \csbal's batching window parameter (see Sections \ref{subsection:pseudocode.csbalclient} and \ref{subsection:pseudocode.csbalbroker}). \textbf{Discussion.} Similarly to broker count, this assumption is made for pedagogical reasons: as long as the rate at which payloads are submitted to each broker is high enough, all derivations in this section still hold true.
    \item \textbf{Limit: infinite broadcasts}. We derive \csbal's complexity at the limit of \emph{infinitely many broadcasting clients}. \textbf{Discussion.} All limits derived in this section converge approximately inversely with the number of broadcasting clients. This means that even a finite, real-world implementation of \csbal can achieve near-oracular efficiency.
    \item \textbf{Limit: sub-double-exponential clients}. We assume that \emph{the number of clients is infinitely small with respect to the exponential of the exponential of the number of broadcasting clients}. \textbf{Discussion.} This limit is only technical, and trivially satisfied by any realistic number of broadcasting clients.
\end{itemize}

\subsubsection{Protocol analysis}

In this section, we establish \csbal's signature and communication complexity at the batching limit.

\begin{theorem}
At the batching limit, a \csbal server delivers a payload $p$ by performing $0$ signature verifications and exchanging at most $\rp{\ceil{\log_2\rp{\clientcount}} + \abs{p}}$ bits.
\begin{proof}
All derivations in this proof follow from the batching limit's assumptions and limits (see Section \ref{subsubsection:csbal.complexity.batchinglimit}). Let $\broker$ denote the only broker in the system. Let $\client_1, \ldots, \client_M$ denote the set of broadcasting clients. We have $M \rightarrow \infty$. For all $j$, let $i_j = \directorymap{\client_j}$, let $p_j = \rp{c_j, m_j}$ identify the payload broadcast by $\client_j$, let $a_j$ be $i_j$'s id assignment certificate. Without loss of generality we assume that for all $j < j'$ we have $i_j < i_{j'}$. The goal of this proof is to compute, for all $j \in 1..M$, $p_j$'s maximum amortized signature complexity $\psi_j$ and $p_j$'s maximum amortized bit complexity $\chi_j$.

\medskip

Let $n \in 1..M$. Without loss of generality, we assume that $\client_n$ triggers $\event{\clin}{Broadcast}$ (line \ref{line:cbrodcastmessage}) between time $0$ and $\csbalbatchingwindow$. Upon doing so, $\client$ produces a signature $s_n$ for $\qp{\texttt{Message}, c_n, m_n}$ (line \ref{line:csignaturemessage}), then sends a $\qp{\texttt{Submission}, a_n, \rp{c_n, m_n, s_n}}$ message to $\broker$ (line \ref{line:csendsubmit}). 

\medskip

$\broker$ receives $\qp{\texttt{Submission}, a_n, \rp{c_n, m_n, s_n}}$ (line \ref{line:bdeliversubmit}) between time $0$ and time $B + 1$: indeed, $\client_n$'s $\texttt{Submission}$ message takes between $0$ and $1$ time units to reach $\broker$. Upon delivering $\client_n$'s \texttt{Submission} message, $\broker$ pushes
\begin{equation*}
    y_n = Submission\cp{context: c_n, message: m_n, signature: s_n}
\end{equation*}
to $pending\qp{i_n}$ (line \ref{line:bpushback}), detects that $pending\qp{i_n}$ is not empty (line \ref{line:bexistsinpanding}), empties $pending\qp{i_n}$ (line \ref{line:bpandingupdate}) and adds $\rp{i_n, y_n}$ to $pool$ (line \ref{line:bpoolupdate}). Initially, we have $collecting = \false$ and $pool = \emptyset$ at $\broker$. Moreover, $\broker$ adds its first element to $pool$ no earlier than time $0$. Upon doing so (line \ref{line:bpoolcollectingcheck}), $\broker$ sets a $\qp{\texttt{Flush}}$ timer to ring at some time $t_f \geq B + 1$ (line \ref{line:btimerset}). 

\medskip

In summary, for all $j \in 1..M$, $\broker$ adds $\rp{i_j, y_j}$ to $pool$ by time $B + 1$, and $\qp{\texttt{Flush}}$ rings at time $t_f \geq B + 1$. When $\qp{\texttt{Flush}}$ rings (line \ref{line:btimerring}), $\broker$ takes $\rp{i_1, p_1}, \ldots, \rp{i_M, p_M}$ from $pool$ (line \ref{line:bsubmissionsequalpool}) and computes $r = \aroot{\join{i}{p}}$ (line \ref{line:brootcostruction}). For each $j \in 1..M$, $\broker$ produces an inclusion proof $q_j$ for $\rp{i_j, c_j, m_j}$ in $r$ (line \ref{line:binclusionproof}), then sends an $\qp{\texttt{Inclusion}, c_j, r, q_j}$ message to $\client_j$. Next, $\broker$ sets
\begin{equation*}
    batches\qp{r} = Reducing \cp{payloads: p, signatures: s, \ldots}
\end{equation*}
(line \ref{line:bbatchesupdate}). Finally, $\broker$ sets a $\qp{\texttt{Reduce}, r}$ to ring at some time $t_r \geq t_f + 2$ (line \ref{line:btimersetreducing}).

\medskip

$\client_n$ receives $\qp{\texttt{Inclusion}, c_n, r, q_n}$ by time $t_f + 1$ (line \ref{line:creciveinclusion}). Upon doing so, $\client_n$ produces a multi-signature $g_n$ for $\qp{\texttt{Reduction}, r}$ (line \ref{line:cmultisignreduction}), then sends a $\qp{\texttt{Reduction}, r, g_n}$ message to $\broker$ (line \ref{line:csendreduction}).

\medskip

$\broker$ receives $\qp{\texttt{Reduction}, r, g_n}$ by time $t_r$ (line \ref{line:brecivereduction}). Because $\qp{\texttt{Reduce}, r}$ has not yet rung, $batches\qp{r}$ is still $\texttt{Reducing}$ at $\broker$ (line \ref{line:bcheckreducingid}). $\broker$ adds $\rp{i_n, g_n}$ to $batches\qp{r}.reductions$ (line \ref{line:bupdatereduction}), then removes $\rp{i_n, s_n}$ from $batches\qp{r}.signatures$ (line \ref{line:bupdateremoveid}).

\medskip

In summary, before $\qp{\texttt{Reduce}, r}$ rings at time $t_r$, for all $j \in 1..M$, $\broker$ added $\rp{i_j, g_j}$ to $batches\qp{r}.reductions$ and removed $\rp{i_j, s_j}$ from $batches\qp{r}.signatures$. This means in particular that, when $\qp{\texttt{Reduce}, r}$, $batches\qp{r}.signatures$ is empty at $\broker$. Upon ringing $\qp{\texttt{Reduce}, r}$ (line \ref{line:breducering}), $\broker$ builds the integer partition $\tilde i: \domains \rightarrowtail \powerset[<\infty]{\mathbb{N}}$ defined by
\begin{equation*}
    \rp{\rp{d, k} \in \tilde i} \Longleftrightarrow \rp{\rp{d, k} \in i}
\end{equation*}
(line \ref{line:bcompressids}). We have $\abs{\tilde i} = \abs{i} = M$. Moreover, by the density of \dir, we have $\max \tilde i \leq \clientcount - 1$. Recalling that $M \rightarrow \infty$ and $\log\rp{\log\rp{\clientcount}} / M \rightarrow \infty$, by Lemmas \ref{lemma:partitionreadinvertspartitionrepr} and \ref{lemma:partitionreprisefficient}, $\tilde i$ can be represented by a string of bits $\hat i$ such that 
\begin{equation*}
    \abs{\hat i} = M \ceil{\log_2\rp{\clientcount}} + o\rp{M}
\end{equation*}
To each server, $\broker$ sends a 
\begin{equation*}
    x^\rp{b} = \qp{\texttt{Batch}, \tilde i, p}
\end{equation*}
message (line \ref{line:bsendbacth}). We have
\begin{align*}
    \abs{x^\rp{b}} &= M \ceil{\log_2\rp{\clientcount}} + \sum_{j = 1}^M \abs{p_j} + o\rp{M} = \\
    &= \sum_{j = 1}^M \rp{\ceil{\log_2\rp{\clientcount}} + \abs{p_j} + o\rp{1}}
\end{align*}
Consequently, for all $j \in 1..M$, $x^\rp{b}$'s amortized size for $p_j$ is
\begin{equation*}
    \abs{x^\rp{b}}_j = \rp{\ceil{\log_2\rp{\clientcount}} + \abs{p_j}}
\end{equation*}
Having sent $x^\rp{b}$ to all servers, $\broker$ updates $batches\qp{r}$ to $\texttt{Witnessing}$ (line \ref{line:bbatchinwitnessing}).

\medskip

Let $\server$ be a server. Upon receiving $\qp{\texttt{Batch}, \tilde i, p}$ (line \ref{line:sbatchrecive}), $\server$ observes to know all ids in $i_1, \ldots, i_M$ (line \ref{line:bcomputeunknowns}), computes $r$ (line \ref{line:scomputeroot}) then sends back to $\broker$ a
\begin{equation*}
    x^\rp{ba} = \qp{\texttt{BatchAcquired}, r, \emptyset}
\end{equation*}
message (line \ref{line:sbatchreceivesendresponse}). Noting that 
\begin{equation*}
    \abs{x^\rp{ba}} = O\rp{1} = \sum_{j = 1}^M o\rp{1}
\end{equation*}
for all $j \in 1..M$, $x^\rp{ba}$'s amortized size for $p_j$ is
\begin{equation*}
    \abs{x^\rp{ba}}_j = 0
\end{equation*}

\medskip

Upon receiving $\server$'s $\qp{\texttt{BatchAcquired}, r, \emptyset}$ (line \ref{line:bdeliverbatchacquired}), $\broker$ exports no assignment (line \ref{line:bbatchacquiredexportassignments}), aggregates all $g_1, \ldots, g_M$ into a single $g$ (line \ref{line:bbatchacquiredaggregate}), observes $batches\qp{r}.signatures$ to be empty, then sends back to $\server$ a
\begin{equation*}
    x^\rp{s} = \qp{\texttt{Signatures}, r, \emptyset, g, \emptyset}
\end{equation*}
message (line \ref{line:bsendsignatures}). Noting that $\abs{x^\rp{s}} = O\rp{1}$, $x^\rp{s}$'s amortized size for $p_j$ is
\begin{equation*}
    \abs{x^\rp{s}}_j = 0
\end{equation*}

\medskip

Upon receiving $\qp{\texttt{Signatures}, r, \emptyset, g, \emptyset}$ (line \ref{line:sdeliversignatures}), $\server$ verifies only $g$ (line \ref{line:sdeterminemultisigners}, the loop at line \ref{line:ssignsturesloop} does no iterations). Noting that $\server$ performs signature verifications only upon receiving a $\texttt{Signatures}$ message, and noting that
\begin{equation*}
    1 = \sum_1^M o\rp{1}
\end{equation*}
we can immediately derive
\begin{equation*}
    \psi_j = 0
\end{equation*}
Next, $\server$ produces a multi-signature $w_\server$ for $\qp{\texttt{Witness}, r}$ (line \ref{line:ssignwitness}), then sends back to $\broker$ a
\begin{equation*}
    x^\rp{ws}_\server = \qp{\texttt{WitnessShard}, r, w_\server}
\end{equation*}
message (line \ref{line:shandlesignaturessendresponse}). Noting that $\abs{x^\rp{ws}_\server} = O(1)$, $x^\rp{ws}_\server$'s amortized size for $p_j$ is
\begin{equation*}
    \abs{x^\rp{ws}_\server}_j = 0
\end{equation*}

\medskip

Upon receiving $f + 1$ \texttt{WitnessShard} messages (lines \ref{line:bdeliverwitnessshard}, \ref{line:bstorewitnessshard}, \ref{line:binwitnessing}), $\broker$ aggregates all witnesses into a certificate $w$ (line \ref{line:baggregatewitnesses}), then sends a 
\begin{equation*}
    x^\rp{w} = \qp{\texttt{Witness}, r, w}
\end{equation*}
message to all servers (line \ref{line:bbroadcastwitnesses}). Noting that $\abs{x^\rp{w}} = O\rp{1}$, $x^\rp{w}$'s amortized size for $p_j$ is
\begin{equation*}
    \abs{x^\rp{w}}_j = 0
\end{equation*}

\medskip

Upon receiving $\qp{\texttt{Witness}, r, w}$ (line \ref{line:sdeliverwitness}), $\server$ observes that, for all $j$, $p_j$ is not equivocated (lines \ref{line:scheckidsummision} and \ref{line:sconflictcheck}) (because all clients are correct, no client equivocates), produces a multi-signature $c_\server$ for $\qp{\texttt{Commit}, r, \emptyset}$ (line \ref{line:ssigncommit}), then sends a
\begin{equation*}
    x^\rp{cs}_\server = \qp{\texttt{CommitShard}, r, \emptyset, c_\server}
\end{equation*}
message (line \ref{line:striggertobroker}). Noting that $\abs{x^\rp{cs}_\server} = O\rp{1}$, $x^\rp{cs}_\server$'s amortized size for $p_j$ is
\begin{equation*}
    \abs{x^\rp{cs}_\server}_j = 0
\end{equation*}

\medskip

Upon receiving $2f + 1$ \texttt{CommitShard} messages (lines \ref{line:bdelivercommitshard}, \ref{line:bstorecommit} and \ref{line:bpluralitycommitingcheck}), $\broker$ builds a map $h: \powerset{\ids} \rightarrowtail \multisignatures$ with only the key $\emptyset$ (lines \ref{line:blooponthecommits}, \ref{line:bpatchesupdate}, \ref{line:patchesaggregation}) (we recall that every server produced an empty exception set for $r$). $\broker$ then sends a
\begin{equation*}
    x^\rp{c} = \qp{\texttt{Commit}, r, h}
\end{equation*}
message to all servers (line \ref{line:bsendtocommittoserver}). Noting that $|x^\rp{c}| = O\rp{1}$, $x^\rp{c}$'s amortized size for $p_j$ is
\begin{equation*}
    \abs{x^\rp{c}}_j = 0
\end{equation*}

\medskip

Upon delivering $\qp{\texttt{Commit}, r, h}$ (line \ref{line:sdelivercommitpatches}), $\server$ delivers $p_1, \ldots, p_M$ (line \ref{line:sdeliver}), produces a multi-signature $z_\server$ for $\qp{\texttt{Completion}, r, \emptyset}$ (line \ref{line:scompletionsign}), then sends a
\begin{equation*}
    x^\rp{zs}_\server = \qp{\texttt{CompletionShard}, r, z_\server}
\end{equation*}
message to $\broker$ (line \ref{line:ssendcompletionshard}). Noting that $\abs{x^\rp{zs}_\server} = O\rp{1}$, $x^\rp{zs}_\server$'s amortized size for $p_j$ is
\begin{equation*}
    \abs{x^\rp{zs}_\server}_j = 0
\end{equation*}
Finally, $\server$ sets an $\qp{\texttt{OfferTotality}}$ timer to ring after $7$ time units (\ref{line:soffertotality}). Summarizing the scheduling of messages we have that: every server delivers the $\texttt{Batch}$ message between time $t_r$ and $t_r + 1$; $\broker$ delivers all $\texttt{BatchAcquired}$ messages between time $t_r$ and $t_r + 2$; every server delivers a $\texttt{Signatures}$ message between time $t_r$ and $t_r + 3$; $\broker$ delivers all $\texttt{WitnessShard}$ messages between time $t_r$ and $t_r + 4$; every server delivers the $\texttt{Witness}$ message between time $t_r$ and $t_r + 5$; $\broker$ delivers all $\texttt{CommitShard}$ messages between time $t_r$ and $t_r + 6$; all servers deliver the $\texttt{Commit}$ message between time $t_r$ and $t_r + 7$. As a result, $\qp{\texttt{OfferTotality}, r}$ rings at $\server$ after all servers delivered $p_1, \ldots, p_M$. Upon ringing $\qp{\texttt{OfferTotality}, r}$ (line \ref{line:soffertotalityring}), $\server$ sends a
\begin{equation*}
    x^\rp{ot} = \qp{\texttt{OfferTotality}, r}
\end{equation*}
message to all servers (line \ref{line:ssendoffertotality}). 

\medskip

Let $\server'$ be a server. Upon receiving $\server$'s $\qp{\texttt{OfferTotality}, r}$ message, $\server'$ verifies to have already delivered a batch with root $r$ and no exclusions (line \ref{line:scommitsdelivercheck}) and ignores the message.

\medskip

In summary, noting that $\server$ exchanges $2\servercount$ $\texttt{OfferTotality}$ messages ($N$ outgoing, $N$ incoming), $\server$'s amortized bit complexity for $p_j$ is
\begin{equation*}
    \psi^\server_j = \abs{x^\rp{b}}_j + \abs{x^\rp{ba}}_j + \abs{x^\rp{s}}_j + \abs{x^\rp{ws}_\server}_j + \abs{x^\rp{w}}_j + \abs{x^\rp{cs}_\server}_j + \abs{x^\rp{c}}_j + \abs{x^\rp{zs}_\server}_j + 2\servercount \abs{x^\rp{ot}}_j
\end{equation*}
from which immediately follows
\begin{equation*}
    \psi^\server_j = \rp{\ceil{\log_2\rp{\clientcount}} + \abs{p_j}}
\end{equation*}
and
\begin{equation*}
    \psi_j = \max_{\server} \psi^\server_j = \rp{\ceil{\log_2\rp{\clientcount}} + \abs{p_j}}
\end{equation*}
which concludes the theorem.
\end{proof}
\end{theorem}
\section{\diral: Full analysis}
\label{appendix:diral}

\subsection{Interface}
\label{subsection:diral.interface}

\begin{notation}[Signatures and multi-signatures]
We use $\signatures$ and $\multisignatures$ to respectively denote the set of signatures and multisignatures.
\end{notation}

\begin{definition}[Id]
An \textbf{id} is an element of $\rp{\ids = \domains \times \mathbb{N}}$, where $\domains$ is a finite set of domains. Let $i = \rp{d, n}$ be an id, we call $d$ and $n$ the \textbf{domain} and \textbf{index} of $i$, respectively.
\end{definition}

The \textbf{\dir} interface (instance $\dirin$) exposes the following procedures and events:
\begin{itemize}
    \item \textbf{Request} $\event{\dirin}{Signup}{}$: requests that an id be assigned to the local process.
    \item \textbf{Indication} $\event{\dirin}{SignupComplete}$: indicates that an id was successfully assigned to the local process.
    \item \textbf{Getter} $\dirin[id]$: returns the process associated with id $id$, if known. Otherwise, returns $\bot$.
    \item \textbf{Getter} $\dirin[process]$: returns the id associated with process $process$, if known. Otherwise, returns $\bot$.
    \item \textbf{Getter} $\dirin.export(id)$: returns the assignment for id $id$, if known. Otherwise, returns $\bot$.
    \item \textbf{Setter} $\dirin.import(assignment)$: imports assignment $assignment$.
\end{itemize}

\begin{notation}[Bijective relations]
Let $X$, $Y$ be sets. We use $\rp{X \leftrightarrow Y}$ to denote the set of \textbf{bijective relations} between $X$ and $Y$.
\end{notation}

\begin{notation}[Tuple binding]
Let $t = \rp{t_1, \ldots, t_n}$ be a tuple. When binding $t$, we use the \textbf{any} symbol $\any$ to mark which elements of $t$ are discarded from the binding. For example,
\begin{equation*}
    \rp{x, \any, \ldots, \any, y, \any} = t
\end{equation*}
binds $x = t_1$, and $y = t_{n - 1}$. We use the \textbf{tail} symbol $\tail$ to indicate that all subsequent elements of $t$ are discarded from the binding. For example,
\begin{equation*}
    \rp{x', y', \tail} = t
\end{equation*}
binds $x' = t_1$ and $y' = t_2$, regardless of $n$. Let $X$ be a set. We use the tuple binding notation to \textbf{filter} the elements $X$. For example, we use $\cp{\rp{x'', \any, y'', \tail}}$ to identify the set
\begin{equation*}
    \cp{t \in X \suchthat \abs{t} \geq 3, \rp{x'', \any, y'', \tail} = t}
\end{equation*}
of tuples in $X$ whose first element is $x''$ and whose third element is $y''$.
\end{notation}

\begin{definition}[Directory record]
\label{definition:directoryrecord}
Let $\directoryrecordsym: \rp{\correctprocesses \times \mathbb{R}} \rightarrow \rp{\ids \leftrightarrow \processes}$. $\directoryrecordsym$ is a \textbf{directory record} if and only if:
\begin{itemize}
    \item For all $\process \in \correctprocesses$, $\directoryrecord{\process}[t]$ is non-decreasing in $t$.
    \item For all $\process, \process' \in \correctprocesses$, $t, t' \in \mathbb{R}$, $\directoryrecord{\process}[t] \cup \directoryrecord{\process'}[t']$ is a a bijective relation.
\end{itemize}
Let $\process$ be a correct process, let $t$ be a time, let $i$ be an id, let $\processsecondary$ be a process. $\process$ \textbf{associates} $i$ and $\processsecondary$ by time $t$ if and only if $\rp{i, \processsecondary} \in \directoryrecord{\process}[t]$. $\process$ \textbf{knows} $i$ (resp., $\processsecondary$) by time $t$ if and only if $\rp{i, \any} \in \directoryrecord{\process}[t]$ (resp., $\rp{\any, \processsecondary} \in \directoryrecord{\process}[t]$).
\end{definition}

\begin{notation}[Directory record]
Let $\process$ be a correct process. Wherever it can be unequivocally inferred from context, we omit the time from the directory record $\directoryrecord{\process}$.
\end{notation}

A \dir satisfies the following properties:
\begin{itemize}
    \item \textbf{Correctness}: Some directory record $\directoryrecordsym$ exists such that, for any id $i$ and process $\processsecondary$, if a process $\process$ invokes $dir[i]$ (resp., $dir[\processsecondary]$), $\process$ obtains $\processsecondary$ (resp., $i$) if and only if $\process$ associates $i$ and $\processsecondary$.
    \item \textbf{Signup Integrity}: A correct process never triggers $\event{\dirin}{SignupComplete}$ before triggering $\event{\dirin}{Signup}$.
    \item \textbf{Signup Validity}: If a correct process triggers $\event{\dirin}{Signup}$, it eventually triggers $\event{\dirin}{SignupComplete}$.
    \item \textbf{Self-knowledge}: Upon triggering $\event{\dirin}{SignupComplete}$, a correct process knows itself.
    \item \textbf{Transferability}: If a correct process invokes $\dirin.export(i)$ to obtain an assignment $a$, then any correct process knows $i$ upon invoking $\dirin.import(a)$.
    \item \textbf{Density}: For every correct process $\process$, for every time $t$, for every $\rp{\rp{\any, n}, \any} \in \directoryrecord{\process}[t]$, we have $n < \abs{\processes}$.
\end{itemize}

\begin{notation}[Directory mapping]
\label{notation:directorymapping}
Let $i$ be an id, let $\processsecondary$ be a process such that, for some $\process \in \correctprocesses$, $t \in \mathbb{R}$, we have $\rp{i, \processsecondary} \in \directoryrecord{\process}[t]$. We say that $i$ and $\processsecondary$ are \textbf{known}, and we use
\begin{equation*}
    i = \directorymap{\processsecondary} \;\;\;\; \processsecondary = \directorymap{i}
\end{equation*}
\end{notation}

We underline the soundness of Notation \ref{notation:directorymapping}: by Definition \ref{definition:directoryrecord}, if $\rp{i, \processsecondary} \in \directoryrecord{\process}[t]$, then no $i' \neq i$ or $\processsecondary' \neq \processsecondary$ exist such that, for some $\process'$, $t'$, we have $\rp{i, \processsecondary'} \in \directoryrecord{\process'}[t']$ or $\rp{i', \processsecondary} \in \directoryrecord{\process'}[t']$. In other words, no $i' \neq i$ or $\processsecondary' \neq \processsecondary$ exist such that $i' = \directorymap{\processsecondary}$ or $\processsecondary' = \directorymap{i}$.
\subsection{Pseudocode}
\label{subsection:diral.pseudocode}

The full code of our \dir implementation \diral can be found can be found in Appendix \ref{appendix:pseudocode}. More in detail, Section \ref{subsection:pseudocode.diral} implements \csbal, Section \ref{subsection:pseudocode.diralserver} implements \csbal's server.
\subsection{Correctness}
\label{subsection:diral.correctness}

In this section, we prove to the fullest extent of formal detail that \diral implements a \dir.

\subsubsection{Correctness}

In this section, we prove that \diral satisfies correctness.

\begin{lemma}
\label{lemma:rankingsaredistinct}
Let $\server$ be a correct server, let $\hat \server$ be a server. All elements of $ranking\qp{\hat \server}$ at $\server$ are distinct.
\begin{proof}
Upon initialization, $rankings\qp{\hat \server}$ is empty at $\server$ (line \ref{line:dsrankingsinit}). Moreover, $\server$ appends $\process$ to $rankings\qp{\hat \server}$ only by executing line \ref{line:dsrankingsupdate}. Because $\server$ does so only if $\process \notin rankings\qp{\hat \server}$ (line \ref{line:dsprocessnotinranking}), the lemma is proved.
\end{proof}
\end{lemma}

\begin{lemma}
\label{lemma:rankingsmatch}
Let $\server$, $\server'$ be correct servers, let $\hat \server$ be a server, let $n \in \mathbb{N}$, let $\process$ be a process. If $rankings\qp{\hat \server}\qp{n} = \process$ at $\server$, then eventually $rankings\qp{\hat \server}\qp{n} = \process$ at $\server'$ as well.
\begin{proof}
Let $\qp{\texttt{Rank}, \process_1}, \ldots, \qp{\texttt{Rank}, \process_R}$ denote the sequence of $\qp{\texttt{Rank}, \any}$ messages $\server$ FIFO-delivered from $\hat \server$. By the totality and FIFO properties of FIFO broadcast, $\server'$ eventually delivers $\qp{\texttt{Rank}, \process_1}, \ldots, \qp{\texttt{Rank}, \process_R}$ as well. Upon initialization, $rankings\qp{\hat \server}$ is empty at both $\server$ and $\server'$ (line \ref{line:dsrankingsinit}). Moreover, $\server$ (resp., $\server'$) adds some process $\hat \process$ to $rankings\qp{\hat \server}$ (line \ref{line:dsrankingsupdate} only) only upon FIFO-delivering a $\qp{\texttt{Rank}, \hat \process}$ message from $\hat \server$ (line \ref{line:dsdeliverrankmessage}), and only if $rankings\qp{\hat \server}$ satisfies a deterministic condition on $\hat \process$ (line \ref{line:dsprocessnotinranking}). As a result, noting $\server$ (resp., $\server'$) never removes elements from $rankings\qp{\hat \server}$, if $rankings\qp{\hat \server}\qp{n} = \process$ at $\server$ as a result of $\server$ delivering $\qp{\texttt{Rank}, \process_1}, \ldots, \qp{\texttt{Rank}, \process_R}$, then eventually $rankings\qp{\hat \server}\qp{n} = \process$ at $\server'$ as well, as it also eventually delivers $\qp{\texttt{Rank}, \process_1}, \ldots, \qp{\texttt{Rank}, \process_R}$.
\end{proof}
\end{lemma}

\begin{lemma}
\label{lemma:nocorrectequivocatesprocesses}
Let $\server$, $\server'$ be correct servers. Let $\hat \server$ be a server, let $n \in \mathbb{N}$, let $\process$, $\process'$ be processes such that $\server$ and $\server'$ respectively sign $\qp{\texttt{Assignment}, \rp{\hat \server, n}, \process}$ and $\qp{\texttt{Assignment}, \rp{\hat \server, n}, \process'}$. We have $\process = \process'$.
\begin{proof}
$\server$ (resp., $\server'$) signs $\qp{\texttt{Assignment}, \rp{\hat \server, n}, \process}$ (resp., $\qp{\texttt{Assignment}, \rp{\hat \server, n}, \process'}$) only by executing line \ref{line:dsmultisihnassignment}. $\server$ (resp., $\server'$) does so only if $n$ is the index of $\process$ (resp., $\process'$) in $rankings\qp{\hat \server}$ at $\server$ (resp., $\server'$) (line \ref{line:dsextractindex}). We underline that, by Lemma \ref{lemma:rankingsaredistinct}, at most one element of $rankings\qp{\hat \server}$ is $\process$ (resp., $\process'$) at $\server$ (resp., $\server'$), hence $n$ is well-defined. By Lemma \ref{lemma:rankingsmatch}, however, if $rankings\qp{\hat \server}\qp{n} = \process$ at $\server$, then $rankings\qp{\hat \server}\qp{n} = \process$ at $\server'$ as well. This proves $\process = \process'$ and concludes the lemma.
\end{proof}
\end{lemma}

\begin{lemma}
\label{lemma:noprocessequivocation}
Let $\hat \server$ be a server, let $n \in \mathbb{N}$, let $\process$, $\process' \neq \process$ be processes. If a quorum certificate for $\qp{\texttt{Assignment}, \rp{\hat \server, n}, \process}$ exists, then no quorum certificate for $\qp{\texttt{Assignment}, \rp{\hat \server, n}, \process'}$ exists. 
\begin{proof}
Let us assume by contradiction that quorum certificates exist for both $\qp{\texttt{Assignment}, \rp{\hat \server, n}, \process}$ and $\qp{\texttt{Assignment}, \rp{\hat \server, n}, \process'}$. Noting that at most $f$ servers are Byzantine, at least one correct server $\server$ (resp., $\server'$) signed $\qp{\texttt{Assignment}, \rp{\hat \server, n}, \process}$ (resp., $\qp{\texttt{Assignment}, \rp{\hat \server, n}, \process'}$). By Lemma \ref{lemma:nocorrectequivocatesprocesses} we then have $\process = \process'$, which contradicts $\process \neq \process'$ and proves the lemma.
\end{proof}
\end{lemma}

\begin{lemma}
\label{lemma:nocorrectequivocatesindices}
Let $\server$, $\server'$ be correct servers. Let $\hat \server$ be a server, let $n, n' \in \mathbb{N}$, let $\process$ be a process such that $\server$ and $\server'$ respectively sign $\qp{\texttt{Assignment}, \rp{\hat \server, n}, \process}$ and $\qp{\texttt{Assignment}, \rp{\hat \server, n'}, \process}$, We have $n = n'$.
\begin{proof}
$\server$ (resp., $\server'$) signs $\qp{\texttt{Assignment}, \rp{\hat \server, n}, \process}$ (resp., $\qp{\texttt{Assignment}, \rp{\hat \server, n'}, \process}$) only by executing line \ref{line:dsmultisihnassignment}. $\server$ (resp., $\server'$) does so only if $n$ (resp., $n'$) is the index of $\process$ in $rankings\qp{\hat \server}$ at $\server$ (resp., $\server'$) (line \ref{line:dsextractindex}). We underline that, by Lemma \ref{lemma:rankingsaredistinct}, at most one element of $rankings\qp{\hat \server}$ is $\process$ at $\server$ (resp., $\server'$), hence $n$ (resp., $n'$) is well-defined. By Lemma \ref{lemma:rankingsmatch}, however, if $rankings\qp{\hat \server}\qp{n} = \process$ at $\server$, then $rankings\qp{\hat \server}\qp{n} = \process$ at $\server'$ as well. Again by Lemma \ref{lemma:rankingsaredistinct}, this proves $n = n'$ and concludes the lemma.
\end{proof}
\end{lemma}

\begin{lemma}
\label{lemma:nocorrectequivocatesassigner}
Let $\server$ be a correct server. Let $\hat \server$, $\hat \server'$ be servers, let $\process$ be a process such that $\server$ signs both an $\qp{\texttt{Assignment}, \rp{\hat \server, \any}, \process}$ and an $\qp{\texttt{Assignment}, \rp{\hat \server', \any}, \process}$ message. We have $\hat \server = \hat \server'$.
\begin{proof}
$\server$ signs an $\qp{\texttt{Assignment}, \rp{\hat \server, \any}, \process}$ (resp., $\qp{\texttt{Assignment}, \rp{\hat \server', \any}, \process}$) message only by executing \ref{line:dsmultisihnassignment}. $\server$ does so only if $\hat \server = assigners\qp{\process}$ (resp., $\hat \server' = assigners\qp{\process}$) (line \ref{line:dsassignersextraxt}). The lemma follows immediately from the observation that $\server$ sets $assigners\qp{\process}$ (line \ref{line:dsaaigneres} only) at most once (line \ref{line:dsprocessnotinassigners}). 
\end{proof}
\end{lemma}

\begin{lemma}
\label{lemma:noidequivocation}
Let $\hat \server$, $\hat \server'$ be servers, let $n, n' \in \mathbb{N}$ such that $\rp{\hat \server, n} \neq \rp{\hat \server', n'}$, let $\process$ be a process. If a quorum certificate for $\qp{\texttt{Assignment}, \rp{\hat \server, n}, \process}$ exists, then no quorum certificate for $\qp{\texttt{Assignment}, \rp{\hat \server', n'}, \process}$ exists.
\begin{proof}
Let us assume that $\hat \server = \hat \server'$. Because $\rp{\hat \server, n} \neq \rp{\hat \server', n'}$, we have $n \neq n'$. Let us assume by contradiction that quorum certificates exist for both $\qp{\texttt{Assignment}, \rp{\hat \server, n}, \process}$ and $\qp{\texttt{Assignment}, \rp{\rp{\hat \server = \hat \server'}, n'}, \process}$. Noting that at most $f$ servers are Byzantine, at least one correct server $\server$ (resp. $\server'$) signed an $\qp{\texttt{Assignment}, \rp{\hat \server, n}, \process}$ (resp., $\qp{\texttt{Assignment}, \rp{\hat \server, n'}, \process}$) message. By Lemma \ref{lemma:nocorrectequivocatesindices} we then have $n = n'$, which contradicts $n \neq n'$.

\medskip

Let us assume that $\hat \server \neq \hat \server'$. We have that at least $f + 1$ correct processes signed $\qp{\texttt{Assignment}, \rp{\hat \server, \any}, \process}$ (resp., $\qp{\texttt{Assignment}, \rp{\hat \server', \any}, \process}$). Because any two sets of $f + 1$ correct processes intersect in at least one element, some correct server $\server^*$ exists that signed both $\qp{\texttt{Assignment}, \rp{\hat \server, \any}, \process}$ and $\qp{\texttt{Assignment}, \rp{\hat \server', \any}, \process}$. By Lemma \ref{lemma:nocorrectequivocatesassigner} we then have $\hat \server = \hat \server'$ which contradicts $\hat \server \neq \hat \server'$ and proves the lemma.
\end{proof}
\end{lemma}

\begin{definition}[Certifiable assignments]
\label{definition:certifiableassignments}
Let $i$ be an id, let $\process$ be a process. The set $\directorycert$ of \textbf{certifiable assignments} contains $\rp{i, \process}$ if and only if a quorum certificate ever exists for $\qp{\texttt{Assignment}, i, \process}$.
\end{definition}

\begin{lemma}
\label{lemma:certifiableassignmentsarebijective}
$\directorycert$ is a bijection.
\begin{proof}
It follows immediately from Definition \ref{definition:certifiableassignments} and Lemmas \ref{lemma:noprocessequivocation} and \ref{lemma:noidequivocation}.
\end{proof}
\end{lemma}

\begin{notation}[Directory record]
\label{notation:diraldirectoryrecord}
Let $\process$ be a correct process, let $t \in \mathbb{R}$. We use $\directoryrecord{\process}[t]$ to denote the value of $directory$ at $\process$ at time $t$.
\end{notation}

As we immediately prove, $\directoryrecordsym$ is a directory record, which makes Notation \ref{notation:diraldirectoryrecord} compatible with Definition \ref{definition:directoryrecord}.

\begin{lemma}
\label{lemma:directoryrecordgrows}
Let $\process$ be a correct process, let $t, t' \in \mathbb{R}$ such that $t' \geq t$. We have $\directoryrecord{\process}[t'] \supseteq \directoryrecord{\process}[t]$.
\begin{proof}
The lemma immediately follows from Notation \ref{notation:diraldirectoryrecord} and the observation that $\process$ never removes elements from $directory$.
\end{proof}
\end{lemma}

\begin{lemma}
\label{lemma:directoryrecordiscertifiable}
Let $\process$ be a correct process, let $t \in \mathbb{R}$. We have $\directoryrecord{\process}[t] \subseteq \directorycert$.
\begin{proof}
Upon initialization, $directory$ is empty at $\process$ (line \ref{line:ddirectoryinit}). Moreover, $\process$ adds $\rp{\hat i, \hat \process}$ only by executing line \ref{line:ddirectoryadd}. $\process$ does so only upon verifying a quorum certificate for $\qp{\texttt{Assignment}, \hat i, \hat \process}$ (line \ref{line:dverifyquorumcertificate}). The lemma immediately follows from Definition \ref{definition:certifiableassignments} and Notation \ref{notation:diraldirectoryrecord}.
\end{proof}
\end{lemma}

\begin{lemma}
\label{lemma:diraldirectoryrecordisdirectoryrecord}
$\directoryrecordsym$ is a directory record.
\begin{proof}
By Lemma \ref{lemma:directoryrecordgrows} we immediately have that, for every correct process $\process$, $\directoryrecord{\process}[t]$ is non-decreasing in $t$. Let $\process$, $\process'$ be correct processes, let $t, t' \in \mathbb{R}$. By Lemma \ref{lemma:directoryrecordiscertifiable} we have
\begin{equation*}
    \directoryrecord{\process}[t] \cup \directoryrecord{\process'}[t'] \subseteq \directorycert
\end{equation*}
By Lemma \ref{lemma:certifiableassignmentsarebijective}, this proves that $\directoryrecord{\process}[t] \cup \directoryrecord{\process'}[t']$ is bijective. The lemma immediately follows from Definition \ref{definition:directoryrecord}.
\end{proof}
\end{lemma}

\begin{theorem}
\diral satisfies correctness.
\begin{proof}
Let $\process$ be a correct process, let $\hat i$ be an id, let $\hat \process$ be a process. By Notation \ref{notation:diraldirectoryrecord}, upon invoking $\dirin\qp{\hat i}$ (resp., $\dirin\qp{\hat \process}$) (line \ref{line:dinvokeprocedure}), $\process$ obtains $\hat \process$ (resp., $\hat i$) (line \ref{line:dreturnkycard}, resp., line \ref{line:dreturnid}) if and only if $\rp{\hat i, \hat \process} \in \directoryrecord{\process}$ (line \ref{line:dcheckkeycardindirectory}, resp., line \ref{line:dcheckidindirectory}). The theorem follows immediately from Lemma \ref{lemma:diraldirectoryrecordisdirectoryrecord}.
\end{proof}
\end{theorem}

\subsubsection{Signup integrity}

In this section, we prove that \diral satisfies signup integrity.

\begin{theorem}
\diral satisfies signup integrity.
\begin{proof}
Let $\process$ be a correct process. Upon initialization, we have $status = \texttt{Outsider}$ at $\process$ (line \ref{line:dsignupoutsider}). Moreover, $\process$ triggers $\ap{\dirin.SignupComplete}$ only by executing line \ref{line:dtriggersignupcomplete}. $\process$ does so only if $status = \texttt{SigningUp}$ (line \ref{line:dchecksigningupstatuscheck}). The theorem follows immediately from the observation that $\process$ sets $status = \texttt{SigningUp}$ (line \ref{line:dstatussigninigup} only) only upon triggering $\ap{\dirin.Signup}$ (line \ref{line:dsignup}).
\end{proof}
\end{theorem}

\subsubsection{Signup validity}

In this section, we prove that \diral satisfies signup validity.

\begin{lemma}
\label{lemma:Signupeventuallyassigner}
Let $\process$ be a correct process that triggers $\ap{\dirin.Signup}$. We eventually have $assigner \neq \bot$ at $\process$.
\begin{proof}
Upon triggering $\ap{\dirin.Signup}$ (line \ref{line:dsignup}), $\process$ sends a $\qp{\texttt{Signup}}$ message to all servers (lines \ref{line:dserverinservers} and \ref{line:dsendsignup}). Let $\hat \server$ be a correct server. Noting that at most $f$ servers are Byzantine, $\hat \server$ is guaranteed to exist. 

\medskip

Upon delivering $\qp{\texttt{Signup}}$ from $\process$ (line \ref{line:dsdeliversignup}), $\hat \server$ FIFO-broadcasts a $\qp{\texttt{Rank}, \process}$ message (line \ref{line:dsbrodcastrank}). Let $\server_1, \ldots, \server_{f + 1}$ be distinct correct servers. Noting that at most $f$ servers are Byzantine, $\server_1, \ldots, \server_{f + 1}$ are guaranteed to exist. 

\medskip

Let $n \leq f + 1$. We start by noting that, upon initialization, $rankings\qp{\hat \server}$ is empty at $\server_n$ (line \ref{line:dsrankingsinit}). Moreover, $\server_n$ adds $\process$ to $rankings\qp{\hat \server}$ (line \ref{line:dsrankingsupdate}) only upon delivering a $\qp{\texttt{Rank}, \process}$ message from $\hat \server$ (line \ref{line:dsdeliverrankmessage}). By the validity and totality of FIFO broadcast, $\server_n$ is eventually guaranteed to FIFO-deliver $\qp{\texttt{Rank}, \process}$ from $\hat \server$. Upon doing so (line \ref{line:dsdeliverrankmessage}), $\server_n$ verifies that $\process \notin rankings\qp{\hat \server}$ (line \ref{line:dsrankingsupdate}), then sends a $\qp{\texttt{Ranked}, \hat \server}$ message to $\process$. 

\medskip

Upon delivering a $\qp{\texttt{Ranked}, \hat \server}$ message from $\server_n$ (line \ref{line:dserverdeliverranked}), $\process$ adds $\server_n$ to $rankings\qp{\hat \server}$ (line \ref{line:daddservertorankings}). In summary, recalling that the above holds true for any $n \leq f + 1$, $\process$ adds  $\server_1, \ldots, \server_{f + 1}$ to $rankings\qp{\hat \server}$. Noting that $\process$ never removes elements from $rankings\qp{\hat \server}$, that initially $assigner = \bot$ at $\process$ (line \ref{line:dassignerinit}), and that $\process$ updates $assigner$ only by executing line \ref{line:dassignerupdate}, $\process$ is eventually guaranteed to detect that, for some $\server^*$, $\abs{rankings\qp{\server^*}} \geq f + 1$ and $assigner = \bot$ (line \ref{line:dcheckrankingquorum}). Upon doing so, $\process$ sets $assigner = \server^*$, and the lemma is proved.
\end{proof}
\end{lemma}

\begin{lemma}
\label{lemma:assignerranksatallcorrectprocesses}
Let $\process$ be a correct process, let $\hat \server$ be a server such that $assigner = \hat \server$ at $\process$. We eventually have $\process \in rankings\qp{\hat \server}$ at all correct processes.
\begin{proof}
We start by noting that, upon initialization, we have $assigner = \bot$ at $\process$ (\ref{line:dassignerinit}). Moreover, $\process$ sets $assigner = \hat \server$ (line \ref{line:dassignerupdate} only) only if $\abs{rankings\qp{\hat \server}} \geq f + 1$ at $\process$ (line \ref{line:dcheckrankingquorum}). Upon initialization, $rankings\qp{\hat \server}$ is empty at $\process$ (line \ref{line:drankingsinit}). Moreover, $\process$ adds $\server$ to $rankings\qp{\hat \server}$ only upon delivering a $\qp{\texttt{Ranked}, \hat \server}$ message from $\server$. In summary, at least $f + 1$ servers sent a $\qp{\texttt{Ranked}, \hat \server}$ message to $\process$. Let $\server$ be a correct server that sent a $\qp{\texttt{Ranked}, \hat \server}$ message to $\process$. Noting that at most $f$ servers are Byzantine, $\server$ is guaranteed to exist.

\medskip

$\server$ sends a $\qp{\texttt{Ranked}, \hat \server}$ to $\process$ (line \ref{line:dssendranked} only) only upon FIFO-delivering a $\qp{\texttt{Rank}, \process}$ message from $\hat \server$. By the totality of FIFO broadcast, eventually every correct process delivers $\qp{\texttt{Rank}, \process}$. Let $\server'$ be a correct process. Upon FIFO-delivering $\qp{\texttt{Rank}, \process}$ from $\hat \server$ (line \ref{line:dsdeliverrankmessage}), $\server'$ either already satisfies $\process \in rankings\qp{\hat \server}$ (line \ref{line:dsprocessnotinranking}) or adds $\process$ to $rankings\qp{\hat \server}$. Noting that no correct process ever removes elements from $rankings\qp{\hat \server}$, eventually $\process \in rankings\qp{\hat \server}$ at all correct processes, and the lemma is proved.
\end{proof}
\end{lemma}

\begin{lemma}
\label{lemma:assigmentsreachquorum}
Let $\process$ be a correct process, let $\hat \server$ be a server such that $assigner = \hat \server$ at $\process$, let $r \in \mathbb{N}$ such that, eventually, $rankings\qp{\hat \server}\qp{r} = \process$ at all correct servers. We eventually have $\abs{assignments\qp{r}} \geq 2f + 1$ at $\process$.
\begin{proof}
We start by noting that, upon initialization, we have $assigner = \bot$ at $\process$ (line \ref{line:dassignerinit}). Moreover, $\process$ updates $assigner$ to $\hat \server$ only by executing line \ref{line:dassignerupdate}. Upon doing so, $\process$ sends a $\qp{\texttt{Assigner}, \hat \server}$ to all servers (lines \ref{line:dassignerloop} and \ref{line:dassignersend}). Moreover, because $\process$ updates $assigner$ to a non-$\bot$ value (line \ref{line:dassignerupdate}) immediately before disseminating $\qp{\texttt{Assigner}, \hat \server}$, and $\process$ never resets $assigner$ back to $\bot$, $\process$ never issues any $\qp{\texttt{Assigner}, \rp{\tilde \server \neq \hat \server}}$ message (see line \ref{line:dcheckrankingquorum}). Let $\server_1, \ldots, \server_{2f + 1}$ be distinct correct servers. Noting that at most $f$ servers are Byzantine, $\server_1, \ldots, \server_{2f + 1}$ are guaranteed to exist.

\medskip

Let $n \leq 2f + 1$. Upon initialization, $assigners$ is empty at $\server_n$ (line \ref{line:dsassignersinit}). Moreover, $\server_n$ adds $\process$ to $assigners$ (line \ref{line:dsaaigneres} only) only upon delivering a $\qp{\texttt{Assigner}, \any}$ message from $\process$ (line \ref{line:dsdeliverassigner}). As a result, upon delivering $\qp{\texttt{Assigner}, \hat \server}$ from $\process$ (line \ref{line:dsdeliverassigner}), $\server_n$ verifies $\process \notin assigners$ (line \ref{line:dsprocessnotinassigners}) and sets $assigners\qp{\process} = \hat \server$ (line \ref{line:dsaaigneres}). Upon initialization, $certified$ is empty at $\server_n$ (line \ref{line:dscertifiedinit}). Moreover, $\server_n$ adds $\process$ to $certified$ only by executing line \ref{line:dsaddtocertified}. As a result, $\server_n$ is eventually guaranteed to observe $assigners\qp{\process} = \hat \server$, $rankings\qp{\hat \server} = r$, and $\process \notin certified$ (line \ref{line:dsreadytocertify}). Upon doing so, $\server_n$ produces a signature $s_n$ for $\qp{\texttt{Assignment}, \rp{\hat \server, r}, \process}$ (lines \ref{line:dsassignersextraxt}, \ref{line:dsextractindex} and \ref{line:dsmultisihnassignment}) and sends an $\qp{\texttt{Assignment}, r, s_n}$ message back to $\process$ (line \ref{line:dssendassignment}).

\medskip

Upon delivering $\qp{\texttt{Assignment}, r, s_n}$ from $\server_n$ (line \ref{line:ddeliverassignment}), $\process$ verifies $s_n$ against $\qp{\texttt{Assignment}, \rp{\hat \server, r}, \process}$ (line \ref{line:dverifyassignmentsignature}), then adds $\server_n$ to $assignments\qp{r}$. In summary, recalling that the above holds true for all $n \leq 2f + 1$, $assignments\qp{r}$ eventually contains $\server_1, \ldots, \server_{2f + 1}$ at $\process$, and the lemma is proved.

\end{proof}
\end{lemma}

\begin{theorem}
\diral satisfies signup validity.
\begin{proof}
Let $\process$ be a correct process that triggers $\event{\dirin}{Signup}$. By Lemma \ref{lemma:Signupeventuallyassigner}, for some $\hat \server$ we eventually have $assigner = \hat \server$ at $\process$. By Lemmas \ref{lemma:assignerranksatallcorrectprocesses} and \ref{lemma:rankingsmatch}, some $r \in \mathbb{N}$ exists such that, eventually, $rankings\qp{\hat \server}\qp{r} = \process$ at all correct processes. By Lemma \ref{lemma:assigmentsreachquorum}, we eventually have $\abs{assignments\qp{r}} \geq 2f + 1$ at $\process$. Upon triggering $\ap{\dirin.Signup}$ (line \ref{line:dsignup}), $\process$ sets $status = \texttt{SigningUp}$ (line \ref{line:dstatussigninigup}). Moreover, if $status = \texttt{SigningUp}$, $\process$ updates $status$ only by executing line \ref{line:dstatussignedup}. As a result, $\process$ is eventually guaranteed to detect that $\abs{assignments\qp{r}} \geq 2f + 1$ and $status = \texttt{SigningUp}$ (line \ref{line:dchecksigningupstatuscheck}). Upon doing so, $\process$ triggers $\event{\dirin}{SignupComplete}$ (line \ref{line:dtriggersignupcomplete}), and the theorem is proved.
\end{proof}
\end{theorem}

\subsubsection{Self-knowledge}

In this section, we prove that \diral satisfies self-knowledge.

\begin{lemma}
\label{lemma:assignmentsarecorrectlysigned}
Let $\process$ be a correct process. Let $\hat \server$ be a server such that $assigner = \hat \server$ at $\process$. Let $r \in \mathbb{N}$, let $\server$ be a server such that, for some $s$, we have $assignments\qp{r}\qp{\server} = s$ at $\process$. We have that $s$ is $\server$'s signature for $\qp{\texttt{Assignment}, \rp{\hat \server, r}, \process}$.
\begin{proof}
We start by noting that $\process$ updates $assigner$ (line \ref{line:dassignerupdate} only) only if $assigner = \bot$ (line \ref{line:ddeliverassignment}). Consequently, at all times we either have $assigner = \bot$ or $assigner = \hat \server$ at $\process$. Upon initialization, $assignments$ is empty at $\process$ (line \ref{line:dassigmentsinit}). Moreover, $\process$ sets $assignments\qp{r}\qp{\server} = s$ only by executing line \ref{line:dassigmentsignaturedefinition}. $\process$ does so only if $s$ is $\server$'s multisignature for $\qp{\texttt{Assignment}, \rp{\hat \server, r}, \process}$ (line \ref{line:dverifyassignmentsignature}).
\end{proof}
\end{lemma}

\begin{theorem}
\diral satisfies self-knowledge.
\begin{proof}
Let $\process$ be a correct process that triggers $\ap{\dirin.SignupComplete}$. $\process$ does so (line \ref{line:dtriggersignupcomplete} only) only if, for some $r$, we have $\abs{assignments\qp{r}} \geq 2f + 1$ at $\process$ (line \ref{line:dchecksigningupstatuscheck}). Let $\hat \server$ denote the value of $assigner$ at $\process$, let $i = \rp{\hat \server, r}$. Immediately before triggering $\event{\dirin}{SignupComplete}$, $\process$ aggregates $assignments\qp{r}$ into a certificate $t$ (line \ref{line:ddefinecertificate}). By Lemma \ref{lemma:assignmentsarecorrectlysigned}, $t$ is a quorum certificate for $\qp{\texttt{Assignment}, i, \process}$. $\process$ then invokes
\begin{equation*}
    \dirin.import\rp{Assignment \cp{id: i, process: \process, certificate: t}}
\end{equation*}
(line \ref{line:ddirinimport}). Upon doing so, $\process$ verifies that $t$ is indeed a certificate for $\qp{\texttt{Assignment}, i, \process}$ (line \ref{line:dverifyquorumcertificate}), and adds $\rp{i, \process}$ to $directory$ (line \ref{line:ddirectoryadd}). By Notation \ref{notation:diraldirectoryrecord} and Definition \ref{definition:directoryrecord}, $\process$ knows itself, and the theorem is proved.
\end{proof}
\end{theorem}

\subsubsection{Transferability}

In this section, we prove that \diral satisfies transferability.

\begin{lemma}
\label{lemma:directoryiscertified}
Let $\process$ be a correct process, let $i$ be an id, be a process such that $\rp{i, \any} \in directory$ at $\process$. We have $i \in certificates$ at $\process$.
\begin{proof}
Upon initialization, $directory$ is empty at $\process$ (line \ref{line:ddirectoryinit}). Moreover, $\process$ adds $\rp{i, \any}$ to $directory$ only by executing line \ref{line:ddirectoryadd}. Immediately after doing so, $\process$ adds $i$ to $certificates$ (line \ref{line:ddaddidtocerticates}).
\end{proof}
\end{lemma}

\begin{lemma}
\label{lemma:directorycertificatesarecorrect}
Let $\process$ be a correct process, let $\rp{\hat i, \hat \process} \in directory$ at $\process$. Let $\hat c = certificates\qp{\hat i}$ at $\process$. We have that $\hat c$ is a quorum certificate for $\qp{\texttt{Assignment}, \hat i, \hat \process}$.
\begin{proof}
We underline that, by Lemma \ref{lemma:directoryiscertified}, $\hat c$ is guaranteed to exist. By Notation \ref{notation:diraldirectoryrecord} and Lemmas \ref{lemma:directoryrecordiscertifiable} and \ref{lemma:certifiableassignmentsarebijective}, for all $\hat \process' \neq \hat \process$ we always have $\rp{\hat i, \hat \process'} \notin directory$ at $\process$. Upon initialization, $certificates$ is empty at $\process$ (line \ref{line:ddcertificatempty}). Moreover, $\process$ adds $\rp{\hat i, \hat c}$ to $certificates$ only by executing line \ref{line:ddaddidtocerticates}. $\process$ does so only if $\hat c$ is a quorum certificate for $\qp{\texttt{Assignment}, \hat i, \tilde \process}$ for some $\tilde \process$ (line \ref{line:dverifyquorumcertificate}). Immediately before adding $\rp{\hat i, \hat c}$ to $certificates$, however, $\process$ adds $\rp{\hat i, \tilde \process}$ to $directory$ (line \ref{line:ddirectoryadd}). This proves that $\tilde \process = \hat \process$ and concludes the lemma.
\end{proof}
\end{lemma}

\begin{theorem}
\diral satisfies transferability.
\begin{proof}
Let $\process, \process'$ be correct processes, let $\hat i$ be an id such that $a = \dirin.export\rp{\hat i}$ at $\process$ is an assignment. Let
\begin{align*}
    \tilde i &= a.id \\
    \hat \process &= a.process \\
    \hat c &= a.certificate
\end{align*}
By lines \ref{line:ddreturnassigment} and \ref{line:ddprocedureexport}, we immediately have $\tilde i = \hat i$. Moreover, by line \ref{line:ddcertificateincerticatesid} we have $\hat c = certificates\qp{\hat i}$ at $\process$. Finally, by line \ref{line:ddequationkeycardid} we have $\rp{\hat i, \hat \process} \in directory$ at $\process$. By Lemma \ref{lemma:directorycertificatesarecorrect}, we then have that $\hat c$ is a quorum certificate for $\qp{\texttt{Assignment}, \hat i, \hat \process}$.

\medskip

Upon invoking $\dirin.import\rp{a}$, $\process'$ verifies that $\hat c$ is indeed a quorum certificate for $\qp{\texttt{Assignment}, \hat i, \hat \process}$ (line \ref{line:dverifyquorumcertificate}), then adds $\rp{i, \any}$ to $directory$ (line \ref{line:ddirectoryadd}). By Notation \ref{notation:diraldirectoryrecord} and Definition \ref{definition:directoryrecord}, $\process'$ knows $\hat i$, and the theorem is proved.
\end{proof}
\end{theorem}

\subsubsection{Density}

In this section, we prove that \diral satisfies density.

\begin{lemma}
\label{lemma:rankingsareboundbyprocesses}
Let $\server$ be a correct server, let $\hat \server$ be a server. We have $\abs{rankings\qp{\hat \server}} < \abs{\processes}$ at $\server$.
\begin{proof}
Upon initialization, $rankings\qp{\hat \server}$ is empty at $\server$ (line \ref{line:drankingsinit}). Moreover, $\server$ adds a process $\process$ to $rankings\qp{\hat \server}$ (line \ref{line:dsrankingsupdate} only) only if $\process$ is not already in $rankings\qp{\hat \server}$ (line \ref{line:dsprocessnotinranking}). This proves that, at $\server$, the elements of $rankings\qp{\hat \server}$ are distinct processes. We then have $\abs{rankings\qp{\hat \server}} < \abs{\processes}$ at $\server$, and the lemma is concluded.
\end{proof}
\end{lemma}

\begin{lemma}
\label{lemma:nondenseidsarenotassigned}
Let $\rp{d, n}$ be an id, let $\process$ be a process. If a quorum certificate exists for $\qp{\texttt{Assignment}, \rp{d, n}, \process}$, then $n < \abs{\processes}$.
\begin{proof}
Let us assume that a quorum certificate exists for $\qp{\texttt{Assignment}, \rp{d, n}, \process}$. Noting that at most $f$ servers are Byzantine, at least one correct server $\server$ signed $\qp{\texttt{Assignment}, \rp{d, n}, \process}$. $\server$ does so (line \ref{line:dsmultisihnassignment} only) only if $n < \abs{rankings\qp{\hat \server}}$, for some server $\hat \server$ (line \ref{line:dsextractindex}). By Lemma \ref{lemma:rankingsareboundbyprocesses} we then have $n < \abs{\processes}$, and the lemma is proved.
\end{proof}
\end{lemma}

\begin{theorem}
\diral satisfies density.
\begin{proof}
Let $\process$ be a correct process, let $t \in \mathbb{R}$, let $\rp{\rp{\hat d, \hat n}, \hat \process} \in \directoryrecord{\process}[t]$. By Lemma \ref{lemma:directoryrecordiscertifiable} we have $\rp{\rp{\hat d, \hat n}, \hat \process} \in \directorycert$. By Definition \ref{definition:certifiableassignments} and Lemma \ref{lemma:nondenseidsarenotassigned} we then have $n < \abs{\processes}$, and the theorem is proved.
\end{proof}
\end{theorem}

\clearpage

\section{Pseudocode}
\label{appendix:pseudocode}

\subsection{\csbal Client}
\label{subsection:pseudocode.csbalclient}
\begin{lstlisting}
implements:
    %\clab%, instance %\clin%


uses:
    %\dirab%, instance %\dirin%
    %PerfectPointToPointLinks%, instance %pl%
    
    
parameters:
    %\csbalbatchingwindow%: Interval // Batching window


struct Submission:
    message: Message,
    signature: Signature,
    submitted_to: {Broker},
    included_in: {Root}


upon <%\clin%.Init>:
    trigger <%\dirin%.Signup>; %\label{line:csignup}%
    await <%\dirin%.SignupComplete | id>; %\label{line:csignupcomplete}%
    
    submissions: {Context: Submission} = {}; %\label{line:csubmissionsinit}%
    completed: {Root} = {}; %\label{line:ccompleteddefine}%


upon <%\clin%.Broadcast | context, message>: %\label{line:cbrodcastmessage}%
    signature = sign([Message, context, message]); %\label{line:csignaturemessage}%
    
    %\StartMultiline%submissions[context] = Submission {
        message, signature, 
        submitted_to: {}, included_in: {}
    };%\EndMultiline% %\label{line:csubmissionsupdate}%
       
    submit(context): %\label{line:csubmitinvoke}%
    
    
procedure submit(context): %\label{line:cproceduresubmit}%
    %\StartMultiline%if let submission = submissions[context] 
      and let %$\broker$% in (%$\brokers$% \ submission.submitted_to):%\EndMultiline% %\label{line:csubmissioncheck}%
        assignment = dir.export(id); %\label{line:cdirexport}%
        %\StartMultiline%trigger <pl.Send | %$\broker$%, 
          [Submission, assignment, (context, submission.message, submission.signature)]>;%\EndMultiline% %\label{line:csendsubmit}%
        
        submission.sumbitted_to.add(%$\broker$%);%\label{line:csubmissionupdate}%
        trigger <timer.Set | [Submit, context], 13 + %\csbalbatchingwindow%>; %\label{line:ctimerset}%
        
        
upon <timer.Ring | [Submit, context]>: %\label{line:ctimerring}%
    submit(context); %\label{line:cinvokesubmitring}%
    
    
upon <pl.Deliver | %$\broker$%, [Inclusion, context, root, proof]>: %\label{line:creciveinclusion}%
    if let submission = submissions[context]: %\label{line:cexistsubmissionforinclusion}%
        if verify(root, proof,(id, context, submission.message)): %\label{line:cverifymerkleproof}%
            submission.included_in.add(root); %\label{line:cupdateincludedin}%
            multisignature = multisign([Reduction, root]); %\label{line:cmultisignreduction}%
            trigger <pl.Send | %$\broker$%, [Reduction, root, multisignature]>; %\label{line:csendreduction}%
		    

upon <pl.Deliver | %$\broker$%, [Completion, root, exclusions, certificate]>:%\label{line:clientdelivercompletion}%
    %\StartMultiline%if verify_plurality(certificate, [Completion, root, exclusions]) 
      and id not in exclusions:%\EndMultiline% %\label{line:cverifyplurality}%
        completed.add(root); %\label{line:ccomplrtrdupdate}%


%\StartMultiline% upon exists (context, submission) in submissions such that
  (submission.included_in %$\cap$% completed) != {}: %\EndMultiline%
    submissions.remove(context); %\label{line:csubmissionsremove}%

\end{lstlisting}

\clearpage

\subsection{\csbal Broker}
\label{subsection:pseudocode.csbalbroker}
\begin{lstlisting}
implements:
    %\bkab%, instance %\bkin%
    

uses:
    %\dirab%, instance %\dirin%
    %PerfectPointToPointLinks%, instance %pl%


parameters:
    %\csbalbatchingwindow%: Interval // Batching window
    

struct Submission:
    context: Context,
    message: Message,
    signature: Signature


enum Batch:
    payloads: {Id: (Context, Message)},
    signatures: {Id: Signature},
    reductions: {Id: MultiSignature},
    
    commit_to: {Server},
    committable: bool,

    variant Reducing:
        (empty)
        
    variant Witnessing:
        witnesses: {Server: MultiSignature}

    variant Committing:
        commits: {Server: ({Id}, MultiSignature)}
        
    variant Completing:
        exclusions: {Id},
        completions: {Server: MultiSignature}


upon <%\bkin%.Init>:
    pending: {Id: [Submission]} (default []) = {}; %\label{line:bpandingdefinition}%
    pool: {Id: Submission} = {}; %\label{line:bpoolinit}%
    collecting: bool = false; %\label{line:bcollectingset}%
    
    batches: {Root: Batch} = {}; %\label{line:bbatchesinit}%
    

upon <pl.Deliver | %$\client$%, [Submission, assignment, (context, message, signature)]>: %\label{line:bdeliversubmit}%
    %\dirin%.import(assignment); %\label{line:bimportsubmit}%
    
    if %$\client$%.verify(signature, [Message, context, message]) and let id = %\dirin%[%$\client$%]: %\label{line:bverifyclientsignature}%
        pending[id].push_back(Submission {context, message, signature}); %\label{line:bpushback}%
        

upon exists id in pending such that pending[id] != [] and id not in pool: %\label{line:bexistsinpanding}%
    submission = pending[id].pop_front(); %\label{line:bpandingupdate}%
    pool[id] = submission; %\label{line:bpoolupdate}%
    
    
upon pool != {} and collecting = false: %\label{line:bpoolcollectingcheck}%
    collecting = true; %\label{line:bcollectingtrue}%
    trigger <timer.Set | [Flush], %\csbalbatchingwindow% + 1>; %\label{line:btimerset}%
    
    
upon <timer.Ring | [Flush]>: %\label{line:btimerring}%
    collecting = false; %\label{line:bcollectingfalse}%
    
    submissions = pool; %\label{line:bsubmissionsequalpool}%
    pool = {}; %\label{line:bemptypool}%
    
    %\StartMultiline%leaves = [(id, context, message )
      for (id, Submission {context, message, ..}) in submissions];%\EndMultiline% %\label{line:bleavescostruction}%
      
    tree = merkle_tree(leaves); %\label{line:btreecostruction}%
    root = tree.root(); %\label{line:brootcostruction}%
    
    for (id, Submission {context, message, ..}) in submissions:
        %$\client$% = %\dirin%[id];
        proof = tree.prove((id, context, message)); %\label{line:binclusionproof}%
        trigger <pl.Send | %$\client$%, [Inclusion, context, root, proof]>;
        
    %\StartMultiline%payloads = {id: (context, message)
      for (id, Submission {context, message, ..}) in submissions};%\EndMultiline% %\label{line:bpayloaddefine}%
    
    signatures = {id: signature for (id, Submission {signature, ..}) in submissions}; %\label{line:bsignaturedefine}%
    
    %\StartMultiline%batches[root] = Reducing {
                        payloads, signatures, 
                        reductions: {}, 
                        commit_to: {}, 
                        committable: false
                    }; %\EndMultiline% %\label{line:bbatchesupdate}%
                    
    trigger <timer.Set | [Reduce, root], 2>; %\label{line:btimersetreducing}%
    

upon <pl.Deliver | %$\client$%, [Reduction, root, multisignature]>: %\label{line:brecivereduction}%
    %\StartMultiline%if let batch alias batches[root] and batch is Reducing 
      and let id = %\dirin%[%$\client$%] and id in batch.payloads:%\EndMultiline% %\label{line:bcheckreducingid}%
        if %$\client$%.multiverify(multisignature, [Reduction, root]): %\label{line:bcheckmultisignereduction}%
            batch.reductions[id] = multisignature; %\label{line:bupdatereduction}%
            batch.signatures.remove(id); %\label{line:bupdateremoveid}%
            
            
upon <timer.Ring | [Reduce, root]>: %\label{line:breducering}%
    batch alias batches[root]; %\label{line:breducefetchbatch}%
    compressed_ids = compress(batch.payloads.keys()); %\label{line:bcompressids}%
    payloads = batch.payloads.values() %\label{line:bpayloads}%
        
    for %$\server$% in %$\servers$%: %\label{line:bserverinserver}%
        trigger <pl.Send | [Batch, compressed_ids, payloads]>; %\label{line:bsendbacth}%
    
    batch = Witnessing {witnesses: {}, ..batch}; %\label{line:bbatchinwitnessing}%
    trigger <timer.Set | [Committable, root], 4>; %\label{line:bcommitabletimer}%
    

upon event <timer.Ring | [Committable, root]>: %\label{line:bcommitabletimerring}%
    if let batch = batches[root]: %\label{line:bhavebacthring}%
        batch.committable = true;
    
    
upon <pl.Deliver | %$\server$%, [BatchAcquired, root, unknowns]>: %\label{line:bdeliverbatchacquired}%
    if let batch = batches[root]: %\label{line:bbatchacquiredgetbatch}%
        if exists unknown in unknowns such that unknown not in %\dirin%: %\label{line:bbatchacquiredcheckids}%
            return;
        
        assignments = %\dirin%.export(unknowns..); %\label{line:bbatchacquiredexportassignments}%
        
        multisignature = aggregate(batch.reductions.values()); %\label{line:bbatchacquiredaggregate}%
        signatures = batch.signatures; %\label{line:bbatchacquiredcopysignatures}%
        
        %\StartMultiline%trigger <pl.Send | %$\server$%, 
          [Signatures, root, assignments, multisignature, signatures]>;%\EndMultiline% %\label{line:bsendsignatures}%


upon <pl.Deliver | %$\server$%, [WitnessShard, root, shard]>: %\label{line:bdeliverwitnessshard}%
    if let batch alias batches[root]: %\label{line:brootandcommitable}%
        batch.commit_to.add(%$\server$%); %\label{line:baddtocommitto}%
        
        if batch is Witnessing: %\label{line:bwitnessshardcheckbatch}%
            if %$\server$%.multiverify(shard, [Witness, root]): %\label{line:bwitnessshardchecksignature}%
                batch.witnesses[%$\server$%] = shard; %\label{line:bstorewitnessshard}%
            
            
%\StartMultiline%upon exists root in batches such that batches[root] is Witnessing 
  and |batches[root].witnesses| >= f + 1:%\EndMultiline% %\label{line:binwitnessing}%
    batch alias batches[root]; %\label{line:binwitnessingfetchbatch}%
    certificate = aggregate(batch.witnesses); %\label{line:baggregatewitnesses}%
    
    for %$\server$% in %$\servers$%: %\label{line:bbroadcastwitnessesloop}%
        trigger <pl.Send | [Witness, root, certificate]>; %\label{line:bbroadcastwitnesses}%
        
    batch = Committing {commits: {}, ..batch}; %\label{line:bcommiting}%


upon <pl.Deliver | %$\server$%, [CommitShard, root, conflicts, commit]>: %\label{line:bdelivercommitshard}%
    if let batch alias batches[root] and batch is Committing: %\label{line:bcommitshardfetchbatch}%
        if !%$\server$%.multiverify(commit, [Commit, root, conflicts.keys()]): %\label{line:bmultiverifycommit}%
            return; %\label{line:bmultiverifycommitreturn}%
            
        %\StartMultiline%for (id, (conflict_root, conflict_witness, conflict_proof, conflict_message)) %\label{line:bcommitshardloop}%
          in conflicts:%\EndMultiline%
            if id not in batch.payloads: %\label{line:bcommitshardcheckid}%
                return;
                
            (context, message) = batch.payloads[id]; %\label{line:bextraxtxontexmessage}%
             
            if not verify_plurality(conflict_witness, [Witness, conflict_root]): %\label{line:bverifyconflictroot}%
                return;
                
            %\StartMultiline%if not verify(conflict_root, conflict_proof, 
              (id, context, conflict_message)):%\EndMultiline% %\label{line:bverifyconflictproof}%
                return;
                
            if message = conflict_message: %\label{line:bverifyconflictmessage}%
                return;
                
        batch.commits[%$\server$%] = (conflicts.keys(), commit); %\label{line:bstorecommit}%
        
        
%\StartMultiline%upon exists root in batches such that batches[root] is Committing 
    and batches[root].committable and |batches[root].commits| >= 2f + 1:%\EndMultiline% %\label{line:bpluralitycommitingcheck}%
   
    batch alias batches[root];
    patches: {{Id}: {MultiSignature}} (default {}) = {}; %\label{line:bpatchesinit}%
    
    for (_, (conflicts, commit)) in batch.commits: %\label{line:blooponthecommits}%
        patches[conflicts].add(commit); %\label{line:bpatchesupdate}%
        
    patches = {ids: aggregate(signatures) for (ids, signatures) in patches}; %\label{line:patchesaggregation}%
    
    for %$\server$% in batch.commit_to: %\label{line:bselectcommittoserver}%
        trigger <pl.Send | %$\server$%, [Commit, root, patches]>; %\label{line:bsendtocommittoserver}%
        
    exclusions = union(patches.keys());
    batch = Completing {exclusions, ..batch}; %\label{line:bcompleting}%


upon <pl.Deliver | %$\server$%, [CompletionShard, root, completion]>:
    if let batch alias batches[root] and batch in Completing:
        if multiverify(completion, [Completion, root, batch.exclusions]):
            batch.completions[%$\server$%] = completion
            
            
%\StartMultiline%upon exists root in batches such that batches[root] is Completing
  and |batches[root].completions| >= f + 1:%\EndMultiline% %\label{line:bbatchcompleting}%
    batch alias batches[root];
    certificate = aggregate(batch.completions);
    
    for id in payloads:
        %$\client$% = %\dirin%[id];
        trigger <pl.Send | %$\client$%, [Completion, root, batch.exclusions, certificate]>;
        
    batches.remove(root); %\label{line:sremoveroot}%
\end{lstlisting}

\clearpage

\subsection{\csbal Server}
\label{subsection:pseudocode.csbalserver}
\begin{lstlisting}
implements:
    %\srab%, instance %\srin%


uses:
    %\dirab%, instance %\dirin%
    %PerfectPointToPointLinks%, instance %pl%
    
    
struct Batch:
    ids: [Id],
    payloads: [(Context, Message)],
    tree: MerkleTree
    
    
upon event <%\srin%.Init>:
    batches: {Root: Batch} = {}; %\label{line:sinitbatches}%
    witnesses: {Root: Certificate} = {}; %\label{line:sinitwitnesses}%
    commits: {(Root, {Id}): {{Id}: Certificate}} = {}; %\label{line:sinitcommits}%
    
    messages: {(Id, Context): (Message, Root)} = {}; %\label{line:smessagesinit}%
    delivered: {(Id, Context)} = {}; %\label{line:sdeliverinit}%
    
    
procedure compress(ids):
    compressed_ids: {Domain: {Id}} (default {}) = {};
    
    for (domain, index) in ids:
        compressed_ids[domain].add(index)

    return compressed_ids

procedure expand(compressed_ids):
    ids = {};
    
    for (domain, indices) in compressed_ids:
        for index in indices:
            ids.push_back((domain, index));
    
    ids.sort();
    return ids;


procedure join(ids, payloads):
    %\StartMultiline%return [(id, context, message) 
      for (id, (context, message)) in zip(ids, payloads)]; %\EndMultiline%


procedure handle_batch(compressed_ids, payloads):
    ids = expand(compressed_ids); %\label{line:sexpandprocedure}%
    
    if ids.has_duplicates() or |ids| != |payloads|: %\label{line:sbatchcheck}%
        return %$\bot$%; %\label{line:sbatchcheckfail}%
        
    unknowns = {id for id in ids such that id not in %\dirin%};%\label{line:bcomputeunknowns}%
        
    leaves = join(ids, payloads); %\label{line:scomputeleaves}%
    tree = merkle_tree(leaves); %\label{line:scomputetree}%
    root = tree.root(); %\label{line:scomputeroot}%
    
    batches[root] = Batch {ids, payloads, tree}; %\label{line:sbatchesupdate}%
    return [BatchAcquired, root, unknowns]; %\label{line:sbatchacquired}%
  
    
upon event <pl.Deliver | %$\process$%, [Batch, compressed_ids, payloads]>: %\label{line:sbatchrecive}%
    response = handle_batch(compressed_ids, payloads);
    
    if response != %$\bot$% and %$\process$% in %$\brokers$%: %\label{line:sbatchreceivecheckresponse}%
        trigger <pl.Send | %$\process$%, response>; %\label{line:sbatchreceivesendresponse}%


procedure handle_signatures(root, assignments, multisignature, signatures):
    %\dirin%.import(assignments..); %\label{line:shandlesignaturesimportassignments}%
    
    if root not in batches: %\label{line:scheckbatcheshandlesignature}%
        return %$\bot$%; %\label{line:scheckbatcheshandlesignaturereturn}%
        
    Batch {ids, payloads, tree} = batches[root]; %\label{line:sbatchload}%
    
    if exists id in ids such that id not in %\dirin%: %\label{line:sdircheck}%
        return %$\bot$%; %\label{line:sdirreturn}%
    
    for (id, signature) in signatures: %\label{line:ssignsturesloop}%
        if id not in ids: %\label{line:ssignatureclientmissingcheck}%
            return %$\bot$%; %\label{line:ssignatureclientmissingreturn}%
            
        (context, message) = payloads[ids.index_of(id)]; %\label{line:stakepayloads}%
        
        if not %\dirin%[id].verify(signature, [Message, context, message]): %\label{line:ssignatureclientverification}%
            return %$\bot$%; %\label{line:ssignatureclientverificationreturn}%
            
    multisigners = ids \ signatures.keys(); %\label{line:sdeterminemultisigners}%
    
    if not %\dirin%[multisigners..].multiverify(multisignature, [Reduction, tree.root()]): %\label{line:smultiverifyreduction}%
        return %$\bot$%; %\label{line:smultiverifyreductionreturn}%
    
    shard = multisign([Witness, root]); %\label{line:ssignwitness}%
    return [WitnessShard, root, shard]; %\label{line:sreturnwitnessshard}%
    

upon <pl.Deliver | %$\broker$%, [Signatures, root, assignments, multisignature, signatures]>: %\label{line:sdeliversignatures}%
    response = handle_signatures(root, assignments, multisignature, signatures);
    
    if response != %$\bot$%: %\label{line:shandlesignaturescheckresponse}%
        trigger <pl.Send | %$\broker$%, response>; %\label{line:shandlesignaturessendresponse}%
        

procedure handle_witness(root, certificate):
    if root not in batches: %\label{line:srootinbatchesw}%
        return %$\bot$%; %\label{line:srootinbatchesreturnw}%
        
    Batch {ids, payloads, tree} = batches[root]; %\label{line:sbatchesnamew}%
    
    if !verify_plurality(certificate, [Witness, root]): %\label{line:switnesssignaturecheck}%
        return %$\bot$%; %\label{line:switnesssignaturecheckreturn}%
        
    witnesses[root] = certificate; %\label{line:sstorewitness}%
    conflicts: {Id: (Root, Certificate, MerkleProof, Message)} = {}; %\label{line:sinitializeconflicts}%
    
    for (id, context, message) in join(ids, payloads): %\label{line:scheckidsummision}%
        %\StartMultiline%if let (original_message, original_root) = messages[(id, context)]
          and original_message != message: %\EndMultiline\label{line:sconflictcheck}%
            original_witness = witnesses[original_root]; %\label{line:sfetchconflictwitness}%
            original_batch = batches[original_root]; %\label{line:soriginalbatch}%
            
            %\StartMultiline%Batch {
                ids: original_ids,
                payloads: original_payloads,
                tree: original_tree
            } = original_batch; %\EndMultiline% %\label{line:sextraxtoriginalbatch}%
            
            original_leaf = (id, context, original_message);%\label{line:soriginalleaf}%
            
            %\StartMultiline%conflicts[id] = (original_root, 
                             original_witness,
                             original_tree.prove(original_leaf), 
                             original_message
                            );%\EndMultiline% %\label{line:sconflictsupdate}%
        else:
            messages[(id, context)] = (message, root); %\label{line:updatevaraiablemessage}%
            
    commit = multisign([Commit, root, conflicts.keys()]); %\label{line:ssigncommit}%
    return [CommitShard, root, conflicts, commit];%\label{line:sreturncommitshard}%
    

upon <pl.Deliver | %$\broker$%, [Witness, root, certificate]>: %\label{line:sdeliverwitness}%
    response = handle_witness(root, certificate);
    
    if response != %$\bot$%: %\label{line:sresponseempty}%
        trigger <pl.Send | %$\broker$%, response>; %\label{line:striggertobroker}%


procedure handle_commit(root, patches):
    if root not in batches: %\label{line:srootinbatches}%
        return %$\bot$%; %\label{line:srootinbatchesreturn}%
        
    Batch {ids, payloads, ..} = batches[root]; %\label{line:sbatchesname}%
    
    if exists id in ids such that id not in %\dirin%: %\label{line:sdircheckcommit}%
        return %$\bot$%; %\label{line:sdirecheckcommitreturn}%
        
    signers: {Server} = {}; %\label{line:ssignerinit}%
    
    for (conflicts, certificate) in patches: %\label{line:spatches}%
        if not certificate.verify([Commit, root, conflicts]): %\label{line:svaliditysignatureconflicts}%
            return %$\bot$%; %\label{line:svaliditysignatureconflictsreturn}%
        
        signers.extend(certificate.signers()); %\label{line:ssignerextend}%
    
    if |signers| < 2f + 1: %\label{line:squorumcsignatureconflicts}%
        return %$\bot$%; %\label{line:squorumcsignatureconflictsreturn}%
        
    exclusions = union(patches.keys()); %\label{line:sunionexclusion}%
    commits[(root, exclusions)] = patches; %\label{line:scommitsupadte}%
    
    for (id, context, message) in join(ids, payloads): %\label{line:sdeliverfrombrokerloop}%
        keycard = %\dirin%[id]; %\label{line:stakekeycard}%
        
        %\StartMultiline%if id not in exclusions 
          and (keycard, context) not in delivered: %\EndMultiline% %\label{line:sdelivercheck}%
            delivered.add((keycard, context)); %\label{line:sdeliveredupdate}%
            trigger <%\srin%.Deliver| keycard, context, message}>; %\label{line:sdeliver}%
            
    trigger <timer.Set | [OfferTotality, root, exclusions], 7>; %\label{line:soffertotality}%
    
    shard = multisign([Completion, root, exclusions]); %\label{line:scompletionsign}%
    return [CompletionShard, root, shard];
    

upon <pl.Deliver | %$\process$%, [Commit, root, patches]>: %\label{line:sdelivercommitpatches}%
    response = handle_commit(root, patches);
    
    if response != %$\bot$% and %$\process$% in %$\brokers$%:
        trigger <pl.Send | %$\process$%, response>; %\label{line:ssendcompletionshard}%
        

upon <timer.Ring | [OfferTotality, root, exclusions]>: %\label{line:soffertotalityring}%
    for %$\server$% in %$\servers$%: %\label{line:sforallserver}%
        trigger <pl.Send | %$\server$%, [OfferTotality, root, exclusions]>; %\label{line:ssendoffertotality}%


upon <pl.Deliver | %$\server$%, [OfferTotality, root, exclusions]>: %\label{line:sdeliveroffertotality}%
    if (root, exclusions) not in commits: %\label{line:scommitsdelivercheck}%
        trigger <pl.Send | %$\server$%, [AcceptTotality, root, exclusions]>; %\label{line:ssendaccepttotality}%


upon <pl.Deliver| %$\server$%, [AcceptTotality, root, exclusions]>: %\label{line:sdeliveraccepttotality}%
    %\StartMultiline%if let batch = batches[root] %\label{line:sbatchinbatches}%
      and let patches = commits[(root, exclusions)]: %\EndMultiline%
        Batch {ids, payloads, ..} = batch;
        
        assignments = %\dirin%.export(ids..); %\label{line:sexportassigments}%
        compressed_ids = compress(ids);
        
        %\StartMultiline%trigger <pl.Send| %$\server$%, [Totality, root, assignments, 
          (compressed_ids, payloads), patches]>; %\EndMultiline\label{line:sserverbacth}%


upon <pl.Deliver | %$\server$%, [Totality, root, assignments, (ids, payloads), patches]>: %\label{line:sdelivertotality}%
        %\dirin%.import(assignments..); %\label{line:simportassigmentstotality}%
        
        handle_batch(ids, payloads); %\label{line:shandlebatchdeliver}%
        handle_commit(root, patches) %\label{line:shandlecommitsdeliver}%
\end{lstlisting}

\clearpage

\subsection{\diral}
\label{subsection:pseudocode.diral}
\begin{lstlisting}
implements:
    %\dirab%, instance %\dirin%
    

uses:
    %AuthenticatedPointToPointLinks%, instance %al%
    
    
struct Assignment:
    id: Id,
    process: Process,
    certificate: Certificate
    

enum Status(Outsider, SigningUp, SignedUp)
    
    
upon <%\dirin%.Init>:
    rankings: {Server: {Server}} (default {}) = {}; %\label{line:drankingsinit}%
    assigner: (Server or %$\bot$%) = %$\bot$%; %\label{line:dassignerinit}%
    assignments: {Integer: {Server: MultiSignature}} (default {}) = {}; %\label{line:dassigmentsinit}%
    status: Status = Outsider; %\label{line:dsignupoutsider}%
    
    directory: {(Id, Process)} = {}; %\label{line:ddirectoryinit}%
    certificates: {Id: Certificate} = {}; %\label{line:ddcertificatempty}%

    
upon <%\dirin%.Signup>: %\label{line:dsignup}%
    status = SigningUp; %\label{line:dstatussigninigup}%
    
    for %$\server$% in %$\servers$%: %\label{line:dserverinservers}%
        trigger <al.Send | %$\server$%, [Signup]>; %\label{line:dsendsignup}%

        
upon <al.Deliver | %$\server$%, [Ranked, source]>: %\label{line:dserverdeliverranked}%
    rankings[source].add(%$\server$%); %\label{line:daddservertorankings}%

    
upon exists source such that |rankings[source]| >= f + 1 and assigner = %$\bot$%: %\label{line:dcheckrankingquorum}%
    assigner = source; %\label{line:dassignerupdate}%
    
    for %$\server$% in %$\servers$%: %\label{line:dassignerloop}%
        trigger <al.Send | %$\server$%, [Assigner, assigner]>; %\label{line:dassignersend}%
        
        
upon event <al.Deliver | %$\server$%, [Assignment, index, signature]>: %\label{line:ddeliverassignment}%
    if %$\server$%.multiverify(signature, [Assignment, (assigner, index), self]): %\label{line:dverifyassignmentsignature}%
        assignments[index][%$\server$%] = signature; %\label{line:dassigmentsignaturedefinition}%
        
        
upon exists index such that |assignments[index]| >= 2f+1 and status = SigningUp: %\label{line:dchecksigningupstatuscheck}%
    status = SignedUp; %\label{line:dstatussignedup}%
    
    id = (assigner, index); %\label{line:dcompositionid}%
    certificate = aggregate(assignments[index]); %\label{line:ddefinecertificate}%
    
    %\dirin%.import(Assignment {id, process: self, certificate}); %\label{line:ddirinimport}%
    trigger <%\dirin%.SignupComplete>; %\label{line:dtriggersignupcomplete}%
    

procedure %\dirin%.[] (query): %\label{line:dinvokeprocedure}%
    if query is Id:
        if exists process such that (query, process) in directory: %\label{line:dcheckkeycardindirectory}%
            return process; %\label{line:dreturnkycard}%
    else if query is Process:
        if exists id such that (id, query) in directory: %\label{line:dcheckidindirectory}%
            return id; %\label{line:dreturnid}%
    
    return %$\bot$%;
    
    
procedure %\dirin%.export(id): %\label{line:ddprocedureexport}%
    if let certificate = certificates[id]: %\label{line:ddcertificateincerticatesid}%
        process = %\dirin%[id]; %\label{line:ddequationkeycardid}%
        return Assignment {id, process, certificate}; %\label{line:ddreturnassigment}%
    else:
        return %$\bot$%; %\label{line:ddreturnassigmentreturn}%
        

procedure %\dirin%.import(assignment): %\label{line:dcallproceduredirimport}%
    %\StartMultiline%if assignment.certificate.verify_quorum(
        [Assignment, assignment.id, assignment.process]
    ):%\EndMultiline% %\label{line:dverifyquorumcertificate}%
        directory.add((assignment.id, assignment.process)); %\label{line:ddirectoryadd}%
        certificates[assignment.id] = assignment.certificate; %\label{line:ddaddidtocerticates}%
\end{lstlisting}

\clearpage

\subsection{\diral Server}
\label{subsection:pseudocode.diralserver}
\begin{lstlisting}
implements:
    %\dirsrab%, instance %\dirsrin%
    

uses:
    FifoBroadcast, instance fb
    %AuthenticatedPointToPointLinks%, instance %al%
    
    
upon <%\dirsrin%.Init>:
    rankings: {Server: [KeyCard]} (default []) = {};  %\label{line:dsrankingsinit}%
    assigners: {KeyCard: Server} = {}; %\label{line:dsassignersinit}%
    certified: {KeyCard} = {}; %\label{line:dscertifiedinit}%
    

upon <al.Deliver | %$\process$%, [Signup]>: %\label{line:dsdeliversignup}%
    trigger <fb.Broadcast | [Rank, %$\process$%]>; %\label{line:dsbrodcastrank}%
    

upon <fb.Deliver | %$\server$%, [Rank, process]>: %\label{line:dsdeliverrankmessage}%
    if process not in rankings[%$\server$%]: %\label{line:dsprocessnotinranking}%
        rankings[%$\server$%].push_back(process); %\label{line:dsrankingsupdate}%
        trigger <al.Send | process, [Ranked, %$\server$%]> %\label{line:dssendranked}%
        

upon <al.Deliver | %$\process$%, [Assigner, assigner]>: %\label{line:dsdeliverassigner}%
    if %$\process$% not in assigners: %\label{line:dsprocessnotinassigners}%
        assigners[%$\process$%] = assigner; %\label{line:dsaaigneres}%
        

%\StartMultiline%upon exists process such that (process in assigners) 
    and (process in rankings[assigners[process]]) 
    and (process not in certified):%\EndMultiline% %\label{line:dsreadytocertify}%
    certified.add(process); %\label{line:dsaddtocertified}%
    
    assigner = assigners[process]; %\label{line:dsassignersextraxt}%
    index = rankings[assigner].index_of(process); %\label{line:dsextractindex}%
    
    signature = multisign([Assignment, (assigner, index), process]); %\label{line:dsmultisihnassignment}%
    trigger <al.Send | process, [Assignment, index, signature]>; %\label{line:dssendassignment}%
    
\end{lstlisting}
\end{appendix}
\end{document}